\documentclass{IEEEtran}
\usepackage{cite}
\usepackage{amsmath,amssymb,amsfonts}
\usepackage{graphicx}
\usepackage{epstopdf}
\usepackage{textcomp}
\usepackage{algorithm}
\usepackage{algpseudocode}
\usepackage{enumitem}
\usepackage{multirow}
\usepackage{verbatimbox}
\usepackage{booktabs}
\usepackage{hyperref}
\usepackage{graphicx}
\usepackage{amsmath}
\DeclareMathOperator{\Tr}{Tr}
\usepackage{amssymb}
\usepackage{amsthm}
\usepackage{color}
\usepackage{caption2}
\theoremstyle{definition}
\newcommand{\tabincell}[2]{\begin{tabular}{@{}#1@{}}#2\end{tabular}}

\usepackage{epsfig}

\newtheorem{theorem}{Theorem}
\newtheorem{lemma}{Lemma}

\def\BibTeX{{\rm B\kern-.05em{\sc i\kern-.025em b}\kern-.08em
    T\kern-.1667em\lower.7ex\hbox{E}\kern-.125emX}}
\begin{document}
	\title{Fast Accurate Beam and Channel Tracking for Two-dimensional Phased Antenna Arrays}
	
	\author{Yu Liu$^{*\S}$, Jiahui Li$^{*\S}$, Xiujun Zhang$^\S$, Shidong Zhou$^{*\S}$\\
	$^*$Dept. of EE,  Tsinghua University, Beijing, 100084, China\\
	$^\S$National Laboratory for Information Science and Technology, Tsinghua University, Beijing, 100084, China}
\maketitle

\begin{abstract}
 The sparsity and the severe attenuation of millimeter-wave (mmWave) channel imply that highly directional communication is needed. The narrow beam produced by large array requires accurate alignment, which is difficult to achieve when serving fast-moving users. In this paper, we focus on accurate two-dimensional (2D) beam and channel tracking problem aiming at minimizing exploration overhead and tracking error. Using a typical frame structure with periodic exploration and communication, a proven minimum overhead of exploration is provided first. Then tracking algorithms are designed for three types of channels with different dynamic properties. It is proved that the algorithms for quasi-static channels and channels in Dynamic Case I are optimal in approaching the minimum Cram\'{e}r-Rao lower bound (CRLB). The computational complexity of our algorithms is analyzed showing their efficiency, and simulation results verify their advantages in both tracking error and tracking speed.
\end{abstract}


\section{Introduction}\label{sec_introductionl}
\vspace{-0mm} 

Millimeter-wave (mmWave) mobile communication is currently a hot topic due to its much wider bandwidth compared with the sub-6GHz spectrum. In mmWave channels, the much higher frequency leads to severe propagation loss, atmospheric absorption, penetration loss and other obstructions \cite{Xiao2017Millimeter}. Fortunately, the shorter wavelength in the mmWave band allows the deployment of a larger antenna array, providing a considerable beamforming gain to compensate for the path loss \cite{Pi2011An,Larsson2014massive,Han2015Large,Heath2016overview,Molisch2017Hybrid}. For a hybrid or analog beamforming (ABF) system as cost-efficient ways to obtain this array gain, misalignment of beam direction may not only degrade the effective receiving power, increasing mutual interference, but also lead to the loss of beam observations due to the users' mobility, especially in fast-varying environments \cite{Gao2017Fast,Zhang19Codebook}. Therefore, accurate beam tracking is crucial for serving fast-moving users in mmWave mobile communication system.

In this paper, we will focus on the problem in ABF. Since only one RF chain connected with the antennas via programmable phase shifters is available in ABF, only one set of phase shifts can be applied (forming a so-called \textbf{exploring beamforming vector} (\textbf{EBV}) in this paper) and one dimension of the multiple-antenna channel can be observed at a certain time. Hence, in order to estimate the direction and the gain of the beam, the transceiver needs to try several different EBVs one by one. These EBVs can have a significant impact on tracking performance \cite{JLiAnalogbeamtracking2017,JLiJoint2018ICCASP,Garcia2018Optimal}. 

Although there already exist some beam tracking methods in \cite{Alkhateeb2015Compressed,IEEE80211ad,Rial2016Hybrid,Vutha2016Tracking,Gao2017Fast,3GPP19}, which utilized historical exploring directions
and observations to obtain current estimates, the EBVs were not optimized in those tracking algorithms. While beamforming resulting in the highest combining signal-to-noise ratio (SNR) is the best for data transmission, it is not the best for tracking accuracy \cite{JLiJoint2018ICCASP,Garcia2018Optimal}. Optimal design of EBVs is necessary to achieve as accurate beam alignment as possible in mmWave mobile communication.

In \cite{JLiAnalogbeamtracking2017}, a beam tracking algorithm was proposed, trying to optimize the EBVs, assuming that the channel gain is known. In \cite{JLiJoint2018ICCASP}, the authors started to jointly track the channel gain and the beam direction with optimal EBVs. In \cite{Garcia2018Optimal}, the optimization of the EBVs is converted to a convex problem and solved by online optimization toolboxes. Despite the progress, only one-dimensional (1D) array is supported for beam tracking optimization in these works. However, in most mobile applications, two-dimensional (2D) arrays are necessary, not only for providing much higher array gain, but also for supporting both horizontal and vertical beam direction variation \cite{Rappaport2013CompactBroadband,Brown2016Promise}. It brings huge challenges when extending the optimal EBVs in \cite{JLiJoint2018ICCASP,Garcia2018Optimal} to 2D arrays. The optimization of the exploring directions in \cite{JLiJoint2018ICCASP} relies on the symmetrical property of the two exploring beams, however, keeping this symmetrical property may result in more overhead when extended to 2D tracking. As for the algorithm in \cite{Garcia2018Optimal}, the objective function versus the EBVs becomes extremely complicated and quite different in 2D tracking compared with 1D tracking, leading to the failure of the previously used optimization method in 1D system. To the best of the authors' knowledge, there is no work on the design of optimal EBVs for 2D array yet.

\begin{figure}[!t]
	\centering
	\vspace{-0mm}
	\includegraphics[width=3.3in]{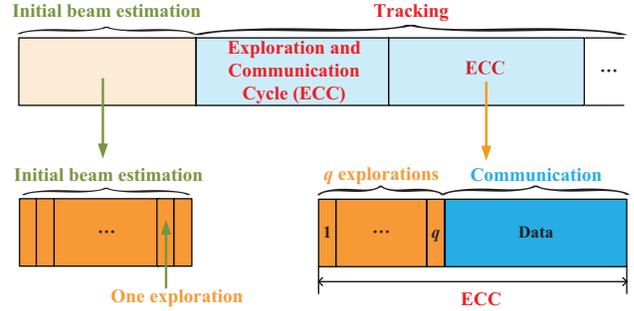}
	\vspace{-0mm}
	\caption{The frame structure for tracking.}
	\vspace{-0mm}
	\label{FrameStrcuctureTVT}
\end{figure}


In this paper, we focus on the design of the optimal EBVs and the accurate single-path tracking algorithms based on 2D phased antenna array. The widely used frame structure \cite{JLiJoint2018ICCASP,Zhang19Codebook,Boljanovic18Tracking} is adopted here. As shown in Fig. \ref{FrameStrcuctureTVT}, the transceiver periodically works in exploration and communication mode. In the exploration stage of each \textbf{exploration and communication cycle (ECC)}, the transmitter sends a pre-defined pilot sequence for $q$ times. At each time, the receiver forms one exploring beam pointing in one direction to observe the channel. Then the channel gain and the direction of the incoming beam are estimated with the $q$ observations of the channel. In the communication stage of each ECC, the beam is aligned in the current estimated direction, and the current estimated channel gain will be used for the subsequent process. Based on this structure, the following questions are to be answered:

1) What is the minimum exploration overhead $q$ in each ECC for 2D tracking?

2) How to determine the $q$ exploring directions based on the channel observations in previous ECCs for 2D tracking?

3) How to track the 2D beam direction and the channel gain for different time-varying channels, e.g., from quasi-static channels to fast-fading channels? 


4) How is the accuracy, convergence and stability of the tracking algorithm?

Following these questions, we summarize the main contributions of this paper as below:

1) Based on a reasonable EBV constraint, it is proved that the minimum exploration overhead counted by the number of exploring directions is $q=3$, for a unique solution of the 2D beam direction and the channel gain within only one ECC, while simple extension from 1D to 2D tracking will need $q=4$.

2)  Dynamic beam and channel tracking strategies for three different time-varying channels (called \textbf{Quasi-static Case}, \textbf{Dynamic Case I} and \textbf{Dynamic Case II} in this paper) are proposed and optimized. The salient advantages of these tracking algorithms are given below:

i) In Quasi-static Case (channels with quasi-static beam direction and channel gain), the optimal exploration offsets are derived. Also, a joint beam direction and channel gain tracking algorithm is proposed, and the tracking error is proved to converge to the minimum \textbf{Cram\'{e}r-Rao lower bound} (\textbf{CRLB}).

ii) In Dynamic Case I (channels with quasi-static beam direction and fast-fading channel gain), the Rayleigh fading channel is studied as a special case in this paper. The optimal exploration offsets are obtained and an algorithm for beam (only) tracking is proposed, which is proved to converge and achieve the minimum CRLB on the beam direction. 

iii) In Dynamic Case II (channels with fast-changing beam direction and channel gain), a joint tracking  algorithm of the beam direction and the channel gain is proposed with faster and more accurate performance.

3) The impact of the antenna pattern on the tracking algorithms and the performance is taken into account, showing that the proposed algorithms are suitable for practical implementations.

Part of this work was presented in our conference paper \cite{YuLiu18}, while the main difference and novelty of this paper lies in the following four aspects: 1) tracking with a general antenna pattern rather than a simple isotropic pattern: we consider a more general direction-dependent antenna element pattern here rather than an isotropic pattern; 2) tracking for different types of time-varying channels: in addition to slow-fading channels, fast-fading channels are also studied in this paper and the corresponding tracking strategy is proposed and optimized, while the prior algorithm does not support fast-fading channels; 3) the complexity analysis:
we analyze the computational complexity of the proposed tracking algorithms while this was missing in our previous work; 4) more rigorous conclusions and more complete proofs: we provide more rigorous lemmas and theorems with more complete proofs in this paper, correcting the corresponding flaws in the previous work.

The remaining part of this paper is organized as follows: the system model is described in Section \ref{sec_model}. In Section \ref{sec_problem}, the tracking problem with some constraints is formulated. Then the minimum exploration overhead of joint 2D beam and channel tracking is given in theory in Section \ref{sec_PilotOverhead}. In Section \ref{sec_Quasi-static_Tracking} and Section \ref{sec_BeamTrackingDI}, the tracking problems for Quasi-static Case (Section \ref{sec_Quasi-static_Tracking}) and Dynamic Case I (Section \ref{sec_BeamTrackingDI}) are studied separately. The tracking performance bounds are derived and corresponding tracking algorithms are developed with convergence and optimality analysis. In Section \ref{sec_BeamtrackingDII}, a tracking algorithm is developed for Dynamic Case II. Then the complexity analysis of these algorithms is given in Section \ref{sec_ComputationComplexity}. Section \ref{sec_simulation} presents numerical results to verify the performance of our proposed algorithms.

\emph{Notations}: We use lower case letters such as $a$ and $\textbf{a}$ to denote scalars and column vectors. Respectively, $\lvert\textbf{a}\rvert$ and $\left\|\textbf{a}\right\|_2$ represent the modulus and 2-norm of the vector $\textbf{a}$. Upper case boldface letters, e.g., $\textbf{A}$, are used to denote matrices. The superscript $\bar{\left(\cdot\right)}$, $\left(\cdot\right)^\text{T}$, $\left(\cdot\right)^\text{H}$ are utilized to denote conjugate, transpose and conjugate-transpose. For a matrix $\textbf{A}$, its inverse, pseudo-inverse and determinant are written as $\textbf{A}^{-1}$, $\textbf{A}^{+}$ and $\lvert \textbf{A}\rvert$. The identity matrix of order $q$ is denoted by $\textbf{J}_q$. Let $\mathcal{CN}(\mu,\sigma^2)$ represent the symmetric complex Gaussian distribution with mean $\mu$ and variance $\sigma^2$, and $\mathcal{N}(\mu,\sigma^2)$ stand for the real Gaussian distribution with mean $\mu$ and variance $\sigma^2$. 	The Kronecker product is represented as $\otimes$. The statistical expectation is denoted by $\mathbb{E}\left[\cdot\right]$. The real (imaginary) part is represented as $\text{Re}\left\{\cdot\right\}$ $\left(\text{Im}\left\{\cdot\right\}\right)$. The natural logarithm of a scalar $y$ is obtained by $\log\left(\cdot\right)$ and the phase angle of a complex number $z$ is written as $\angle z$. The main acronyms used in this paper are summarized in TABLE \ref{Tab_acronyms}.

\renewcommand\arraystretch{1.2}
\begin{table}[!t!]
	\centering
	\vspace{-0mm}
	\caption{Summary of the main acronyms.}
	\vspace{2mm}
	\begin{tabular} {c|c}
		\hline
		\hline
         Cram\'{e}r-Rao lower bound  & CRLB\\
        \hline
        analog beamforming  & ABF\\
        \hline
        exploring beamforming vector  & EBV\\
        \hline
		exploration and communication cycle & ECC\\
		\hline
        direction parameter vector  & DPV\\
		\hline
		exploring beamforming matrix  & EBM\\
		\hline
		joint beam and channel tracking  & JBCT\\
		\hline
		recursive beam tracking & RBT\\
		\hline
	\end{tabular}\label{Tab_acronyms}
	\vspace{-0mm}
\end{table}

\vspace{-0mm}
\section{System Model}\label{sec_model}
\vspace{-0mm}
\subsection{System Configuration}\label{subsec_sys_config}
\begin{figure}[!t]
	\centering
	\vspace{-0mm}
\includegraphics[width=3.4in]{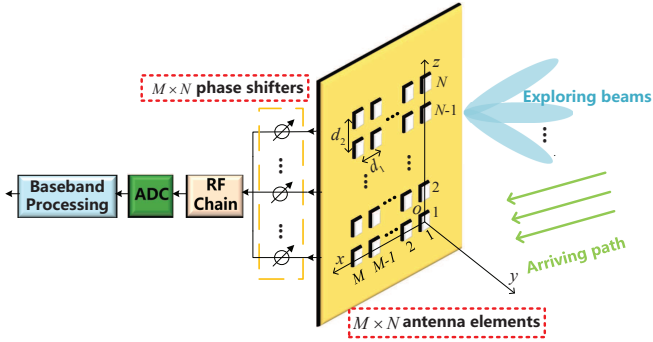}
	\vspace{-0mm}
	\caption{MmWave receiver with 2D phased antenna array.}
	\vspace{-0mm}
	\label{antenna}
\end{figure}

We consider a mmWave receiver\footnote{\,Note that tracking is needed at both the transmitter and the receiver. However, considering the transmitter-receiver reciprocity, the tracking of both sides have similar designs. Hence, we focus on beam and channel tracking at the receiver side.} equipped with a planar phased antenna array, as shown in Fig. \ref{antenna}. The planar array consists of $M \times N$ antenna elements that are placed in a rectangular area, where $M$($N$) antenna elements are evenly distributed along $x$-axis ($z$-axis) with a distance $d_1$ ($d_2$) between neighboring elements. These antenna elements are connected to the same RF chain via programmable phase shifters. 

Single RF chain of ABF makes a constraint that only one beam can be formed at any time and hence the receiver has to work alternatively in exploration and communication mode, resulting in a frame structure of periodic ECC. The angle of arrival (AoA) and the channel gain are assumed to be constant in each ECC and
may change in different ECCs. In the exploration stage of one ECC, the transmitter sends a pre-defined pilot sequence $\textbf{s}$ for $q$ times, where $\textbf{s}=\left[s_1,\cdots,s_{L_s}\right]\in \mathbb{C}^{1\times L_s}$ contains $L_s$ same symbols. At each time, the receiver forms one exploring beam pointing in one direction to observe the channel. Then the channel gain and the direction of the incoming beam are estimated according to the $q$ observations obtained in the current and previous ECCs. In the communication stage of each ECC, the beam is aligned in current estimated direction, and the current estimated channel gain will be used for the subsequent process.

\vspace{-0mm}
\subsection{Channel Model}\label{subsec_channel}
\vspace{-0mm}
In mmWave outdoor communication, the scattering is not rich and the number of effective propagation paths is usually limited \cite{Xiao2017Millimeter,Akdeniz14Millimeter}. Besides, the beam formed by a large array in the mmWave system is quite narrow and the interaction between multi-path is relatively weak \cite{zhu2017auxiliary}. In other words, the incoming paths are usually sparse in space, making it possible to track each path independently. Hence, we focus on the method for tracking one path, while different paths can be tracked separately by using the same method.
In $k$-th ECC, the direction of the incoming beam is denoted by ($\theta_k, \phi_k$), where $\theta_k\in[-\frac{\pi}{2},\frac{\pi}{2})$ is the elevation AoA and $\phi_k\in[0,\pi)$ is the azimuth AoA. Then the channel vector of this path during $k$-th ECC is
\vspace{-0mm}
\begin{equation}\label{eq_channelvectorini}
\textbf{h}_k = \eta\left(\theta_k,\phi_k\right)\beta_k^c \textbf{a}(\textbf{x}_k),
\end{equation}\vspace{-0mm}where $\eta\left(\theta,\phi\right)$ is the direction-dependent antenna gain (antenna pattern) of each element,\footnote{\,Note that the antenna patterns of different elements in the array may not be exactly the same and need to be carefully calibrated. However, considering that it is not the focus of this paper, we assume the same patterns for all the elements here.} $\beta_k^c$ is the complex channel gain, 
$\textbf{x}_k\triangleq\left[x_{k,1},x_{k,2} \right]^\text{T}=\big[\frac{M d_1 \cos(\theta_k) \cos(\phi_k)}{\lambda},$ $\frac{N d_2 \sin(\theta_k) }{\lambda} \big]^\text{T}$is the \textbf{direction parameter vector (DPV)} determined by ($\theta_k, \phi_k$),
\vspace{-0mm}
\begin{equation}\label{eq_steeringvector}
\textbf{a}(\textbf{x}_k)= \textbf{a}_1\left(x_{k,1}\right)\otimes\textbf{a}_2\left(x_{k,2}\right)
\end{equation}\vspace{-0mm}is the 2D steering vector with 
\vspace{-0mm}\begin{align}\label{eq_steeringvector1D}
&\textbf{a}_1({x}_{k,1})\triangleq\left[1,e^{j 2 \pi \frac{x_{k,1}}{M} },\cdots,e^{j 2 \pi \frac{M-1}{M} x_{k,1}}\right]^\text{T}\\
&\textbf{a}_2({x}_{k,2})\triangleq\left[1,e^{j 2 \pi \frac{x_{k,2}}{N} },\cdots,e^{j 2 \pi \frac{N-1}{N} x_{k,2}}\right]^\text{T},
\end{align}\vspace{-0mm}and $\lambda$ is the wavelength. For the convenience of expression, the antenna gain $\eta\left(\theta_k,\phi_k\right)$ is denoted by $\eta\left(\textbf{x}_k\right)$ hereinafter.

Define the equivalent channel gain in $k$-th ECC as below:
\vspace{-0mm}\begin{equation}\label{eq_beta}
\beta\left(\textbf{x}_k\right) \triangleq \eta\left(\textbf{x}_k\right)\beta_k^c,
\end{equation}\vspace{-0mm}then the channel vector in \eqref{eq_channelvectorini}
can be rewritten as
\begin{equation}\label{eq_channelvector}
\textbf{h}_k = \beta\left(\textbf{x}_k\right) \textbf{a}(\textbf{x}_k).
\end{equation}

\vspace{-0mm}
\subsection{RF and Baseband Preprocessing}\label{subsec_preprocessing}
\vspace{-0mm}
Synchronization in both carrier frequency
and symbol timing is necessary in mmWave wireless communications. In the initial beam estimation stage in Fig. \ref{FrameStrcuctureTVT}, the carrier frequency synchronization information can be obtained and estimated, the residual error of which can be converted to the time-varying phase of the equivalent channel gain in \eqref{eq_beta}. As for the symbol timing, since it changes much slower, it can be estimated and tracked much more easily both in the initial beam estimation stage and the tracking stage. There already exists a lot of work on the synchronization algorithms in mmWave communication systems \cite{Meng18Omnidirectional,Zhu19Directional,Zhu20Double}. To make the research goals more focused, we assume perfect synchronization in this paper. Future work may be needed to further study the impact of residual synchronization error on the beam tracking performance.


Next, we will focus on the receiving beamforming based on the perfect synchronization assumption above. Let $\textbf{w}_{k,i} \in \mathbb{C}^{MN \times 1}$ be the EBV for receiving the pilot sequence $\textbf{s}$ the $i$-th $(i=1,\cdots,q)$ time in $k$-th ECC. The entries of $\textbf{w}_{k,i}$ are of the same amplitude with $\bigg|\left[\textbf{w}_{k,i}\right]_{l}\bigg|=\frac{1}{\sqrt{MN}}$, where $\left[\textbf{w}_{k,i}\right]_{l}$ denotes the $l$-th element of $\textbf{w}_{k,i}$. After phase shifting and combining, the $i$-th received sequence in $k$-th ECC at the baseband output of the RF chain is given by
\vspace{-0mm}
\begin{equation}\label{eq_observationb}
\boldsymbol{\nu}_{k,i} =\beta\left(\textbf{x}_k\right) \textbf{w}_{k,i}^\text{H} \textbf{a}(\textbf{x}_k)\textbf{s}+\boldsymbol{\zeta}_{k,i}.
\end{equation}\vspace{-0mm}where $\boldsymbol{\zeta}_{k,i} \in \mathbb{C}^{1\times L_s}$ is the receiving noise vector.

By match filtering on the sequence $\boldsymbol{\nu}_{k,i}$, the $i$-th observation in $k$-th ECC is given below:
\vspace{-0mm}
\begin{equation}\label{eq_observation}
\begin{aligned}
{y}_{k,i} = \boldsymbol{\nu}_{k,i} \frac{\textbf{s}^\text{H}}{\lvert \textbf{s}\rvert}=& \beta\left(\textbf{x}_k\right) \textbf{w}_{k,i}^\text{H} \textbf{a}(\textbf{x}_k)\textbf{s}\frac{\textbf{s}^\text{H}}{\lvert \textbf{s}\rvert} + \boldsymbol{\zeta}_{k,i}\frac{\textbf{s}^\text{H}}{\lvert \textbf{s}\rvert} \\= &\lvert\textbf{s}\rvert\beta\left(\textbf{x}_k\right) \textbf{w}_{k,i}^\text{H} \textbf{a}(\textbf{x}_k)+{z}_{k,i},
\end{aligned}
\end{equation}\vspace{-0mm}where $z_{k,i} \triangleq \boldsymbol{\zeta}_{k,i}\frac{\textbf{s}^\text{H}}{\lvert \textbf{s}\rvert}$ is an additive noise, which is modeled as i.i.d. Gaussian distributed in this paper, i.e., $z_{k,i} \sim \mathcal{CN}\left(0,\sigma_z^2\right)$. This assumption is certainly held when the receiving noise vector $\boldsymbol{\zeta}_{k,i}$ is i.i.d. Gaussian distributed. Besides, even when the noise vector $\boldsymbol{\zeta}_{k,i}$ is non-Gaussian, if the real and imaginary parts of the elements in $\boldsymbol{\zeta}_{k,i}$ are i.i.d, the observation noise $z_{k,i}$ can also be regarded as i.i.d. Gaussian distributed as long as the pilot sequence length $L_s$ is sufficiently large, according to the central limit theorem \cite{LindleyLimit}.

Let $\textbf{W}_k\triangleq \left[\textbf{w}_{k,1},\ldots,\textbf{w}_{k,q}\right]$, $\textbf{z}_k \triangleq \left[z_{k,1},\ldots,z_{k,q}\right]^\text{T}$ and $\textbf{y}_k\triangleq \left[y_{k,1},\ldots,y_{k,q}\right]^\text{T}$ denote the \textbf{exploring beamforming matrix} (\textbf{EBM}), the noise vector and the observation vector respectively. Then we can rewrite \eqref{eq_observation} as follows:
\vspace{-0mm}
\begin{equation}\label{eq_observation_vector}
\textbf{y}_k =\lvert\textbf{s}\rvert \beta\left(\textbf{x}_k\right) \textbf{W}_{k}^\text{H} \textbf{a}(\textbf{x}_k)+\textbf{z}_{k}.
\end{equation}
\vspace{-0mm}

\subsection{Tracking Loop}\label{subsec_loop}

As shown in Fig. \ref{FrameStrcuctureTVT}, an initial estimate $\hat{\beta}_0 = \hat{\beta}_0^\text{re}+j\hat{\beta}_0^\text{im}$ and $\hat{\textbf{x}}_{0} = \left[\hat{x}_{0,1},\hat{x}_{0,2}\right]^\text{T}$ can be obtained in the beam estimation stage. {It is assumed in this paper that the initial beam estimator can output an estimate $\hat{\textbf{x}}_{0}$ falling within the main lobe of $\textbf{x}_0$, i.e., $\hat{\textbf{x}}_{0} \in \mathcal{B}\left(\textbf{x}_0\right)$, where $\mathcal{B}\left(\textbf{x}_t\right)$ denotes the main lobe of an arbitrary DPV $\textbf{x}_t = \big[x_{t,1},x_{t,2}\big]^\text{T}$, given by
\vspace{-0mm}
\begin{equation}\label{eq_MainLobe}
\begin{aligned}  \mathcal{B}\!\left(\textbf{x}_t\right) \!\triangleq\! \left(x_{t,1}-1,x_{t,1}+1\right) \!\times\! \left(x_{t,2}-1,x_{t,2}+1\right)\,.\,
\end{aligned}
\vspace{-0mm}
\end{equation}Then our tracking starts from this initial estimate $\hat{\textbf{x}}_{0}$ to find more accurate beam directions. It is worth pointing out that the main lobe in $\textbf{x}$ domain in \eqref{eq_MainLobe} has been normalized to a square with twice the unit length of each side and centered at the DPV $\textbf{x}_t$ after the transformation from the angle domain to \textbf{x} domain. Hence, the main lobe size in $\textbf{x}$ domain remains unchanged even if the antenna size $M,\,N$ scale.


In the exploration stage of $k$-th ECC, the receiver needs to choose an EBM $\textbf{W}_k$ based on historical observation vectors $\textbf{y}_{1}, \cdots, \textbf{y}_{k-1}$ along with the corresponding EBMs $\textbf{W}_{1}, \cdots, \textbf{W}_{k-1}$. The new observation $\textbf{y}_k$ can be obtained by applying $\textbf{W}_{k}$. Then the estimate $\hat {\boldsymbol{\psi}}_k \triangleq \left[\hat{\beta}_k^\text{re},\hat{\beta}_k^\text{im},\hat x_{k,1},\hat {x}_{k,2}\right]^\text{T}$ of the channel parameter vector $\boldsymbol{\psi}_k\triangleq \big[\text{Re}\left\{\beta\left(\textbf{x}_k\right)\right\},\text{Im}\left\{\beta\left(\textbf{x}_k\right)\right\},$
$ x_{k,1},{x}_{k,2}\big]^\text{T}$ is obtained by using all observation vectors available and the corresponding EBMs. The whole tracking loop is given in Procedure \ref{alg_trackingloop} and the focus of this paper lies in Step 3 and Step 6.

\floatname{algorithm}{Procedure}
\begin{algorithm}[!t]
	\caption{Tracking Loop}
	\label{alg_trackingloop}
	\begin{algorithmic}[1]
		\Require Array size $M,N$ and the pilot sequence $\textbf{s}$.
		\Ensure  The estimate of the channel parameter vector $\hat {\boldsymbol{\psi}}_k$.
		\State Initialize $\hat {\boldsymbol{\psi}}_0 = \left[\hat{\beta}_0^\text{re},\hat{\beta}_0^\text{im},\hat x_{0,1},\hat {x}_{0,2}\right]^\text{T}$;
		\For{$k=1,2,\cdots$}
		 
		\State Calculate $\textbf{W}_k$ based on $\hat {\boldsymbol{\psi}}_0$, $\textbf{W}_1,\cdots,\textbf{W}_{k-1}$, \indent $\!\!\!$$\textbf{y}_1,\cdots,\textbf{y}_{k-1}$;
		\State Apply $\textbf{W}_k$ in the exploring stage of $k$-th ECC;
		\State Obtain the observation vector $\textbf{y}_k$ in $k$-th ECC;
		\State Estimate $\hat {\boldsymbol{\psi}}_k$ based on $\hat {\boldsymbol{\psi}}_0$, $\textbf{W}_1,\cdots,\textbf{W}_{k}$, \indent $\!\!\!$$\textbf{y}_1,\cdots,\textbf{y}_{k}$;
		\State Point to $\hat{\textbf{x}}_k$ in the communication stage of $k$-th ECC;
		\State use $\hat{\beta}_k$ for receiving in the communication stage of \indent $\!\!\!$$k$-th ECC.
		\EndFor 
	\end{algorithmic}
\end{algorithm}

From a control system perspective, ${\boldsymbol{\psi}}_k$ is the system state, $\hat{{\boldsymbol{\psi}}}_k$ is the estimate of the system state, the EBM $\textbf{W}_k$ is the control action and $\textbf{y}_k$ is a noisy observation non-linearly determined by the system state and the control action. Hence, the task of a tracking design is to find the following strategy:
\vspace{-1mm}
\begin{align}
\label{eq_Fc}\textbf{W}_k = &\textbf{F}_{k}^c\left(\hat {\boldsymbol{\psi}}_0,\textbf{W}_{1},\cdots,\textbf{W}_{k-1}, \textbf{y}_1,\cdots,\textbf{y}_{k-1}\right)\\
\label{eq_Fe}\hat{\boldsymbol{\psi}}_k =& \textbf{F}_{k}^e\left(\hat {\boldsymbol{\psi}}_0,\textbf{W}_{1},\cdots,\textbf{W}_{k}, \textbf{y}_1,\cdots,\textbf{y}_{k}\right),
\end{align}\vspace{-0mm}where $\textbf{F}_{k}^c$ denotes the control function and $\textbf{F}_k^e$ denotes the estimation function in $k$-th ECC.

\vspace{-0mm}
\section{Problem Formulation}\label{sec_problem}
\vspace{-0mm}
Let $\Xi_k = \left\{\textbf{F}_k^{c}, \textbf{F}_k^e\right\}$ denote the set of beam and channel tracking schemes in $k$-th ECC. Then the optimal beam and channel tracking problem minimizing the mean square error (MSE) of the channel vector estimate is formulated as:
	\vspace{-1mm}
	\begin{align}\label{eq_problemIni}
	\underset{\Xi_k}{\min} ~& \frac {1}{MN} \,\mathbb{E} \left[{\left\|\hat{\textbf{h}}_{k} - \textbf{h}_k\right\|}_2^2 \right] \\\vspace{-0mm}
	\label{eq_constrant1} \text{s.t.} ~& \mathbb{E}\left[\hat{\textbf{h}}_{k}\right] = \textbf{h}_k,\\
~&	\eqref{eq_observation_vector},\eqref{eq_Fc},\eqref{eq_Fe},\nonumber	
\vspace{-5mm}
	\end{align}where the constraint \eqref{eq_constrant1} ensures that $\hat{\textbf{h}}_{k} \triangleq \hat{\beta}_k \textbf{a} \left(\hat {\textbf{x}}_k\right)$ is an unbiased estimate of the channel vector $\textbf{h}_k = \beta\left(\textbf{x}_k\right) \textbf{a}\left(\textbf{x}_k\right)$. It is worth explaining the following two points. First, an unbiased estimator may not be the best estimator that achieves the minimum MSE. Nevertheless, such an optimal estimator with no constraints is hard to obtain and hence we add this unbiasedness constraint. Second, we only need to guarantee the unbiasedness of $\hat{\textbf{h}}_k$, as the objective function in \eqref{eq_problemIni} is the MSE of the channel vector. The estimate of the equivalent channel gain and the DPV, i.e., $\hat{\beta}_k$ and $\hat{\textbf{x}}_k$, can be biased.

Problem \eqref{eq_problemIni} is challenging to be solved optimally due to the following reasons:

1) It is a partially observed Markov decision process (POMDP) which generally has not been solved optimally \cite{Lovejoy1991Computationally,Hauskrecht2000Value}.

2) There are $M \times N$ phase shifts to adjust in each EBV $\textbf{w}_{k,i}$. This makes the optimization of the EBV too complicated due to the joint design of so many phase shifts, especially when the array size $M \times N$ grows large.

3) To obtain $\hat{\boldsymbol{\psi}}_k$ in $k$-th ECC, $k$ EBMs, i.e., $\textbf{W}_1,\cdots,\textbf{W}_{k}$, need to be designed, making it difficult to optimize so many beamforming matrices simultaneously as $k$ increases.

4) The time-varying features of the channel vector in \eqref{eq_channelvector} restrict the tracking algorithm and the system performance. Thus, it is hard to design an optimal tracking method for a general channel model.

These challenges above make it extremely difficult to solve this problem optimally. Hence, we add some reasonable constraints in this paper to take the first step of the optimal tracking policy:
\vspace{-0mm}
\subsection{The EBV constraint}
\vspace{-0mm}
As it is complicated to obtain the optimal $M \times N$ phase shifts in general for each EBV, we use steering vectors to design the EBVs,
\vspace{-0mm}
\begin{equation}\label{eq_bf}
\textbf{w}_{k,i}= \frac{1}{\sqrt{MN}}\textbf{a}\left(\boldsymbol{\omega}_{k,i}\right),
\end{equation}\vspace{-0mm}where $\boldsymbol{\omega}_{k,i} \triangleq \left[{\omega}_{k,i1},{\omega}_{k,i2}\right]^\text{T}$ denotes the $i$-th \textbf{exploring direction vector} in $k$-th ECC. This ensures that only two variables need to be designed for each EBV.
\vspace{-0mm}
\subsection{The exploring direction constraint}
\vspace{-0mm}
Although the exploring direction vector $\boldsymbol{\omega}_{k,i}$ in \eqref{eq_bf} can be of any form, however, considering the tracking accuracy, it is better to make sure that $\boldsymbol{\omega}_{k,i}$ falls within the main lobe of the DPV $\textbf{x}_k$ in \eqref{eq_MainLobe}. Thus, it is reasonable to choose exploring directions near the recently estimated direction $\hat{\textbf{x}}_{k-1}$. For this purpose, we use such an architecture in this paper. That is, the $i$-th exploring direction vector in $k$-th ECC, i.e., $\boldsymbol{\omega}_{k,i}$, is determined by the previous estimate of the DPV plus an \textbf{exploration offset} $\boldsymbol{\Delta}_{k,i}$. Considering the design of the offsets that change in different ECCs is also very complicated, we adopt fixed exploration offsets $\boldsymbol{\Delta}_{i} (i=1,\cdots,q)$ in this paper:
\vspace{-0mm}\begin{equation}\label{eq_ed}
\boldsymbol{\omega}_{k,i}  = \hat{\textbf{x}}_{k-1}+\boldsymbol{\Delta}_{i},\,i=1,\cdots,q.
\vspace{-0mm}
\end{equation}Therefore, the EBV in \eqref{eq_bf} can be rewritten as
\vspace{-0mm}\begin{equation}\label{eq_sv}
\textbf{w}_{k,i}= \frac{1}{\sqrt{MN}}\textbf{a}\left(\hat{\textbf{x}}_{k-1}+\boldsymbol{\Delta}_{i}\right),\,i=1,\cdots,q.
\end{equation}
\vspace{-0mm}
\subsection{The time-varying channel constraint}
The time-varying channel vector in \eqref{eq_channelvectorini} is determined by three parts: the antenna gain $\eta\left(\textbf{x}_k\right)$, the channel gain $\beta_{k}^c$ and the DPV $\textbf{x}_k$. Since the change of the antenna gain $\eta\left(\textbf{x}_k\right)$ depends on the DPV $\textbf{x}_k$ for a given antenna element pattern, we only consider the change of the DPV $\textbf{x}_k$ and the channel gain $\beta_{k}^c$ when exploring the properties of the time-varying channels. As the user motion characteristics can be quite different in various situations \cite{Xiao15Iterative,Ishikawa19Differential,Ke19Position}, both of the DPV $\textbf{x}_k$ and the channel gain $\beta_{k}^c$ may change slowly or fast. Therefore, four possible cases exist, which correspond to four different practical scenarios and can be modeled as follows:

%

\begin{itemize}
\item{\textbf{Quasi-static Case}: $\textbf{x}_k \approx \textbf{x}, \beta_k^c \approx \beta^c$}

When both $\textbf{x}_k$ and $\beta_k^c$ change slowly, e.g., the user keeps static or quasi-static in a room, the antenna gain $\eta\left(\textbf{x}_k\right)$ and the equivalent channel gain $\beta\left(\textbf{x}_k\right)$ defined in \eqref{eq_beta} also change slowly. The channel in this case can be seen as approximately fixed. For the sake of convenience, we assume that $\beta\left(\textbf{x}_k\right) = \beta=\beta^\text{re}+j\beta^\text{im}$, $\textbf{x}_k = \textbf{x}=\left[x_{1},x_{2}\right]^\text{T}$ in this case. 

\item{\textbf{Dynamic Case}: $\textbf{x}_k \approx \textbf{x}, \beta_{k+1}^c \neq \beta_{k}^c$}

For channels that $\textbf{x}_k$ changes slowly while $\beta_k^c$ changes fast, e.g., a person walks at a fast pace in a room, the beam direction can be seen as approximately fixed, i.e., $\textbf{x}_k = \textbf{x}$\cite{Xiao15Iterative,Ishikawa19Differential}. To distinguish from other dynamic scenarios, this case is called \textbf{Dynamic Case I}.

\item{\textbf{Dynamic Case}: $\textbf{x}_{k+1} \neq \textbf{x}_{k}, \beta_{k}^c \approx \beta^c$}

This case requires that the beam direction changes fast while the channel gain keeps static or quasi-static. However, in real mmWave channels, the fast change of the beam direction usually leads to the fast change of the channel gain since the propagation paths change. This case exists only when the user rotates around the base station (BS) exactly in a circle in line of sight (LOS) channels. This is not the usual case and not studied in this paper.

\item{\textbf{Dynamic Case}: $ \textbf{x}_{k+1} \neq \textbf{x}_{k},\beta_{k+1}^c \neq \beta_{k}^c$}

Both the beam direction and the channel gain in this case change fast, which happens in most fast-moving scenarios except Dynamic Case I, e.g., an unmanned aerial vehicle (UAV) flies in the sky \cite{Ke19Position}. To distinguish from Dynamic Case I, we call it \textbf{Dynamic Case II}. 
\end{itemize}

It would be helpful to explain the following two aspects. First of all, extra algorithms need to be introduced to efficiently classify the channels according to the time-varying features. However, to make the research goals more focused in this paper, we leave the details of this classification in future work. Second, this paper only exploits the independent variation properties of $\beta_k^c$ and $\textbf{x}_k$ in these four cases to obtain theoretical results. While in real mmWave channels, the variation of the channel gain $\beta_k^c$ and the DPV $\textbf{x}_k$ might be interrelated \cite{Shafi2018Microwave}, which are supposed to be jointly taken into account in future work.

With the above-mentioned \textbf{EBV constraint}, \textbf{the exploring direction constraint} and \textbf{the time-varying channel constraint}, the beam and channel tracking problem in \eqref{eq_problemIni} can be reformulated as:
\vspace{-0mm}
\begin{align}\label{eq_problem}
\underset{\Xi}{\min} ~& \frac {1}{MN} \,\mathbb{E} \left[{\left\|\hat{\textbf{h}}_{k} - \textbf{h}_k\right\|}_2^2 \right] \\
\vspace{-0mm}
\text{s.t.} ~& 
\eqref{eq_observation_vector},\eqref{eq_Fc},\eqref{eq_Fe},	\eqref{eq_constrant1}, \eqref{eq_sv}.\nonumber
\end{align}\vspace{-0mm}
\vspace{-3mm}
\section{How Many Explorations Are Needed In each ECC?}\label{sec_PilotOverhead}
\vspace{-0mm}
Before delving into the detailed tracking process in \eqref{eq_problem}, we will first study the number of explorations needed in this section.


To estimate ${\boldsymbol{\psi}}_k$, sufficient measurements from different exploring directions are required. For Quasi-static Case where ${\boldsymbol{\psi}}_k$ remains unchanged, i.e., ${\boldsymbol{\psi}}_k=\boldsymbol{\psi}\triangleq \left[\beta^\text{re},\beta^\text{im},x_1,x_2\right]^\text{T}$, one exploration in each ECC is enough since sufficient measurements are available after quite a number of ECCs. Nevertheless, in dynamic case, only using one exploration in each ECC does not work well as ${\boldsymbol{\psi}}_k$ may change fast. Hence, it is necessary to ensure that the estimate can be obtained even by using the explorations in a single ECC. Then the question becomes: under the condition above, how many explorations are needed in each ECC?

With the constraint in \eqref{eq_bf}, two explorations in each ECC are sufficient to jointly track the equivalent channel gain and the 1D beam direction according to \cite{JLiJoint2018ICCASP}. When tracking the 2D direction, it is straight forward that four explorations are feasible by separately using two explorations to track each dimension of the 2D direction. However, using four explorations will lower the system efficiency since it will cost time resources for each exploration. Hence, we may ask that can we reduce the times of exploration, or what is the minimum number of explorations required?

Then the following lemma is proposed to help determine the minimum exploration overhead $q$ in each ECC:
\vspace{-0mm}
\begin{lemma}\label{IndObservations}
	\emph{If the EBVs are of the steering vector forms, i.e.,} $\textbf{w}_{k,i} = \frac{1}{\sqrt{MN}}\textbf{a}\!\left(\boldsymbol{\omega}_{k,i}\right)$\emph{, and the observation vector in \eqref{eq_observation_vector} is noiseless, then} 
	
	\emph{1) to obtain the unique solution of the channel parameter vector $\boldsymbol{\psi}_k$ within one ECC, the minimum exploration overhead is $q=3$ in each ECC;}
	
    \emph{2) to obtain the unique solution of the DPV} $\textbf{x}_k$ \emph{within one ECC, the minimum exploration overhead is $q=3$ in each ECC.}

\end{lemma}
\begin{proof}
	See Appendix \ref{proof_IndObservations}.
\end{proof}\vspace{-0mm}

Lemma \ref{IndObservations} reveals that it is impossible to obtain the unique solution within one ECC when only using two explorations, whether we want to jointly estimate $\beta\left(\textbf{x}_k\right)$ and $\textbf{x}_k$ or just estimate $\textbf{x}_k$. If we use three explorations and design three appropriate exploring directions in each ECC, then the unique solution of the channel parameter vector $\boldsymbol{\psi}_k$ can be obtained. Hence, we set $q = 3$ in this paper, i.e., the EBM $\textbf{W}_{k} = \left[\textbf{w}_{k,1},\textbf{w}_{k,2},\textbf{w}_{k,3}\right]$.

\vspace{-0mm}
\section{Quasi-static Tracking: Performance Bound, Convergence and Optimality}\label{sec_Quasi-static_Tracking}
\vspace{-0mm}
In this section, we will focus on Quasi-static Case. As mentioned in Section \ref{sec_problem}, in Quasi-static Case, $\boldsymbol{\psi}_k=\boldsymbol{\psi}= \left[\beta^\text{re},\beta^\text{im},x_1,x_2\right]^\text{T}$ and $\textbf{h}_k=\textbf{h} \triangleq {\beta} \textbf{a} \left({\textbf{x}}\right)$. For a given channel parameter vector $\boldsymbol{\psi}$ and EBM $\textbf{W}_{k}$, the observation vector satisfies normal distribution with $\textbf{y}_k \sim \mathcal{CN}\left(\lvert \textbf{s} \rvert \beta \textbf{W}_k^\text{H} \textbf{a}(\textbf{x}),\sigma_z^2\textbf{J}_3\right)$. Hence, the conditional probability density function of $\textbf{y}_k$ is given by
\vspace{-0mm}
\begin{equation}\label{eq_static_pdf}
p_S(\textbf{y}_k| \boldsymbol{\psi}, \textbf{W}_k) = {\frac{1}{\pi^{3} \sigma_z^{6}} e^{- \frac {{\left\| \textbf{y}_k-\lvert \textbf{s} \rvert \beta \textbf{W}_k^\text{H} \textbf{a}(\textbf{x}) \right\|}_2^2} {\sigma_z^2}}}.
\end{equation}\vspace{-0mm}In this section, we will first provide the lower bound of the tracking error in Quasi-static Case. Then we develop a tracking algorithm and prove it can converge to the minimum CRLB with time. 
\vspace{-0mm}
\subsection{Cram\'{e}r-Rao Lower Bound of The Tracking Error}\label{subsec_static_Bound}
\vspace{-0mm}
The Cram\'{e}r-Rao lower bound theory gives the lower bound of the unbiased estimation error \cite{Sengijpta1993Fundamental}. Based on this, we introduce the following lemma to obtain the lower bound of the tracking error in Quasi-static Case:
\vspace{-0mm}
\begin{lemma}\label{MSEOpt}
	\emph{In Quasi-static Case, given $\textbf{W}_1,\cdots,\textbf{W}_k$}, \emph{the MSE of the channel vector estimate in \eqref{eq_problem} is lower bounded as follows:}
	\vspace{-0mm}
	\begin{align}\label{MSELB}
	&\,\frac{1}{{MN}} \mathbb{E}\left[\left\| \hat{\textbf{h}}_{k} - \textbf{h} \right\|_2^2 \right]
	\\\ge&\, \frac{1}{{MN}}\Tr \left\{ {{{\left(\sum\limits_{l = 1}^k {\textbf{I}_S(\boldsymbol{\psi} ,{{\bf{W}}_{l}})} \right)}^{-1}} {\left({\textbf{V}^\text{H}}{\textbf{V}}\right)}  } \right\}\nonumber\\
	\triangleq &\,{C}_S^t\left(\boldsymbol{\psi},\textbf{W}_1,\cdots,\textbf{W}_k\right),\nonumber
	\end{align}
	\vspace{-0mm}\emph{where} $\textbf{V}$ \emph{is the Jacobian matrix given by}
	   \begin{equation}
		\begin{aligned}\label{eq_Jacobian}
		\textbf{V} \triangleq \frac{{\partial \textbf{h}}}{{\partial { {\boldsymbol{\psi}}^\text{T}}}}=& \left[\frac{\partial {\textbf{h}}}{\partial {\beta^\text{re}}},\frac{\partial {\textbf{h}}}{\partial{\beta^\text{im}}},\frac{\partial {\textbf{h}}}{\partial {x_1}},\frac{\partial {\textbf{h}}}{\partial {x_2}}\right]
		\\=&\left[\textbf{a}\left(\textbf{x}\right),j\textbf{a}\left(\textbf{x}\right),\beta \frac {\partial \textbf{a}\left(\textbf{x}\right)}{\partial x_1},\beta \frac {\partial \textbf{a}\left(\textbf{x}\right)}{\partial x_2}\right]
		\end{aligned}\end{equation}\vspace{-0mm}\emph{and the Fisher information matrix} $\textbf{I}_S(\boldsymbol{\psi}, \textbf{W}_{l})$ \emph{is given by}
		\begin{align}\label{eq_fisher}
		\!\textbf{I}_S(\boldsymbol{\psi}, \textbf{W}_{l})&\triangleq    \mathbb{E}\left[\frac {\partial \text{log} \, p_S \left(\textbf{y}_{l} |\boldsymbol{\psi},\textbf{W}_{l} \right)}{\partial \boldsymbol{\psi}} \!\cdot\! \frac {\partial \text{log}\, p_S \left(\textbf{y}_{l} |\boldsymbol{\psi},\textbf{W}_{l} \right)}{\partial \boldsymbol{\psi}^\text{T}}\right]
		\nonumber\\&=\frac {2 {\lvert \textbf{s} \rvert}^2} {{\sigma}_z^2}  \text{Re}\left\{\textbf{V}^\text{H} \textbf{W}_{l} \textbf{W}_{l}^\text{H} \textbf{V} \right\}.
		\end{align}
    \end{lemma}
\vspace{-0mm}
\begin{proof}
See Appendix \ref{proof_MSEOpt}.
\end{proof}

The CRLB in \eqref{MSELB} is a function of the EBMs $\textbf{W}_{1},\ldots,\textbf{W}_{k}$. 
Since it is hard to optimize so many EBMs simultaneously, we will first try to find a lower bound of the CRLB under the constraint \eqref{eq_sv}, and later design a tracking algorithm approaching this lower bound.

Consider any tracking algorithm under the constraint \eqref{eq_sv} that can converge to the DPV $\textbf{x}$, i.e.,
\vspace{-0mm}
\begin{equation}
{\lim\limits_{k \to +\infty}} \hat{\textbf{x}}_k=\textbf{x}.
\vspace{-0mm}
\end{equation}
Then the EBM $\textbf{W}_k$ also converges,
\vspace{-0mm}
\begin{equation}
{\lim\limits_{k \to +\infty}} {\textbf{W}}_k=\textbf{W}=\left[\textbf{w}_1,\textbf{w}_2,\textbf{w}_3\right]^\text{T},
\vspace{-0mm}
\end{equation}where $\textbf{w}_i$ is given by
\vspace{-0mm}
\begin{equation}\label{eq_svr}
\textbf{w}_{i} \triangleq \frac{1}{\sqrt{MN}}\textbf{a}\left(\textbf{x}+\boldsymbol{\Delta}_{S,i}\right),i = 1,2,3
\vspace{-0mm}
\end{equation}with $\left\{\boldsymbol{\Delta}_{S,1},\boldsymbol{\Delta}_{S,2},\boldsymbol{\Delta}_{S,3}\right\}$ denoting the fixed set of exploration offsets in Quasi-static Case. Hence, the normalized CRLB (by multiplying $k$) converges as $k \to +\infty$:
\begin{small}
\vspace{-0mm}
\begin{align}\label{eq_CMMSETemp}
	&{\lim\limits_{k \to +\infty}}k{C}_S^t(\boldsymbol{\psi},\textbf{W}_1,\cdots,\textbf{W}_k)\nonumber\\
	=&{\lim\limits_{k \to +\infty}}\frac{k}{{MN}}\Tr \left\{ {{{\left(\sum\limits_{l = 1}^k {\textbf{I}_S(\boldsymbol{\psi} ,{{\bf{W}}_{l}})} \right)}^{-1}} {\!\left({\textbf{V}^\text{H}}{\textbf{V}}\right)}  } \right\} \\
	=&	\frac{1}{{MN}}\Tr \left\{ {{{ {\textbf{I}_S(\boldsymbol{\psi},{{\bf{W}}})} }^{-1}} {\left({\textbf{V}^\text{H}}{\textbf{V}}\right)}  } \right\},\nonumber
	\end{align}\end{small}which is a function of $\boldsymbol{\psi},\,\textbf{W}$ and will be denoted as ${C}_S(\boldsymbol{\psi},\textbf{W})$.

According to \eqref{eq_CMMSETemp}, for a given channel (direction and gain), there exists an optimal EBM, which leads to the minimum normalized CRLB as a function of the channel parameter vector $\boldsymbol{\psi}$:

\vspace{-0mm}
\begin{align}\label{eq_CMMSE}
{C}_S^{\min}(\boldsymbol{\psi})=&\min_{\textbf{W}} {C}_S(\boldsymbol{\psi},\textbf{W})={C}_S(\boldsymbol{\psi},\textbf{W}_S^*).
\end{align}Solving problem \eqref{eq_CMMSE} yields $\textbf{W}_S^*=\left[\textbf{w}_{S,1}^*,\textbf{w}_{S,2}^*,\textbf{w}_{S,3}^*\right]$ with
\vspace{-0mm}
\begin{equation}\label{eq_bfo}
\textbf{w}_{S,i}^* = \frac{1}{\sqrt{MN}}\textbf{a}\left(\textbf{x}+ \boldsymbol{\Delta}_{S,i}^*\right),i=1,2,3,
\vspace{-0mm}
\end{equation}where $\left\{\boldsymbol{\Delta}_{S,1}^*,\boldsymbol{\Delta}_{S,2}^*,\boldsymbol{\Delta}_{S,3}^*\right\}$ denotes the optimal set of exploration offsets for a given array size and a given $\boldsymbol{\psi}$.

\vspace{-0mm}
\subsection{Asymptotically Optimal Set of Exploration Offsets}\label{sbsec_OptBeamforming}
\vspace{-0.5mm}
In general, the minimum CRLB in \eqref{eq_CMMSE} is a function of a set of system parameters including the equivalent channel gain $\beta$, the DPV $\textbf{x}$ and the array size $M,\,N$. Hence, the optimal set of 2D exploration offsets should also be a function of these parameters. Since it is very hard to obtain the expression of this optimal set, we adopt numerical search to deal with this issue. However, as many parameters in \eqref{eq_CMMSE} may affect the optimal result, numerical search has to be reconducted for different parameter sets, resulting in high complexity. 

Fortunately, through our investigation, some useful properties of the minimum CRLB and the optimal set of exploration offsets are given to simplify the numerical search, as described in the following lemma:

\vspace{-0mm}
\begin{lemma}\label{UnifiedOptShift}
\emph{In Quasi-static Case, the minimum CRLB ${C}_S^{\min}(\boldsymbol{\psi})$ and the optimal set of exploration offsets $\left\{\boldsymbol{\Delta}_{S,1}^{*},\boldsymbol{\Delta}_{S,2}^{*},\boldsymbol{\Delta}_{S,3}^{*}\right\}$ have the following three properties}:

\emph{1)} ${C}_S^{\min}(\boldsymbol{\psi})$, $\left\{\boldsymbol{\Delta}_{S,1}^{*},\boldsymbol{\Delta}_{S,2}^{*},\boldsymbol{\Delta}_{S,3}^{*}\right\}$ \emph{are invariant to the equivalent channel gain} $\beta$;

\emph{2)} ${C}_S^{\min}(\boldsymbol{\psi})$, $\left\{\boldsymbol{\Delta}_{S,1}^{*},\boldsymbol{\Delta}_{S,2}^{*},\boldsymbol{\Delta}_{S,3}^{*}\right\}$ \emph{are invariant to the DPV} $\textbf{x}$;

\emph{3)}
${C}_S^{\min}(\boldsymbol{\psi})$ \emph{converges as} $\emph{M},\,\emph{N} \to +\infty$ \emph{and there exists a fixed set of exploration offsets that are unrelated to the array size} $M,N$, \emph{denoted as} $\left\{\widetilde{\boldsymbol{\Delta}}_{S,1}^{*},\,\widetilde{\boldsymbol{\Delta}}_{S,2}^{*},\widetilde{\boldsymbol{\Delta}}_{S,3}^{*}\right\}$, \emph{such that} 
\vspace{-0mm}
\begin{align}\label{eq_asymt} {\lim\limits_{M,N \to +\infty}}{C}_S(\boldsymbol{\psi},\widetilde{\textbf{W}}_S^*) = {\lim\limits_{M,N \to +\infty}}{C}_S^{\min}(\boldsymbol{\psi}),
 \nonumber
\vspace{-0mm}
\end{align}\emph{where} $\widetilde{\textbf{W}}_S^{*}=[\tilde{\textbf{w}}_{S,1}^{*},\tilde{\textbf{w}}_{S,2}^{*},\tilde{\textbf{w}}_{S,3}^{*}]$ \emph{is obtained with}
\vspace{-0mm}
\begin{equation}\label{eq_obf}
\tilde{\textbf{w}}_{S,i}^{*} \triangleq \frac{1}{\sqrt{MN}}\textbf{a}\left(\textbf{x}+\widetilde{\boldsymbol{\Delta}}_{S,i}^{*}\right),i=1,2,3.
\vspace{-0mm}
\end{equation}
\end{lemma}
\begin{proof}
See Appendix \ref{proof_UnifiedOptShift}.
\end{proof}
\vspace{-0mm}
\vspace{-1mm}

Lemma \ref{UnifiedOptShift} reveals that $\left\{\boldsymbol{\Delta}_{S,1}^{*},\boldsymbol{\Delta}_{S,2}^{*},\boldsymbol{\Delta}_{S,3}^{*}\right\}$ is only related to the array size $M,\,N$. Hence, the numerical search times can be reduced to one for a particular array size $M,\,N$. Numerically, we find later that even if $\left\{\boldsymbol{\Delta}_{S,1}^{*},\boldsymbol{\Delta}_{S,2}^{*},\boldsymbol{\Delta}_{S,3}^{*}\right\}$ may change for different array sizes, $\left\{\widetilde{\boldsymbol{\Delta}}_{S,1}^{*},\,\widetilde{\boldsymbol{\Delta}}_{S,2}^{*},\widetilde{\boldsymbol{\Delta}}_{S,3}^{*}\right\}$ can be used to take the place of $\left\{\boldsymbol{\Delta}_{S,1}^{*},\boldsymbol{\Delta}_{S,2}^{*},\boldsymbol{\Delta}_{S,3}^{*}\right\}$ as long as $M$ and $N$ are sufficiently large. Therefore, the numerical search times is reduced to one in the end. As $\left\{\widetilde{\boldsymbol{\Delta}}_{S,1}^{*},\,\widetilde{\boldsymbol{\Delta}}_{S,2}^{*},\widetilde{\boldsymbol{\Delta}}_{S,3}^{*}\right\}$ can be used to achieve the minimum CRLB when $M,N\to +\infty$, it is called \textbf{the asymptotically optimal set of exploration offsets} in Quasi-static Case in this paper.

By numerical search in the main lobe in \eqref{eq_MainLobe}, we can obtain one asymptotically optimal set of exploration offsets $\left\{\widetilde{\boldsymbol{\Delta}}_{S,1}^{*},\,\widetilde{\boldsymbol{\Delta}}_{S,2}^{*},\widetilde{\boldsymbol{\Delta}}_{S,3}^{*}\right\}$ in TABLE \ref{Tab_asymptotically optimal_parameters} and Fig. \ref{fig_offsets}. It can be seen that the three exploring direction vectors do not form a regular triangle as the radiation pattern produced by \eqref{eq_svr} is not isotropic from different angles.
\renewcommand\arraystretch{1.4}
\begin{table}[!t!]
\centering
\vspace{-0mm}
\caption{The asymptotically optimal set of exploration offsets in Quasi-static Case.}
\vspace{2mm}
\begin{tabular} {c|c|c}
\hline
\hline
$\widetilde{\boldsymbol{\Delta}}_{S,1}^{*}$ & $\widetilde{\boldsymbol{\Delta}}_{S,2}^{*}$ & $\widetilde{\boldsymbol{\Delta}}_{S,3}^{*}$\\
\hline
$\left[-0.0963, 0.5098\right]^\text{T}$ & $\left[-0.2906,-0.2906\right]^\text{T}$ & $\left[0.5098,-0.0963\right]^\text{T}$\\
\hline
\hline
\end{tabular}\label{Tab_asymptotically optimal_parameters}
\vspace{-0mm}
\end{table}
\begin{figure}[!t!]
	\vspace{-0mm}
	\centering
	\includegraphics[width=7.5cm]{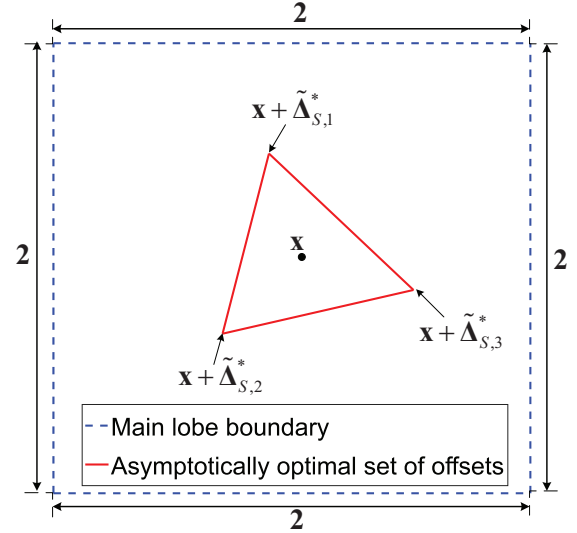}
	\vspace{-0mm}
	\caption{The asymptotically optimal set of exploration offsets in Quasi-static Case.}
	\vspace{-0mm}
	\label{fig_offsets}
\end{figure}With this set in TABLE \ref{Tab_asymptotically optimal_parameters}, a general way to generate the EBM $\widetilde{\textbf{W}}_S^{*}$ is obtained by \eqref{eq_obf}.

The set of exploration offsets $\left\{\widetilde{\boldsymbol{\Delta}}_{S,1}^{*},\,\widetilde{\boldsymbol{\Delta}}_{S,2}^{*},\widetilde{\boldsymbol{\Delta}}_{S,3}^{*}\right\}$ may become sub-optimal when the antenna size $M \times N$ is finite. To evaluate the robustness of this set of exploration offsets to finite array size, we adopt $\left\{\widetilde{\boldsymbol{\Delta}}_{S,1}^{*},\,\widetilde{\boldsymbol{\Delta}}_{S,2}^{*},\widetilde{\boldsymbol{\Delta}}_{S,3}^{*}\right\}$ to antenna arrays of limited size and compare the minimum CRLB with the CRLB achieved by $\left\{\widetilde{\boldsymbol{\Delta}}_{S,1}^{*},\,\widetilde{\boldsymbol{\Delta}}_{S,2}^{*},\widetilde{\boldsymbol{\Delta}}_{S,3}^{*}\right\}$ in TABLE \ref{Tab_asymptotically optimal_parameters}.
As illustrated in Fig. \ref{fig_Extensibility.eps}, when the antenna number $M=N \geq 8$, we can approach the minimum CRLB with a relative error less than $0.1\%$ by using $\left\{\widetilde{\boldsymbol{\Delta}}_{S,1}^{*},\,\widetilde{\boldsymbol{\Delta}}_{S,2}^{*},\widetilde{\boldsymbol{\Delta}}_{S,3}^{*}\right\}$. 

As a conclusion, it is practical to apply this asymptotically optimal set of exploration offsets to any antenna array with $M=N\geq8$, any channel gain and any direction.

\begin{figure}[!t]
	\centering
	\vspace{-0mm}
	\includegraphics[width=8.5cm]{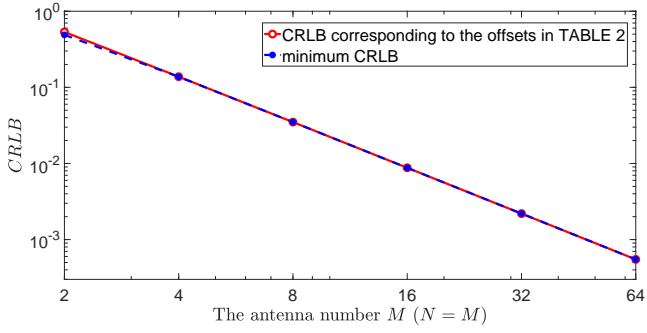}
	\vspace{-0mm}
	\caption{The performance of the offsets in TABLE \ref{Tab_asymptotically optimal_parameters}.}
	\vspace{-0mm}
	\label{fig_Extensibility.eps}
\end{figure}
\vspace{-0mm}
\subsection{Joint Beam and Channel Tracking}\label{sbsec_Tracking}
\vspace{-0mm}
In the above subsections, we have provided a low-complexity numerical method to design the optimal exploration offsets and obtain the minimum CRLB, given that the DPV $\textbf{x}$ is known. However, in a real tracking problem, the DPV $\textbf{x}$ is unknown and the EBMs need to be adjusted dynamically. In addition, a sequence of optimal beamforming matrices only tells us what the minimum CRLB is, but it cannot tell us which tracking algorithm can achieve the minimum CRLB. In this subsection, we propose a specific tracking algorithm to approach the minimum CRLB.



The proposed tracker is motivated by the following maximum likelihood problem:
\begin{small}
\vspace{-0mm}
\begin{align}\label{eq_MLEIni}
\underset{\textbf{W}_k}{\max}\! &\left\{\!\underset{\hat{\boldsymbol{\psi}}_k}{\max}\,\text{log} \, p \!\left(\!\!\begin{matrix}\textbf{y}_1,
\!\cdots,\textbf{y}_{k}
\end{matrix} \Bigg|\begin{matrix}{\boldsymbol{\psi}},
\textbf{W}_1,\cdots,\textbf{W}_{k}
\end{matrix} \!\!\right)\Bigg|_{{\boldsymbol{\psi}}=\hat{\boldsymbol{\psi}}_k}\!\!\right\}\!\!\\
\text{s.t.} ~& 
\eqref{eq_observation_vector},\eqref{eq_Fc},\eqref{eq_Fe},	\eqref{eq_constrant1}, \eqref{eq_sv}.\nonumber
\end{align}\end{small}\vspace{-0mm}Since $\textbf{y}_1,\cdots,\textbf{y}_k$ are independently observed vectors, we can convert \eqref{eq_MLEIni} as follows:
	\begin{align}\label{eq_MLE}
	\underset{\textbf{W}_k}{\max} &\left\{\!\underset{\hat{\boldsymbol{\psi}}_k}{\max}\,
	\sum\limits_{l=1}^k\left[\!\text{log} \, p_S \left(\!\textbf{y}_l \Bigg|\begin{matrix}\boldsymbol{\psi},\textbf{W}_{l}\end{matrix}\right)\Bigg|_{\boldsymbol{\psi} = \hat{\boldsymbol{\psi}}_k}\right]
	\right\}\!\!\\
	\text{s.t.} ~& 
	\eqref{eq_observation_vector},\eqref{eq_Fc},\eqref{eq_Fe},	\eqref{eq_constrant1}, \eqref{eq_sv}.\nonumber
	\end{align}This problem is somewhat similar to a kind of estimation problem with control in \cite[Section 10.2]{nevel1973stochastic}, where a two-layer recursive algorithm is proposed and proved to converge to the minimum CRLB under several requirements. The difference is that the observation vector in \cite[Section 10.2]{nevel1973stochastic} is given by $\textbf{y} = \textbf{f}^{o}\left(\boldsymbol{\psi},\textbf{W}\right) +\textbf{z}$, where $\textbf{z}$ is an i.i.d. Gaussian noise vector and $\textbf{f}^{o}\left(\boldsymbol{\psi},\textbf{W}\right)$ is a convex function with respect to $\boldsymbol{\psi}$. However, in our problem, $\textbf{f}^{o}\left(\boldsymbol{\psi},\textbf{W}\right) = \lvert s\rvert\beta\textbf{W}^\text{H}\textbf{a}\left(\textbf{x}\right)$ is non-convex, leading to the failure of directly applying the algorithm and the theoretical results of \cite[Section 10.2]{nevel1973stochastic}.

Despite this, we can still design a two-layer nested optimization algorithm inspired by \cite[Section 10.2]{nevel1973stochastic}. Furthermore, our proposed algorithm can be proved to converge to the minimum CRLB under some necessary requirements, as will be explained in Section \ref{subsec_asy_opt}.

The proposed algorithm is based on iterative maximization in the inner layer and the outer layer of \eqref{eq_MLE}. In the \emph{inner layer} of \eqref{eq_MLE}, we use the{\tiny } stochastic Newton's method \cite{James2003Introduction} to update the estimate, given by
	\begin{align}\label{eq_stoNew}
	\hat {\boldsymbol{\psi}}_k =\hat {\boldsymbol{\psi}}_{k-1}+  b_{S,k} \boldsymbol{\varsigma}_{k},
	\vspace{-0mm}
	\end{align}where $b_{S,k}$ is the tracking step-size in Quasi-static Case, and $\boldsymbol{\varsigma}_{k}$ is the updating direction vector. This updating direction vector is a function of the observation vector $\textbf{y}_k$ and the latest estimated value of the channel parameter vector $\hat{\boldsymbol{\psi}}_{k-1}$, and is defined as below:
\vspace{-0mm}
\begin{align}\label{eq_ud}
\boldsymbol{\varsigma}_{k} \!\triangleq\! \textbf{I}_S\left(\hat{\boldsymbol{\psi}}_{k-1}, \textbf{W}_k\right)^\text{-1} \frac {\partial \text{log} \, p_S \left(\textbf{y}_k |\boldsymbol{\psi},\textbf{W}_k \right)}{\partial \boldsymbol{\psi}}\Bigg|_{{\boldsymbol{\psi}} \!=\! \hat{\boldsymbol{\psi}}_{k\!-\!1}}.
\end{align}\vspace{-0mm}
And it is derived that
\vspace{-0mm}
\begin{align}\label{eq_udcal}
\boldsymbol{\varsigma}_{k}\!=\! \left(\!\text{Re}\!\left\{\!\hat{\textbf{V}}_k^\text{H} \textbf{W}_{k} \textbf{W}_{k}^\text{H} \hat{\textbf{V}}_k \!\right\}\!\right)^{\!-\!1}\!\left[\!
\begin{matrix}
{\text{Re}\left\{ \textbf{e}_k^\text{H}\left(\textbf{y}_k-\hat{\textbf{y}}_k \right)\right\}}\\
{\text{Im}\left\{ \textbf{e}_k^\text{H}\left(\textbf{y}_k-\hat{\textbf{y}}_k \right)\right\}}\\
{\text{Re}\left\{ \tilde{\textbf{e}}_{k1}^\text{H}\left(\textbf{y}_k-\hat{\textbf{y}}_k \right)\right\}}\\
{\text{Re}\left\{ \tilde{\textbf{e}}_{k2}^\text{H}\left(\textbf{y}_k-\hat{\textbf{y}}_k \right)\right\}}
\end{matrix}
\!\right]\!,\!
\vspace{-0mm}
\end{align}\vspace{-0mm}where ${\textbf{e}}_k = \textbf{W}_k^\text{H} \textbf{a}\left(\hat{\textbf{x}}_{k-1}\right)$, $\hat{\textbf{y}}_k = \lvert \textbf{s} \rvert \hat{\beta}_{k-1}\textbf{W}_k^\text{H} \textbf{a}\left(\hat{\textbf{x}}_{k-1}\right)$, $\tilde{\textbf{e}}_{k1} = \hat{\beta}_{k-1} \textbf{W}_k^\text{H} \frac{\partial \textbf{a}\left(\hat{\textbf{x}}_{k-1}\right)}{\partial x_1}$, $\tilde{\textbf{e}}_{k2} = \hat{\beta}_{k-1} \textbf{W}_k^\text{H} \frac{\partial \textbf{a}\left(\hat{\textbf{x}}_{k-1}\right)}{\partial x_2}$ and $\hat{\textbf{V}}_k$ is given by
\begin{align}\label{eq_JacobianVk}
\hat{\textbf{V}}_k = \left[\textbf{a}\left(\textbf{x}\right),j\textbf{a}\left(\textbf{x}\right),\beta \frac {\partial \textbf{a}\left(\textbf{x}\right)}{\partial x_1},\beta \frac {\partial \textbf{a}\left(\textbf{x}\right)}{\partial x_2}\right]\Bigg|_{{\boldsymbol{\psi}} = \hat{\boldsymbol{\psi}}_{k-1}}.
\end{align}\vspace{-0mm}

In the outer layer of \eqref{eq_MLE}, assuming that the estimate of the channel parameter vector is accurate, i.e., ${{\boldsymbol{\psi}} = \hat{\boldsymbol{\psi}}_{k-1}}$, the EBM $\textbf{W}_k=\left[\textbf{w}_{k,1},\textbf{w}_{k,2},\textbf{w}_{k,3}\right]$ is obtained with
\begin{equation}
\begin{aligned}\textbf{w}_{k,i} = \frac {1}{\sqrt{MN}} \textbf{a} \left(\hat{\textbf{x}}_{k-1} + \widetilde{\boldsymbol{\Delta}}_{S,i}^{*}\right), i =1,2,3,
\end{aligned}
\vspace{-1mm}
\end{equation}

Finally, the proposed tracking algorithm is summarized in Algorithm \ref{alg_Static}.

\floatname{algorithm}{Algorithm}
\setcounter{algorithm}{0}
\begin{algorithm}[!t]
	\caption{\textbf{Joint Beam and Channel Tracking} (\textbf{JBCT}) for Quasi-static Case}
	\label{alg_Static}
	\begin{algorithmic}[0]
\State 1) \textbf{Exploring and Receiving} (Step 3 in Procedure \ref{alg_trackingloop}): Transmit 3 pilot sequences in each ECC. The corresponding EBV for receiving the $i$-th pilot sequence in $k$-th ECC is given below:
\vspace{-0mm}
\begin{equation}\label{eq_BF}
\begin{aligned}\textbf{w}_{k,i} = \frac {1}{\sqrt{MN}} \textbf{a} \left(\hat{\textbf{x}}_{k-1} + \widetilde{\boldsymbol{\Delta}}_{S,i}^{*}\right), i =1,2,3,
\end{aligned}
\vspace{-0mm}
\end{equation}where $\left\{\widetilde{\boldsymbol{\Delta}}_{S,1}^{*},\,\widetilde{\boldsymbol{\Delta}}_{S,2}^{*},\widetilde{\boldsymbol{\Delta}}_{S,3}^{*}\right\}$ is given by TABLE \ref{Tab_asymptotically optimal_parameters}. After match filtering, the observation vector $\textbf{y}_k$ is obtained via \eqref{eq_observation_vector}. 

\State 2) \textbf{Updating Estimate} (Step 6 in Procedure \ref{alg_trackingloop}): The estimate of the channel parameter vector in $k$-th ECC, i.e., $\hat {\boldsymbol{\psi}}_k = \left[\hat{\beta}_k^\text{re},\hat{\beta}_k^\text{im},\hat x_{k,1},\hat {x}_{k,2}\right]^\text{T}$, is updated by
\vspace{-0mm}
\begin{equation}\label{eq_Tracking}
\begin{aligned} \hat {\boldsymbol{\psi}}_{k} = \hat {\boldsymbol{\psi}}_{k-1}+b_{S,k} \boldsymbol{\varsigma}_{k},
\end{aligned}
\vspace{-0mm}
\end{equation}where $\boldsymbol{\varsigma}_{k}$ is the updating direction vector given by \eqref{eq_udcal} and $b_{S,k}$ is the step-size that will be specified after.
    \end{algorithmic}
\end{algorithm}
\vspace{-0mm}
\subsection{Asymptotic Optimality Analysis}\label{subsec_asy_opt}
\vspace{-0mm}

In this subsection, the convergence and the optimality of our proposed algorithm will be discussed. Since the entire proofs are very long, we will provide the main statements and ideas or clues here, and leave the proofs in the appendices.

The convergence and optimality will be stated in three steps as follows:

i) We prove that the estimate of the proposed tracking algorithm converges to a unique point with probability one given appropriate sequences of step-sizes.

ii) We prove that if the initial estimate is within the main lobe, i.e., $\hat {\textbf{x}}_0 \in \mathcal{B}\left(\textbf{x}\right)$ and the step-size is appropriate, then the convergence point will be exactly the real channel parameter vector $\boldsymbol{\psi}$, with probability approaching one.

iii) Finally, if $\hat{\boldsymbol{\psi}}_k \to {\boldsymbol{\psi}}$ and the step-size is appropriate, then 
the tracking error of our algorithm converges to the minimum CRLB.

\emph{\textbf{1) Convergence to a unique point}} 

Since the observation vector $\textbf{y}_k$ is corrupted by the Gaussian noise vector, the updating direction vector $\boldsymbol{\varsigma}_{k}$ in \eqref{eq_ud} is also a random vector, and can be expressed as follows:
\vspace{-0mm}
\begin{equation}\label{eq_udp}
\boldsymbol{\varsigma}_{k} = \textbf{f}_{\boldsymbol{\psi}} \left(\hat {\boldsymbol{\psi}}_{k-1}\right) + \hat{\textbf{z}}_k,
\vspace{-0mm}
\end{equation}where $\textbf{f}_{\boldsymbol{\psi}} \left(\hat {\boldsymbol{\psi}}_{k-1}\right)$ is the deterministic part of $\boldsymbol{\varsigma}_{k}$ defined as below:
\vspace{-0mm}
\begin{align}\label{eq_f}
&\textbf{f}_{\boldsymbol{\psi}} \left(\hat {\boldsymbol{\psi}}_{k-1}\right)  \triangleq \mathbb{E} \left[\boldsymbol{\varsigma}_{k} \right],
\vspace{-0mm}
\end{align}which is a function of $\hat{\boldsymbol{\psi}}_{k-1}$ that takes $\boldsymbol{\psi}$ as a parameter vector. The zero-mean random part of $\boldsymbol{\varsigma}_{k}$, i.e.,  $\hat{\textbf{z}}_k$, is given by
\vspace{-0mm}
\begin{align}\label{eq_zini}
\hat{\textbf{z}}_k  \triangleq \boldsymbol{\varsigma}_{k} - \textbf{f}_{\boldsymbol{\psi}} \left(\hat {\boldsymbol{\psi}}_{k-1}\right).
\end{align}\vspace{-0mm}

The randomness of $\boldsymbol{\varsigma}_{k}$ might cause the proposed algorithm to diverge. However, if we adopt the diminishing step-size as that in \cite{nevel1973stochastic,borkar2008stochastic,kushner2003stochastic}, i.e., 
\vspace{-0mm}
\begin{equation}\label{eq_stepsize}
\begin{aligned}
b_{S,k}=\frac{\epsilon_{S}}{k+K_{S,0}},k=1,2,\cdots
\end{aligned}
\vspace{-0mm}
\end{equation}where $K_{S,0} \geq 0$ and $\epsilon_S > 0$, then some convergence property can be obtained, as described in the following theorem:

\begin{theorem}[\textbf{Convergence to a Unique Stable Point}]\label{Converge to unique stable point}
	\emph{If we adopt the iterative method in \eqref{eq_BF}, \eqref{eq_Tracking} and $b_{S,k}$ is given by \eqref{eq_stepsize} with $\epsilon_S > 0$ and $K_{S,0} \geq 0$, then $\hat {\boldsymbol{\psi}}_k$ converges to a unique stable point of} $\textbf{f}_{\boldsymbol{\psi}} \left(\hat {\boldsymbol{\psi}}_{k-1}\right)$ \emph{with probability one.}
\end{theorem}
\vspace{-0mm}
A point $\hat {\boldsymbol{\psi}}_{k-1}$ is called a stable point of $\textbf{f}_{\boldsymbol{\psi}} \left(\hat {\boldsymbol{\psi}}_{k-1}\right)$ when it satisfies two conditions: 1) $\textbf{f}_{\boldsymbol{\psi}} \left(\hat {\boldsymbol{\psi}}_{k-1}\right) = \textbf{0}$ and 2) $\frac{\partial \textbf{f}_{\boldsymbol{\psi}} \left(\hat {\boldsymbol{\psi}}_{k-1}\right)}{\partial \hat {\boldsymbol{\psi}}_{k-1}^\text{T} }$ is negative definite. Hence, the stable points set is defined as below:
\vspace{-0mm}
\begin{equation}\label{eq_stable_point}
\mathcal{S} \triangleq \left\{\!\hat {\boldsymbol{\psi}}_{k-1}:\textbf{f}_{\boldsymbol{\psi}} \left(\hat {\boldsymbol{\psi}}_{k-1}\right) = 0,\frac{\partial \textbf{f}_{\boldsymbol{\psi}} \left(\hat {\boldsymbol{\psi}}_{k-1}\right)}{\partial \hat {\boldsymbol{\psi}}_{k-1}^\text{T}} \prec \textbf{0} \!\right\},
\vspace{-0mm}
\end{equation}where $\textbf{A} \prec \textbf{0}$ denotes that the matrix $\textbf{A}$ is negative definite. In our problem, $\textbf{f}_{\boldsymbol{\psi}} \left(\hat {\boldsymbol{\psi}}_{k-1}\right)$ defined in \eqref{eq_f} is given by
\vspace{-0mm}\begin{align}\label{eq_fd}
&\textbf{f}_{\boldsymbol{\psi}} \left(\hat {\boldsymbol{\psi}}_{k-1}\right)\\
\!=\!&\frac{2 \lvert \textbf{s} \rvert^2}{\sigma_z^2}\textbf{I}_S\left(\!\hat{\boldsymbol{\psi}}_{k-1}, \textbf{W}_k\right)^\text{\!-\!1}
\!\left[\!
\begin{matrix}
{\text{Re}\left\{\textbf{e}_k^\text{H}\left(\beta \textbf{W}_k^\text{H} \textbf{a}\left(\textbf{x}\right) \!-\!\hat{\beta}_{k\!-1} \textbf{e}_k \right)\!\right\}}\\
{\text{Im}\left\{\textbf{e}_k^\text{H}\!\left(\beta \textbf{W}_k^\text{H} \textbf{a}\left(\textbf{x}\right)\!-\!\hat{\beta}_{k\!-\!1} \textbf{e}_k \right)\right\}}\\
{\text{Re}\left\{\tilde{\textbf{e}}_{k1}^\text{H}\!\left(\beta \textbf{W}_k^\text{H} \textbf{a}\left(\textbf{x}\right)\!-\!\hat{\beta}_{k\!-\!1} \textbf{e}_k \right)\right\}}\\
{\text{Re}\left\{\tilde{\textbf{e}}_{k2}^\text{H}\!\left(\beta \textbf{W}_k^\text{H} \textbf{a}\left(\textbf{x}\right)\!-\!\hat{\beta}_{k\!-\!1} \textbf{e}_k \right)\right\}}
\end{matrix}
\!\right]\!.\nonumber
\end{align}\vspace{-0mm}

\vspace{-2mm}
\noindent
\emph{Proof of Theorem} \ref{Converge to unique stable point}. 
See Appendix \ref{proof_Converge to unique stable point}.$\hfill \blacksquare$
\vspace{2mm}

By Theorem \ref{Converge to unique stable point}, for the general step-size in \eqref{eq_stepsize}, $\hat {\boldsymbol{\psi}}_k$ converges to a unique stable point in $\mathcal{S}$.

\emph{\textbf{2) Convergence to the channel parameter vector $\boldsymbol{\psi}$}}

According to \eqref{eq_stable_point} and \eqref{eq_fd}, it is easy to verify that the channel parameter vector ${\boldsymbol{\psi}}$ is a stable point by the following two points:

1) $\beta \textbf{W}_k^\text{H} \textbf{a}\left(\textbf{x}\right) =\hat{\beta}_{k-1} \textbf{e}_k$ in \eqref{eq_fd} when ${\hat {\boldsymbol{\psi}}_{k-1}= {\boldsymbol{\psi}}}$. Hence, $\textbf{f}_{\boldsymbol{\psi}} \left( {\boldsymbol{\psi}}\right) = \textbf{0}$;
\vspace{2mm}

2)$\frac{\partial \textbf{f}_{\boldsymbol{\psi}} \left(\hat {\boldsymbol{\psi}}_{k-1}\right)}{\partial \hat {\boldsymbol{\psi}}_{k-1}^\text{T} }\big|_{\hat {\boldsymbol{\psi}}_{k-1}= {\boldsymbol{\psi}}}=-\textbf{J}_4$ by derivation, where $\textbf{J}_4$ is the $4$-order identity matrix. Thus,
$\frac{\partial \textbf{f}_{\boldsymbol{\psi}} \left(\hat {\boldsymbol{\psi}}_{k-1}\right)}{\partial \hat {\boldsymbol{\psi}}_{k-1}^\text{T} }\big|_{\hat {\boldsymbol{\psi}}_{k-1}= {\boldsymbol{\psi}}}$ is negative definite.

Therefore, ${\boldsymbol{\psi}}$ is a stable point, i.e., ${\boldsymbol{\psi}} \in \mathcal{S}$. 

Other stable points in $\mathcal{S}$ correspond to the local optimal points of the beam and channel tracking problem, which are out of the main lobe $\mathcal{B}(\textbf{x})$ in \eqref{eq_MainLobe}. Except for the channel parameter vector ${\boldsymbol{\psi}}$, the antenna array gain of other stable points in $\mathcal{S}$ is quite low, resulting in low tracking accuracy. Unfortunately, the estimate of the DPV $\textbf{x}$ may 
jump out of the main lobe in the tracking process and converge to other local optimal points due to the existence of observation noise. Hence, one key challenge is to ensure that the tracking algorithm converges to ${\boldsymbol{\psi}}$ rather than other stable points. Then we develop the following theorem to deal with this challenge:

\vspace{-0mm}
\begin{theorem}[\textbf{Convergence to the DPV x}]\label{Converge to real beam direction}
	\emph{If we adopt the iterative method in \eqref{eq_BF}, \eqref{eq_Tracking} and (i) the initial estimate} of $\textbf{x}$ \emph{is within the main lobe, i.e.,} $\hat {\textbf{x}}_0 \in \mathcal{B}\left(\textbf{x}\right)$; \emph{(ii) $b_{S,k}$ is given by \eqref{eq_stepsize} with $\epsilon_S > 0$, then there exist some $K_{S,0} \geq 0$ and $R > 0$ such that}
	\vspace{-0mm}
	\begin{equation}\label{eq_ConvergeProbablity}
	\begin{aligned}
	P\left(\hat {\textbf{x}}_k \to \textbf{x} \mid \hat {\textbf{x}}_0 \in \mathcal{B}\left(\textbf{x}\right)\right) \geq 1-8e^{-\frac{R \lvert \textbf{s} \rvert^2}{\epsilon_S^2 \sigma_z^2}}.
	\end{aligned}
	\vspace{-0mm}
	\end{equation}
\end{theorem}
\vspace{-0mm}
\begin{proof}
	See Appendix \ref{proof_Converge to real beam direction}.
\end{proof}
\vspace{0mm}

We have assumed that the beam estimator in Fig. \ref{FrameStrcuctureTVT} can output an initial estimate $\hat {\textbf{x}}_0$ within the main lobe $\mathcal{B}\left(\textbf{x}\right)$. Under the condition $\hat {\textbf{x}}_0 \in \mathcal{B}\left(\textbf{x}\right)$, Theorem \ref{Converge to real beam direction} tells us the probability of $\hat {\textbf{x}}_k \to \textbf{x}$ is related to $\frac{ \lvert \textbf{s} \rvert^2}{\epsilon_S^2 \sigma_z^2}$. Hence, we can reduce the step-size or increase the transmit SNR $\frac{\lvert \textbf{s} \rvert^2}{ \sigma_z^2}$ to make sure that $\hat {\textbf{x}}_k \to \textbf{x}$ approaching probability one. 

According to Theorem \ref{Converge to unique stable point}, $\hat{\boldsymbol{\psi}}_k$ converges to a \emph{unique} stable point corresponding to a local optimal point. Hence, this unique stable point will be exactly ${\boldsymbol{\psi}}$ when $\hat {\textbf{x}}_k \to \textbf{x}$, i.e., $\hat{\boldsymbol{\psi}}_k \to {\boldsymbol{\psi}}$.

\emph{\textbf{3) Convergence with the minimum CRLB}}

Finally, the following theorem is developed to tell us the tracking error of the proposed algorithm:
\vspace{-0mm}
\begin{theorem}[\textbf{Convergence to $\boldsymbol{\psi}$ with the minimum CRLB}]\label{Converge to with minimum CRLB}
	\emph{If we adopt the iterative method in \eqref{eq_BF}, \eqref{eq_Tracking} and (i) $\hat {\boldsymbol{\psi}}_k \to {\boldsymbol{\psi}}$; (ii) $b_{S,k}$ is given by \eqref{eq_stepsize} with $\epsilon_S = 1$ and any $K_{S,0} \geq 0$, then $\hat{\textbf{h}}_k - \textbf{h}$ is asymptotically Gaussian and}
	\vspace{-0mm}
	\begin{equation}\label{eq_ConvergeWithMinCRLB}
	\begin{aligned}
	\mathop {\lim }\limits_{k \to +\infty } \frac{k}{MN} \mathbb{E} \left[{\left\| \hat{\textbf{h}}_k - \textbf{h} \right\|}_2^2 \bigg| \hat{\boldsymbol{\psi}}_k \to \boldsymbol{\psi} \right] = {C}_{S}^{\min}(\boldsymbol{\psi}).
	\end{aligned}
	\vspace{-0mm}
	\end{equation}
\end{theorem}
\vspace{-0mm}
\begin{proof}
	See Appendix \ref{proof_Converge to with minimum CRLB}.
\end{proof}
\vspace{-0mm}

By Theorem \ref{Converge to unique stable point}, Theorem \ref{Converge to real beam direction} and Theorem \ref{Converge to with minimum CRLB}, if $\hat {\textbf{x}}_0 \in \mathcal{B}\left(\textbf{x}\right)$ and we adopt the step-size $b_{S,k}$ in \eqref{eq_stepsize} with $\epsilon_S = 1$ and $K_{S,0}\geq 0$, then the minimum CRLB is achieved asymptotically with high probability.


\vspace{-0mm}
\section{Recursive Beam Tracking for Dynamic Case I : Performance Bound, Convergence and Optimality} \label{sec_BeamTrackingDI}
\vspace{-0mm}
In Dynamic Case I, the channel gain changes fast while the beam direction changes slowly. We assume that the beam direction keeps static, i.e., $\textbf{x}_k = \textbf{x}=\left[x_1,x_2\right]^\text{T}$. Hence, the antenna gain in the direction of the arriving path also keeps static, i.e., $\eta\left(\textbf{x}_k\right) = \eta\left(\textbf{x}\right)$. When the channel gain $\beta_k^c$ changes fast, it is very difficult to establish theorems of tracking the channel gain and beam direction simultaneously, as in Section \ref{sec_Quasi-static_Tracking}. Fortunately, acquiring the beam direction information is sufficient for alignment in mmWave mobile communication with analog beamforming. Hence, we only focus on beam direction tracking in Dynamic Case I.

The channel gains of adjacent ECCs in this section are assumed to be independent of each other. In addition, different distributions of the channel gain $\beta_k^c$ can lead to different suitable tracking strategies. The tracking strategy designed for one distribution of the channel gain may deteriorate sharply when applied to other distributions. Hence, each type of channel gain distribution deserves studying, of which Rayleigh fading channel is a special case that is easier to be analyzed. This special case happens when quite a number of rays existing in a cluster are indistinguishable. In this section, we choose Rayleigh fading channels to study for Dynamic Case I, i.e., $\beta_k^c \sim \mathcal{CN}\big(0,\big(\sigma_{\beta}^c\big)^2\big)$. Although the theoretical results in this section are only applicable for Rayleigh fading channels, the proposed algorithm is found robust for other types of time-varying channels according to the numerical results in Section \ref{subsec_simulation_accuracy}.


When the channel gain $\beta_k^c$ is Gaussian distributed, the equivalent channel gain $\beta\left(\textbf{x}\right) = \eta\left(\textbf{x}\right)\beta_k^c$ also satisfies Gaussian distribution with the variance given below:
\vspace{-1mm}
\begin{equation}\label{eq_covariance_beta}
\mathbb{E}\left[\lvert \beta\left(\textbf{x}\right) \rvert^2\right]=\lvert\eta\left(\textbf{x}\right)\rvert^2\left(\sigma_{\beta}^c\right)^2 \triangleq \sigma_{\beta}^2.
\end{equation}\vspace{-0mm}Correspondingly, the observation vector $\textbf{y}_k$ in \eqref{eq_observation_vector} satisfies Gaussian distribution for a given DPV $\textbf{x}$ and EBM $\textbf{W}_{k}$, i.e., $\textbf{y}_k \sim \mathcal{CN}\left(\textbf{0},\boldsymbol{\Sigma}_{\textbf{y},k}\right)$, where $\boldsymbol{\Sigma}_{\textbf{y},k}$ is the covariance matrix of $\textbf{y}_k$ defined as follows:
\vspace{-0mm}
\begin{small}
\begin{equation}\label{eq_SigmaDI}
\boldsymbol{\Sigma}_{\textbf{y},k} \!\triangleq\! \mathbb{E}\left[\textbf{y}_k\textbf{y}_k^\text{H}\right]\\
\!=\!\lvert \textbf{s} \rvert^2 \sigma_{\beta}^2 \textbf{W}_{k}^\text{H}\textbf{a}\left(\textbf{x}\right)\!\left(\textbf{W}_{k}^\text{H}\textbf{a}\left(\textbf{x}\right)\right)^\text{H}\!+\!\sigma_z^2\textbf{J}_3.\!
\vspace{-0mm}
\end{equation}\end{small}According to \eqref{eq_SigmaDI}, we can obtain the determinant of
 $\boldsymbol{\Sigma}_{\textbf{y},k}$:
\begin{equation}\label{eq_SigmaD}
\vspace{-0mm}
\lvert\boldsymbol{\Sigma}_{\textbf{y},k} \rvert = \sigma_z^{4}\left(\sigma_z^2+\lvert\textbf{s}\rvert^2\sigma_{\beta}^2 \lvert\textbf{W}_{k}^\text{H}\textbf{a}\left(\textbf{x}\right)\rvert^2\right).
\vspace{-0mm}
\end{equation}Then the conditional probability density function of $\textbf{y}_k$ is given by
\vspace{-0mm}
\begin{equation}\label{eq_pdfDI}
{p_{DI}(\textbf{y}_k| \textbf{x}, \textbf{W}_k) = {\frac{1}{\pi^{3} \lvert \boldsymbol{\Sigma}_{\textbf{y},k}\rvert} e^{-\textbf{y}_k^\text{H}\boldsymbol{\Sigma}_{{\textbf{y}},k}^{-1}\textbf{y}_k}},
}
\vspace{-0mm}
\end{equation}

The following structure of this section is similar to Section \ref{sec_Quasi-static_Tracking}: we first formulate the beam tracking problem and provide the lower bound of it. Then we develop a tracking algorithm and prove this algorithm can converge to the minimum CRLB.

\vspace{-0mm}
\subsection{Problem Formulation}\label{sbsec_ProblemDI}
\vspace{-0mm}
Since we only track the beam direction in Dynamic Case I, the  estimation function in \eqref{eq_Fe} is reformulated as follows:
\vspace{-0mm}
\begin{align}
\label{eq_FeDI}\hat{\textbf{x}}_k = \textbf{F}_{DI,k}^e\left(\hat {\boldsymbol{\psi}}_0,\textbf{W}_{1}, \cdots,\textbf{W}_{k}, \textbf{y}_1,\cdots,\textbf{y}_{k}\right).
\vspace{-0mm}
\end{align}Let $\Xi_{DI,k}=\left\{\textbf{F}_k^c,\textbf{F}_{DI,k}^e\right\}$ denote a beam tracking scheme set in $k$-th ECC: based on historical observation vectors $\textbf{y}_{1}, \cdots, \textbf{y}_{k-1}$ along with the corresponding EBMs $\textbf{W}_{1}, \cdots, \textbf{W}_{k-1}$, choose an appropriate EBM $\textbf{W}_{k}$, apply it to obtain $\textbf{y}_{k}$ and make an estimation of the DPV $\textbf{x}$ in $k$-th ECC by using all EBMs and observations available. Hence, in $k$-th ECC, the tracking problem is formulated as:
\vspace{-0mm}
\begin{align}\label{eq_problemDI}
\underset{ \Xi_{DI,k}}{\min} ~&\mathbb{E} \left[{\left\|\hat{\textbf{x}}_k - \textbf{x}\right\|}_2^2 \right] \\\vspace{-2mm}
\label{eq_constrant1DI} \text{s.t.} ~& \mathbb{E}\left[\hat{\textbf{x}}_k\right] = \textbf{x},\\\vspace{-2mm}
~&  \eqref{eq_observation_vector},\eqref{eq_Fc}, \eqref{eq_sv},  \eqref{eq_FeDI},\nonumber\vspace{-4mm}
\end{align}where the constraint \eqref{eq_constrant1DI} ensures that $\hat{\textbf{x}}_k$ is an unbiased estimate of the DPV $\textbf{x}$.

Before providing a specific tracking algorithm, we will first explore the performance bound of the problem in \eqref{eq_problemDI}.

\vspace{-0mm}
\subsection{Cram\'{e}r-Rao Lower Bound of Tracking Error}\label{subsec_BoundDI}
\vspace{-0mm}
We now perform some theoretical analysis on the beam tracking problem. Based on the CRLB theory in \cite{Sengijpta1993Fundamental}, we introduce the following lemma to obtain the lower bound of the tracking error:
\vspace{-0mm}
\begin{lemma}\label{MSEOptDI}
\emph{In Dynamic Case I, given $\textbf{W}_1,\cdots,\textbf{W}_k$}, \emph{the MSE of the DPV estimate in \eqref{eq_problemDI} is lower bounded as follows:}
\vspace{-0mm}
		\begin{align}\label{MSELBDI}
		\mathbb{E}\left[\left\| \hat{\textbf{x}}_k - \textbf{x} \right\|_2^2 \right]\ge  \Tr \left\{\left(\sum\limits_{l=1}^k\textbf{I}_{DI}\left(\textbf{x},{{\bf{W}}_{l}}\right)\right)^{-1}\right\},
		\end{align}
	\vspace{-0mm}\emph{where the Fisher information matrix } $\textbf{I}_{DI}\left(\textbf{x} ,{{\bf{W}}_{l}}\right)$ \emph{is given by}\vspace{-0mm}
		\begin{small}\begin{align}\label{eq_fisherDI}
			\textbf{I}_{DI}\!\left(\textbf{x} ,{{\bf{W}}_{l}}
			\right)\!\triangleq\!   \mathbb{E}\left[\!\frac {\partial \text{log} \, p_{DI} \left(\textbf{y}_k |\textbf{x},\textbf{W}_{l} \right)}{\partial \textbf{x}} \!\cdot\! \frac {\partial \text{log}\, p_{DI} \left(\textbf{y}_k |\textbf{x},\textbf{W}_{l} \right)}{\partial \textbf{x}^\text{T}}\!\right]\!,\vspace{-0mm}
			\end{align}\end{small}\emph{and the} $p$\emph{-th row,} $j$\emph{-th column} $\left(p=1,2;j=1,2\right)$ \emph{of} $\textbf{I}_{DI}\left(\textbf{x} ,{{\bf{W}}_{l}}\right)$  \emph{is derived by \eqref{eq_fisherDI_ele}}
		      \begin{figure*}[!t]		
			\normalsize
			\vspace{-0mm}
			\begin{equation}\label{eq_fisherDI_ele}
			\left[\textbf{I}_{DI}\left(\textbf{x} ,{{\bf{W}}_{l}}\right)\right]_{p,j} =\frac {\sigma_z^6\lvert \textbf{s} \rvert^6 \sigma_{\beta}^6} {{\lvert\boldsymbol{\Sigma}_{\textbf{y},k}\rvert}^2}\left\{
			-2  \lvert\textbf{g}_l\rvert^2 \tilde{{g}}_{l,p}\tilde{{g}}_{l,j}+\frac{\sigma_z^2}{\lvert \textbf{s} \rvert^2 \sigma_{\beta}^2}\Tr\left\{\textbf{G}_{l,p}\textbf{G}_{l,j}\right\}+ \textbf{g}_{l}^\text{H}\left(\textbf{G}_{l,p}\textbf{G}_{l,j}+\textbf{G}_{l,j}\textbf{G}_{l,p}\right)\textbf{g}_{l}
			\right\}
			\end{equation}
			\vspace{-0mm}
			\hrulefill
			\vspace*{-0mm}
		\end{figure*}with $\textbf{g}_{l}$, $\tilde{{g}}_{l,p}$ and $\textbf{G}_{l,p}$ defined below:
	\vspace{-0mm}
		\begin{equation}\label{eq_partinfDI}
		\left\{
		\begin{aligned}
		&\textbf{g}_{l} \triangleq  \textbf{W}_{l}^\text{H}\textbf{a}\left(\textbf{x}\right) \\\vspace{0mm}
		&\tilde{{g}}_{l,p}\triangleq\frac{\partial {\lvert\textbf{g}_l\rvert^2}}{\partial {x_{p}}},p=1,2 \\\vspace{0mm}	
		&\textbf{G}_{l,p}\triangleq\frac{\partial {\textbf{g}_l \textbf{g}_l^\text{H}}}{\partial {x_{p}}},p=1,2
		\end{aligned}
		\right..
		\end{equation}
	\vspace{-0mm}
\end{lemma}
\vspace{-8mm}
\begin{proof}
	See Appendix \ref{proof_MSEOptDI}.
\end{proof}
\vspace{-0mm}
 
The CRLB in \eqref{MSELBDI} is a function of the EBMs $\textbf{W}_1,\ldots,\textbf{W}_k$. Similar to that in Quasi-static Case, we consider the normalized CRLB (by multiplying $k$):
\begin{align}\label{eq_CMMSETempDI}
\vspace{-0mm}
{C}_{\begin{small}DI\end{small}}(\textbf{x},\textbf{W}) &\triangleq \Tr \left\{\textbf{I}_{DI}\left(\textbf{x},{{\bf{W}}}\right)^{-1}\right\}.
\vspace{-0mm}
\end{align}By optimizing only one EBM $\textbf{W}$, we can further get the \emph{minimum CRLB}, given by
\vspace{-0mm}
\begin{align}\label{eq_CMMSEDI}
{C}_{\begin{small}DI\end{small}}^{\min}(\textbf{x})=&\min_{\textbf{W}} {C}_{\begin{small}DI\end{small}}(\textbf{x},\textbf{W})={C}_{\begin{small}DI\end{small}}(\textbf{x},\textbf{W}_{\begin{small}DI\end{small}}^*).
\vspace{-0mm}
\end{align}Solving problem \eqref{eq_CMMSEDI} yields the optimal EBM $\textbf{W}_{\begin{small}DI\end{small}}^*=\big[\textbf{w}_{\begin{small}DI\end{small},1}^*,\textbf{w}_{\begin{small}DI\end{small},2}^*, \textbf{w}_{\begin{small}DI\end{small},3}^*\big]$ with
\vspace{-0mm}
\begin{equation}\label{eq_bfo}
\textbf{w}_{\begin{small}DI\end{small},i}^* = \frac{1}{\sqrt{MN}}\textbf{a}\left(\textbf{x}+ \boldsymbol{\Delta}_{\begin{small}DI\end{small},i}^*\right),i=1,2,3,
\vspace{-0mm}
\end{equation}where $\left\{\boldsymbol{\Delta}_{\begin{small}DI\end{small},1}^*,\boldsymbol{\Delta}_{\begin{small}DI\end{small},2}^*,\boldsymbol{\Delta}_{\begin{small}DI\end{small},3}^*\right\}$ denotes the optimal set of exploration offsets in Dynamic Case I.

\vspace{-0mm}
\subsection{Asymptotically Optimal Set of Exploration Offsets}\label{sbsec_OptBeamforming}
\vspace{-0mm}

In general, the CRLB in \eqref{eq_CMMSEDI} is a function of a set of system parameters including the equivalent channel gain parameter $\sigma_{\beta}^2$, the DPV $\textbf{x}$ and the array size $M,\,N$. Hence, the optimal set of 2D exploration offsets should also be a function of these parameters. Since it is very hard to obtain the expression of this optimal set, we adopt numerical search to deal with this issue. However, as many parameters in \eqref{eq_CMMSEDI} may affect the optimal result, numerical search has to be reconducted for different parameter sets, resulting in high complexity.

Fortunately, through our investigation, some useful properties of the minimum CRLB and the optimal set of exploration offsets are given to simplify the numerical search, as described in the following lemma:
\vspace{-0mm}
\begin{lemma}\label{UnifiedOptShiftDI}
		
		\emph{In Dynamic Case I, the minimum CRLB} ${C}_{DI}^{\min}(\boldsymbol{\psi})$ \emph{and the optimal set of exploration offsets $\left\{\boldsymbol{\Delta}_{\begin{small}DI\end{small},1}^*,\boldsymbol{\Delta}_{\begin{small}DI\end{small},2}^*,\boldsymbol{\Delta}_{\begin{small}DI\end{small},3}^*\right\}$ have the following three properties}:
		
		\emph{1)} ${C}_{DI}^{\min}(\boldsymbol{\psi})$ \emph{and} $\left\{\boldsymbol{\Delta}_{\begin{small}DI\end{small},1}^*,\boldsymbol{\Delta}_{\begin{small}DI\end{small},2}^*,\boldsymbol{\Delta}_{\begin{small}DI\end{small},3}^*\right\}$ \emph{are invariant to the DPV} $\textbf{x}$;
		
		\emph{2)}  $\frac{\lvert \textbf{s} \rvert^2  \sigma_{\beta}^2}{\sigma_z^2}{C}_{DI}^{\min}(\boldsymbol{\psi})$ \emph{converges to constant values as} $ \frac{\lvert \textbf{s} \rvert^2  \sigma_{\beta}^2}{\sigma_z^2} \to +\infty$;
		
		\emph{3)}
		${C}_{DI}^{\min}(\boldsymbol{\psi})$ \emph{converges as} $\emph{M},\,\emph{N} \to +\infty$ \emph{and there exists a fixed set of exploration offsets that are unrelated to the array size and} $\frac{\lvert \textbf{s} \rvert^2  \sigma_{\beta}^2}{\sigma_z^2}$, \emph{denoted as} $\left\{\widetilde{\boldsymbol{\Delta}}_{DI,1}^{*},\,\widetilde{\boldsymbol{\Delta}}_{DI,2}^{*},\widetilde{\boldsymbol{\Delta}}_{DI,3}^{*}\right\}$, \emph{such that} 
		\vspace{-0mm}
		\begin{align}\label{eq_asymtDI} {\lim\limits_{M,N \to +\infty}}{C}_{DI}(\boldsymbol{\psi},\widetilde{\textbf{W}}_{DI}^*) = {\lim\limits_{M,N \to +\infty}}{C}_{DI}^{\min}(\boldsymbol{\psi}),
		\nonumber
		\vspace{-0mm}
		\end{align}\emph{where} $\widetilde{\textbf{W}}_{DI}^{*}=[\tilde{\textbf{w}}_{DI,1}^{*},\tilde{\textbf{w}}_{DI,2}^{*},\tilde{\textbf{w}}_{DI,3}^{*}]$ \emph{is obtained with}
		\vspace{-0mm}
		\begin{equation}\label{eq_obfDI}
		\tilde{\textbf{w}}_{DI,i}^{*} \triangleq \frac{1}{\sqrt{MN}}\textbf{a}\left(\textbf{x}+\widetilde{\boldsymbol{\Delta}}_{DI,i}^{*}\right),i=1,2,3.
		\vspace{-0mm}
		\end{equation} 
		\vspace{-0mm}
\end{lemma}
\vspace{-0mm}
\begin{proof}
See Appendix \ref{proof_UnifiedOptShiftDI}.
\end{proof}
\vspace{-0mm}

\renewcommand\arraystretch{1.4}
\begin{table}[!t!]
	\vspace{-0mm}
	\centering
	\caption{ {The asymptotically optimal set of exploration offsets in Dynamic Case I.}}
	\vspace{2mm}
	\begin{tabular} {c|c|c}
		\hline
		\hline
		$\widetilde{\boldsymbol{\Delta}}_{\begin{small}DI\end{small},1}^{*}$ & $\widetilde{\boldsymbol{\Delta}}_{\begin{small}DI\end{small},2}^{*}$ & $\widetilde{\boldsymbol{\Delta}}_{\begin{small}DI\end{small},3}^{*}$\\
		\hline
		$\left[0.5486,0.2451\right]^\text{T}$ &
		$\left[-0.5462, 0.2482\right]^\text{T}$ & $\left[-0.0012,-0.6837\right]^\text{T}$\\
		\hline
		\hline
	\end{tabular}\label{Tab_asymptotically optimal_parametersDI}
\end{table}
\begin{figure}[!t!]
	\centering
	\vspace{-0mm}
	\includegraphics[width=7.5cm]{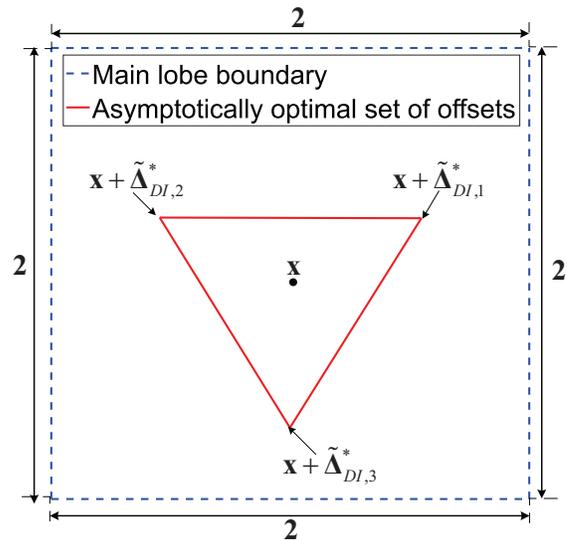}
	\vspace{-0mm}
	\caption{The asymptotically optimal set of exploration offsets in Dynamic Case I.}
	\label{fig_offsetsDI}
\end{figure}
Lemma \ref{UnifiedOptShiftDI} reveals that $\left\{\boldsymbol{\Delta}_{\begin{small}DI\end{small},1}^*,\boldsymbol{\Delta}_{\begin{small}DI\end{small},2}^*,\boldsymbol{\Delta}_{\begin{small}DI\end{small},3}^*\right\}$ is only related to the array size $M,\,N$ and $\frac{\lvert \textbf{s} \rvert^2  \sigma_{\beta}^2}{\sigma_z^2}$. Hence, the numerical search times can be reduced to one for a particular array size $M,\,N$ and a particular $\frac{\lvert \textbf{s} \rvert^2  \sigma_{\beta}^2}{\sigma_z^2}$. Numerically, we find later that even if $\left\{\boldsymbol{\Delta}_{\begin{small}DI\end{small},1}^*,\boldsymbol{\Delta}_{\begin{small}DI\end{small},2}^*,\boldsymbol{\Delta}_{\begin{small}DI\end{small},3}^*\right\}$ may change for different array sizes and $\frac{\lvert \textbf{s} \rvert^2  \sigma_{\beta}^2}{\sigma_z^2}$, $\left\{\widetilde{\boldsymbol{\Delta}}_{\begin{small}DI\end{small},1}^{*},\,\widetilde{\boldsymbol{\Delta}}_{\begin{small}DI\end{small},2}^{*},\widetilde{\boldsymbol{\Delta}}_{\begin{small}DI\end{small},3}^{*}\right\}$ can be used to take the place of $\left\{\boldsymbol{\Delta}_{\begin{small}DI\end{small},1}^*,\boldsymbol{\Delta}_{\begin{small}DI\end{small},2}^*, \boldsymbol{\Delta}_{\begin{small}DI\end{small},3}^*\right\}$ as long as the antenna size $M,N$ and $\frac{\lvert \textbf{s} \rvert^2  \sigma_{\beta}^2}{\sigma_z^2}$ are sufficiently large. Therefore, the numerical search times is reduced to one in the end. Similar to that in Quasi-static Case,  $\left\{\widetilde{\boldsymbol{\Delta}}_{DI,1}^{*},\,\widetilde{\boldsymbol{\Delta}}_{DI,2}^{*},\widetilde{\boldsymbol{\Delta}}_{DI,3}^{*}\right\}$ is called the asymptotically optimal set of exploration offsets in Dynamic Case I in this paper.

By numerical search in the main lobe in \eqref{eq_MainLobe}, one asymptotically optimal set of exploration offsets $\left\{\widetilde{\boldsymbol{\Delta}}_{\begin{small}DI\end{small},1}^{*},\,\widetilde{\boldsymbol{\Delta}}_{\begin{small}DI\end{small},2}^{*},\widetilde{\boldsymbol{\Delta}}_{\begin{small}DI\end{small},3}^{*}\right\}$ can be obtained in TABLE \ref{Tab_asymptotically optimal_parametersDI} and Fig. \ref{fig_offsetsDI}. With this set of exploration offsets, a general way to generate the EBM $\widetilde{\textbf{W}}_{\begin{small}DI\end{small}}^{*}$ is obtained by \eqref{eq_obfDI} to achieve the minimum CRLB.

By adopting $\left\{\widetilde{\boldsymbol{\Delta}}_{\begin{small}DI\end{small},1}^{*},\,\widetilde{\boldsymbol{\Delta}}_{\begin{small}DI\end{small},2}^{*},\widetilde{\boldsymbol{\Delta}}_{\begin{small}DI\end{small},3}^{*}\right\}$ to smaller size antenna arrays when $\frac{\lvert \textbf{s} \rvert^2  \sigma_{\beta}^2}{\sigma_z^2} = 0\,\text{dB}$, we compare the minimum CRLB and the CRLB achieved by $\left\{\widetilde{\boldsymbol{\Delta}}_{\begin{small}DI\end{small},1}^{*},\,\widetilde{\boldsymbol{\Delta}}_{\begin{small}DI\end{small},2}^{*},\widetilde{\boldsymbol{\Delta}}_{\begin{small}DI\end{small},3}^{*}\right\}$ in TABLE \ref{Tab_asymptotically optimal_parametersDI}. As illustrated in Fig. \ref{fig_ExtensibilityDI_MN}, when antenna number $M=N \geq 8$, we can approach the minimum CRLB with a relative error less than $0.1\%$ by using $\left\{\widetilde{\boldsymbol{\Delta}}_{\begin{small}DI\end{small},1}^{*},\,\widetilde{\boldsymbol{\Delta}}_{\begin{small}DI\end{small},2}^{*},\widetilde{\boldsymbol{\Delta}}_{\begin{small}DI\end{small},3}^{*}\right\}$. 

By applying $\left\{\widetilde{\boldsymbol{\Delta}}_{\begin{small}DI\end{small},1}^{*},\,\widetilde{\boldsymbol{\Delta}}_{\begin{small}DI\end{small},2}^{*},\widetilde{\boldsymbol{\Delta}}_{\begin{small}DI\end{small},3}^{*}\right\}$ to different $\frac{\lvert \textbf{s} \rvert^2  \sigma_{\beta}^2}{\sigma_z^2}$ when $M=N=8$, we compare the minimum CRLB and the CRLB achieved by $\left\{\widetilde{\boldsymbol{\Delta}}_{\begin{small}DI\end{small},1}^{*},\,\widetilde{\boldsymbol{\Delta}}_{\begin{small}DI\end{small},2}^{*},\widetilde{\boldsymbol{\Delta}}_{\begin{small}DI\end{small},3}^{*}\right\}$ in TABLE \ref{Tab_asymptotically optimal_parametersDI}. As illustrated in Fig. \ref{fig_ExtensibilityDI_SNR.eps}, when $\frac{\lvert \textbf{s} \rvert^2  \sigma_{\beta}^2}{\sigma_z^2} \geq 0\,\text{dB}$, we can approach the minimum CRLB with a relative error less than $0.1\%$ by using $\left\{\widetilde{\boldsymbol{\Delta}}_{\begin{small}DI\end{small},1}^{*},\,\widetilde{\boldsymbol{\Delta}}_{\begin{small}DI\end{small},2}^{*},\widetilde{\boldsymbol{\Delta}}_{\begin{small}DI\end{small},3}^{*}\right\}$.

As a conclusion, it is practical to apply this asymptotically optimal set of exploration offsets to any antenna array with $M=N\geq8$, any channel gain with $\frac{\lvert \textbf{s} \rvert^2  \sigma_{\beta}^2}{\sigma_z^2}\geq 0\text{dB}$ and any direction.


\begin{figure}[!t]
	\centering
	\vspace{-0mm}
	\includegraphics[width=3.4in]{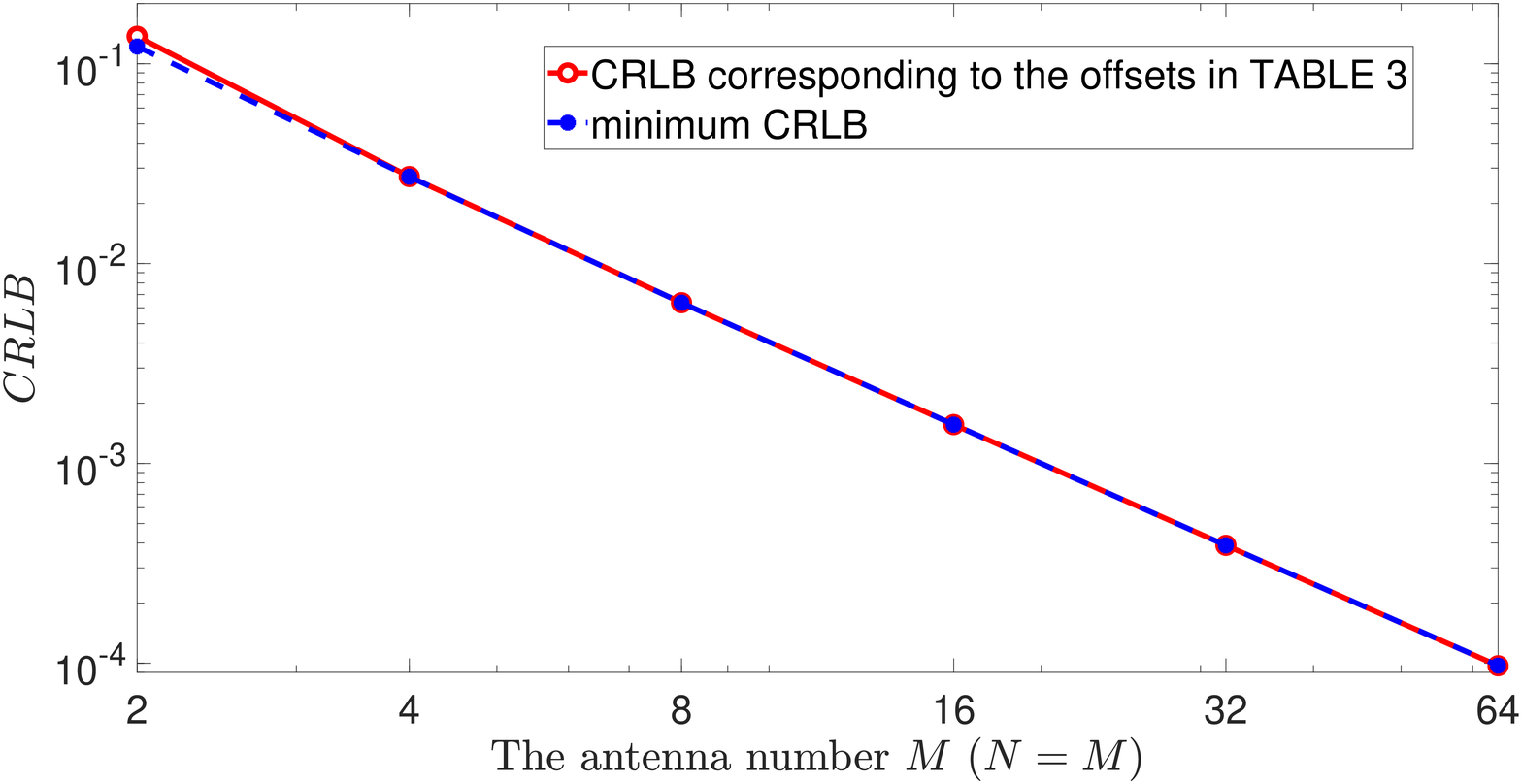}
	\vspace{-0mm}
	\caption{\small{The performance of the offsets in TABLE \ref{Tab_asymptotically optimal_parametersDI} when $\frac{\lvert \textbf{s} \rvert^2  \sigma_{\beta}^2}{\sigma_z^2} = 0\,\text{dB}$.}}
	\label{fig_ExtensibilityDI_MN}
	\vspace{-0mm}
\end{figure}
\begin{figure}[!t]
	\centering
	\vspace{-0mm}
	\includegraphics[width=3.4in]{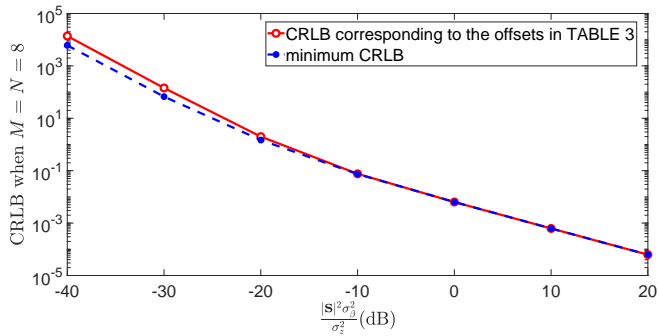}
	\vspace{-0mm}
	\caption{\small{The performance of the offsets in TABLE \ref{Tab_asymptotically optimal_parametersDI} when $M=N=8$.}}
	\label{fig_ExtensibilityDI_SNR.eps}
\end{figure}

\vspace{-0mm}
\subsection{Recursive Beam Tracking with Asymptotic Optimality Analysis}\label{sbsec_Tracking}
\vspace{-0mm}
For the Rayleigh fading channels, it is crucial to acquire the variance of the equivalent channel gain in \eqref{eq_covariance_beta}, while it is hindered by the unknown antenna gain $\eta\left(\textbf{x}\right)$. Fortunately, the estimate of $\eta\left(\textbf{x}\right)$ and $\sigma_{\beta}^2$ can be seen as approximately accurate for a given antenna element pattern as the estimate of the DPV $\hat{\textbf{x}}_k$ approach $\textbf{x}$. Hence, we assume a perfectly-known  $\sigma_{\beta}^2$ here to design the algorithm in Dynamic Case I. The deterioration of the tracking performance caused by the estimation error of the antenna gain will be evaluated in Section \ref{subsec_simulation_pattern}.

The proposed tracker is motivated by the following maximum likelihood problem:
\vspace{-0mm}
\begin{align}\label{eq_MLEDI}
\underset{\textbf{W}_k}{\max}&\left\{\!\underset{\hat{\textbf{x}}_k}{\max}\sum\limits_{l=1}^k\left[\text{log} \, p_{DI} \left(\textbf{y}_l \big|{\textbf{x}},\textbf{W}_{l}
 \right)\bigg|_{\textbf{x}=\hat{\textbf{x}}_k}\right]\right\}\\
 \text{s.t.} ~&  \eqref{eq_observation_vector},\eqref{eq_Fc}, \eqref{eq_sv}, \eqref{eq_FeDI},\eqref{eq_constrant1DI}.\nonumber
\end{align}\vspace{-0mm}Similar to that in Section \ref{sec_Quasi-static_Tracking}, we propose a two-layer nested optimization algorithm to find the solution of \eqref{eq_MLEDI}. Finally, the proposed tracking algorithm is given in Algorithm \ref{alg_DI}.

We now perform the asymptotic optimality analysis. According to  \cite{nevel1973stochastic,borkar2008stochastic,kushner2003stochastic}, the diminishing step-size is adopted as follows: 
\vspace{-0mm}
\begin{equation}\label{eq_stepsizeDI}
\begin{aligned}
b_{\begin{small}DI\end{small},k}=\frac{\epsilon_{{\begin{small}DI\end{small}}}}{k+K_{DI,0}},k=1,2,\cdots
\end{aligned}
\vspace{-0mm}
\end{equation}where $K_{DI,0} \geq 0$ and $\epsilon_{{\begin{small}DI\end{small}}} > 0$. Then we can prove that if the initial estimate $\hat{\textbf{x}}_0$ is within the main lobe and $\epsilon_{{\begin{small}DI\end{small}}} = 1$, the proposed algorithm can converge to $\textbf{x}$ with the minimum CRLB with high probability, i.e.,
\begin{equation}\label{eq_ConvergeWithMinCRLBDI}
\vspace{-0mm}
\begin{aligned}
\mathop {\lim }\limits_{k \to +\infty } k \mathbb{E} \left[{\left\| \hat{\textbf{x}}_k - \textbf{x} \right\|}_2^2\right] = {C}_{\begin{small}DI\end{small}}^{\min}(\textbf{x}).
\end{aligned}
\end{equation}The proof is similar to that in Section \ref{sec_Quasi-static_Tracking} and the details are omitted here since nothing new is provided in the proof.

\begin{algorithm}[!t]
	\caption{\textbf{Recursive Beam Tracking} (\textbf{RBT}) for Dynamic Case I}
	\label{alg_DI}
	\begin{algorithmic}[0]
		\State 1) \textbf{Exploring and Receiving} (Step 3 in Procedure \ref{alg_trackingloop}): Transmit 3 pilot sequences in each ECC. The corresponding EBV for receiving the $i$-th pilot sequence in $k$-th ECC is given below:
		\vspace{-0mm}
		\begin{equation}\label{eq_BFDI}
		\begin{aligned}\textbf{w}_{k,i} = \frac {1}{\sqrt{MN}} \textbf{a} \left(\hat{\textbf{x}}_{k-1} + \widetilde{\boldsymbol{\Delta}}_{DI,i}^{*}\right), i =1,2,3,
		\end{aligned}
		\vspace{-0mm}
		\end{equation}where $\hat{\textbf{x}}_{k} = \left[\hat{x}_{k,1},\hat{x}_{k,2}\right]^\text{T}$ and $\left\{\widetilde{\boldsymbol{\Delta}}_{DI,1}^{*},\,\widetilde{\boldsymbol{\Delta}}_{DI,2}^{*},\widetilde{\boldsymbol{\Delta}}_{DI,3}^{*}\right\}$ is given by TABLE \ref{Tab_asymptotically optimal_parametersDI}. After match filtering, the observation vector $\textbf{y}_k$ is obtained via \eqref{eq_observation_vector}. 
		
		\State 2) \textbf{Updating Estimate} (Step 6 in Procedure \ref{alg_trackingloop}): The estimate $\hat {\textbf{x}}_k = \left[\hat x_{k,1},\hat {x}_{k,2}\right]^\text{T}$ is updated by
		\vspace{-0mm}
		\begin{small}
		\begin{equation}\label{eq_TrackingDI}
		\begin{aligned} {\hat{\textbf{x}}}_k \!=\!  {\hat{\textbf{x}}}_{k\!-1}\!+\! b_{\begin{small}DI\end{small},k} \textbf{I}_{DI}\left( {\hat{\textbf{x}}}_{k\!-\!1}, \textbf{W}_k\right)^\text{\!-1} \!\frac {\partial \text{log} \, p_{DI} \left(\textbf{y}_k |\hat{\textbf{x}}_{k\!-\!1},\textbf{W}_{k} \right)}{\partial \hat{\textbf{x}}_{k\!-\!1}}\!,\!
		\end{aligned}
		\vspace{-0mm}
		\end{equation}\end{small}where $\textbf{I}_{DI}\left( {\hat{\textbf{x}}}_{k-1}, \textbf{W}_k\right)$ is defined in \eqref{eq_fisherDI} and $b_{\begin{small}DI\end{small},k}$ is the step size that will be specified later.
	\end{algorithmic}
\end{algorithm}
\vspace{-0mm}
\section{Joint Beam and Channel Tracking for Dynamic Case II}\label{sec_BeamtrackingDII}
\vspace{-0mm}
\begin{algorithm}[!t]
	\caption{{\textbf{Joint Beam and Channel Tracking} (\textbf{JBCT}) for Dynamic Case II}}
	\label{alg_DII}
	\begin{algorithmic}[0]
		\State 1) \textbf{Exploring and Receiving} (Step 3 in Procedure \ref{alg_trackingloop}): Transmit 3 pilot sequences in each ECC. The corresponding EBV for receiving the $i$-th pilot sequence in $k$-th ECC is given below:
		\vspace{-0mm}
		\begin{equation}\label{eq_BFDII}
		\begin{aligned}\textbf{w}_{k,i} = \frac {1}{\sqrt{MN}} \textbf{a} \left(\hat{\textbf{x}}_{k-1} + {\boldsymbol{\Delta}}_{DII,i}\right), i =1,2,3,
		\end{aligned}
		\end{equation}where $\hat{\textbf{x}}_{k} = \left[\hat{x}_{k,1},\hat{x}_{k,2}\right]^\text{T}$ and ${\boldsymbol{\Delta}}_{DII,i}=\widetilde{\boldsymbol{\Delta}}_{S,i}^{*}\,(i=1,2,3)$ are given by TABLE \ref{Tab_asymptotically optimal_parameters}. After match filtering, the observation vector $\textbf{y}_k$ is obtained via \eqref{eq_observation_vector}. 
		
		\State 2) \textbf{Updating Estimate} (Step 6 in Procedure \ref{alg_trackingloop}): The estimate of the channel parameter vector in $k$-th ECC, i.e., $\hat {\boldsymbol{\psi}}_k = \left[\hat{\beta}_k^\text{re},\hat{\beta}_k^\text{im},\hat x_{k,1},\hat {x}_{k,2}\right]^\text{T}$, is updated by
		\vspace{-0mm}
	\begin{equation}\label{eq_TrackingDII}
	\begin{aligned} \hat {\boldsymbol{\psi}}_{k} = \hat {\boldsymbol{\psi}}_{k-1}+ b_{DII,k} \boldsymbol{\varsigma}_{k},
	\end{aligned}
	\vspace{-0mm}
	\end{equation}where $\boldsymbol{\varsigma}_{k}$ is the updating direction vector given by \eqref{eq_udcal} and $b_{\begin{small}DII\end{small},k}$ is the step size for Dynamic Case II.
	\end{algorithmic}
\end{algorithm}
In Dynamic Case II where both the channel gain $\beta_k^c$ and the DPV $\textbf{x}_k$ change fast, the observation vector $\textbf{y}_k$ satisfies normal distribution with $\textbf{y}_k \sim \mathcal{CN}\left(\lvert \textbf{s} \rvert \beta\left(\textbf{x}_k\right) \textbf{W}_k^\text{H} \textbf{a}(\textbf{x}_k),\sigma_z^2\textbf{J}_3\right)$ for a given channel parameter vector $\boldsymbol{\psi}_k$ and EBM $\textbf{W}_{k}$. Hence, the conditional probability density function of the observation vector $\textbf{y}_k$ is given by
\vspace{-0mm}
\begin{equation}\label{eq_pdfDII}
p_{DII}(\textbf{y}_k| \boldsymbol{\psi}_k, \textbf{W}_k) = {\frac{1}{\pi^{3} \sigma_z^{6}} e^{- \frac {{\left\| \textbf{y}_k-\lvert\textbf{s}\rvert \beta\left(\textbf{x}_{k}\right) \textbf{W}_k^\text{H} \textbf{a}(\textbf{x}_{k}) \right\|}_2^2} {\sigma_z^2}}}.
\vspace{-0mm}
\end{equation}
Establishing theorems of tracking, as in Section \ref{sec_Quasi-static_Tracking} and Section \ref{sec_BeamTrackingDI}, is very difficult in Dynamic Case II. Even if the theoretical analysis is not conducted in this section, we  still provide a tracking algorithm in this section.

Inspired by the asymptotically optimal tracking algorithm in Section \ref{sec_Quasi-static_Tracking} and Section \ref{sec_BeamTrackingDI}, we design a similar joint beam and channel tracking algorithm in Algorithm \ref{alg_DII}.

Different from the step-size in Quasi-static Case and Dynamic Case I, we adopt constant step-size in Dynamic Case II as the diminishing step-size cannot track the fast-changing $\textbf{x}_k$ and $\beta_k$. The constant step-size $b_{\begin{small}DII\end{small},k}$ will be specified later.

\vspace{-0mm}
\section{Computational Complexity}\label{sec_ComputationComplexity}
In this section, we evaluate the computational complexity of the proposed tracking algorithms in Quasi-static Case, Dynamic Case I and Dynamic Case II. We focus on the complex arithmetic operations in the tracking stage including complex multiplication and division, while complex addition and subtraction are omitted since they require much fewer operations. It seems that Algorithm \ref{alg_Static}, Algorithm \ref{alg_DI} and Algorithm \ref{alg_DII} require a huge number of complex arithmetic operations due to the Fisher information matrix inversion in each ECC. However, most of these calculation work can be finished off-line, by which the complex operations are greatly reduced. The following lemma is proposed to tell us the specific computational complexity:
\vspace{-0mm}
\begin{lemma}\label{lemma_complexity}
	\emph{If the number of offline complex arithmetic operations is 
     ignored since it is much smaller than the online ones as the tracking process lasts, then} 
	
	\emph{1) for Algorithm \ref{alg_Static} in Quasi-static Case and Algorithm \ref{alg_DII} in Dynamic Case II, 45 complex arithmetic operations are required in each ECC;}
	
	\emph{2) for Algorithm \ref{alg_DI} in Dynamic Case I, 28 complex arithmetic operations are required in each ECC.}
	
\end{lemma}
\vspace{-0mm}
\begin{proof}
	See Appendix \ref{proof_complexity}.
\end{proof}\vspace{-0mm}

According to Lemma \ref{lemma_complexity}, our algorithms can efficiently work without high complexity.

\vspace{-0mm}
\section{Numerical Results}\label{sec_simulation}
\vspace{-0mm}
In this section, some numerical results will be provided to verify the performance of our proposed tracking algorithms for Quasi-static Case, Dynamic Case I and Dynamic Case II. Based on the model in Section \ref{sec_model}, the parameters are set as: $M\!=\!N\!=\!8$, the antenna spacing $d_1\!=\!d_2\!=\!\frac{\lambda}{2}$, and the transmit $\text{SNR}$ is $\frac{ \lvert \textbf{s} \rvert^2}{\sigma_z^2}=0 \,\text{dB}$. The antenna element pattern is based on the 3GPP model \cite{3GPP19Study}. The vertical cut and the horizontal cut of the radiation power pattern (normalized with 0 dB at the central direction) for each element are given below:
\vspace{-1mm}
\begin{align}
\eta_{\text{dB}}\left(\theta,\phi=\frac{\pi}{2}\right) &= -\min {\left\{12\left(\frac{\theta}{\theta_{\text{3\,dB}}}\right)^2,\eta_{\max}\right\}}\\
\eta_{\text{dB}}\left(\theta=0,\phi\right) &= -\min {\left\{12\left(\frac{\phi-\frac{\pi}{2}}{\phi_{\text{3\,dB}}}\right)^2,\eta_{\max}\right\}},
\end{align}where $\theta_{\text{3\,dB}} = \frac{13\pi}{36}$ is the 3 dB beamwidth in the vertical direction, $\phi_{\text{3\,dB}} = \frac{13\pi}{36}$ is the 3 dB beamwidth in the horizontal direction and $\eta_{\max}=30\,\text{dB}$ is the maximum attenuation. \begin{figure}[!t]
	\centering
	\vspace{-0mm}
	\includegraphics[width=8.5cm]{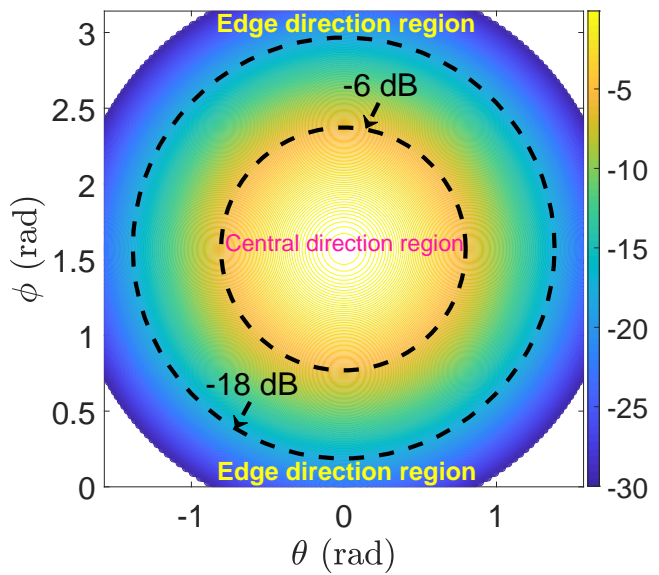}
	\vspace{-0mm}
	\caption{An example of the normalized antenna radiation power pattern (dB) versus the AoA $\theta,\phi$ (rad) for each element.}
	\label{fig_pattern3D}
\end{figure}
The combined radiation power pattern (normalized) of each antenna element in \cite{3GPP19Study} is given by
\vspace{-1mm}
\begin{small}
\begin{align}\label{eq_pattern}
\eta_{\text{dB}}\!\left(\!\theta\!,\phi\!\right) \!=\! -\!\min\! {\left\{\!-\!\left(\!\eta_{\text{dB}}\left(\!\theta,\phi\!=\!\frac{\pi}{2}\!\right)\!+\!\eta_{\text{dB}}\left(\theta\!=\!0,\phi\right)\!\right)\!,\eta_{\max}\!\right\}}\!.\!
\end{align}
\end{small}As can be seen in Fig. \ref{fig_pattern3D}, we define two direction regions: central direction region ($\eta_{\text{dB}}\left(\theta,\phi\right) \ge -6\,\text{dB}$) and edge direction region ($\eta_{\text{dB}}\left(\theta,\phi\right) \le -18\,\text{dB}$). We will separately evaluate the tracking performance in these two direction regions afterwards.

\vspace{-0mm}
\subsection{Reference Algorithms}\label{subsec_refalg}
Reference algorithms include the \emph{compressed sensing} algorithm in \cite{Rial2016Hybrid}, the \emph{3GPP New Radio (NR)} tracking algorithm in \cite{3GPP19}, the \emph{extended Kalman filter (EKF)} algorithm in
\cite{Vutha2016Tracking} and the \emph{recursive beam and channel tracking (RBCT)}  algorithm in \cite{JLiJoint2018ICCASP}. For the compressed sensing algorithm in \cite{Rial2016Hybrid}, we randomly choose phase shifts from $\left\{\pm{1},\pm{j}\right\}$ and then use the sparse recovery algorithm to estimate the DPV, where a discrete Fourier transform (DFT) dictionary with a size of 1024 is utilized. As for the 3GPP NR tracking algorithm, the last estimated beam direction and its adjacent beam directions are probed, then the system determines whether to switch the estimate in current ECC according to the strength of the received signals. For the EKF algorithm, we extend the method in \cite{Vutha2016Tracking} to the 2D array and the three exploring directions used in each ECC form a regular triangle within half the main lobe of the DPV estimate. The original RBCT algorithm in \cite{JLiJoint2018ICCASP} is designed for 1D system and cannot directly support 2D tracking. To compare with our algorithms in the 2D case, we use two symmetrical explorations to track each dimension of the 2D beam for the RBCT algorithm.

%

For the initial beam estimation stage in Fig. \ref{FrameStrcuctureTVT}, an exhaustive beam sweeping is conducted. Then an initial estimate is obtained by using the orthogonal matching pursuit method in \cite{Tony2011Orthogonal}. This ensures that the initial estimate of the DPV, i.e., $\hat{\textbf{x}}_0$, is within the main lobe in \eqref{eq_MainLobe}. 

In the tracking stage, three explorations are conducted in each ECC for all the algorithms. For the RBCT algorithm in \cite{JLiJoint2018ICCASP}, we use a buffer to store the received observations and update the estimate when receiving four new observations. 

It is worth pointing out that the compressed sensing algorithm in \cite{Rial2016Hybrid} does not require an initial estimate. To ensure the fairness of all the algorithms, we compare the tracking performance versus the total number of explorations (i.e., assuming the compressed sensing algorithm uses the same total number of explorations) used both in the initial beam estimation stage and the tracking stage.

\subsection{Performance bound of the tracking algorithms}\label{subsec_simulation_pefromance bound}

As revealed in Section \ref{subsec_static_Bound} and Section \ref{subsec_BoundDI}, the CRLB is a function of the adopted EBMs. Since the EBMs of our algorithms and the reference algorithms are quite different, the corresponding achieved CRLBs are also different. In this subsection, we will compare the CRLBs achieved by the EBMs of these different algorithms to verify the superiority of our optimal EBMs.

The fixed AoA ($\theta$,$\,\phi$) in Quasi-static Case and Dynamic Case I is chosen evenly and randomly in $\theta \in \left[-\frac{\pi}{6},\frac{\pi}{6}\right],\phi \in \left[\frac{\pi}{3},\frac{2\pi}{3}\right]$. The corresponding antenna gain of each element varies from -5.2 dB to 0 dB via \eqref{eq_pattern}. For Quasi-static case, the fixed channel gain $\beta^c$ is modeled as Rician fading with a K-factor $\kappa$=15 dB, according to the channel model in \cite{Samimi2016FadingModel}. For Dynamic Case I, the channel gains between different ECCs are independent of each other and modeled as Rayleigh fading with $\frac{\lvert \textbf{s} \rvert^2 \left(\sigma_{\beta}^c\right) ^2}{\sigma_z^2} = 0\,\text{dB}$. All simulation results in this section are averaged over 1000 random system realizations.


Fig. \ref{fig_bound_static_0dB} and Fig. \ref{fig_bound_DI_0dB} shows the CRLBs achieved by the EBMs of different algorithms. It can be observed that our optimal EBMs both in Quasi-static Case and Dynamic Case I can achieve lower CRLB compared with the EBMs used by the reference algorithms, which results from the fact that the EBMs of our algorithms are carefully optimized and proven to be optimal in theory. Since the CRLB illustrates the performance bound, the other four algorithms cannot perform better than our algorithms in potential tracking accuracy.
\begin{figure}[!t]
	\centering
	\vspace{-0mm}
	\includegraphics[width=8.6cm]{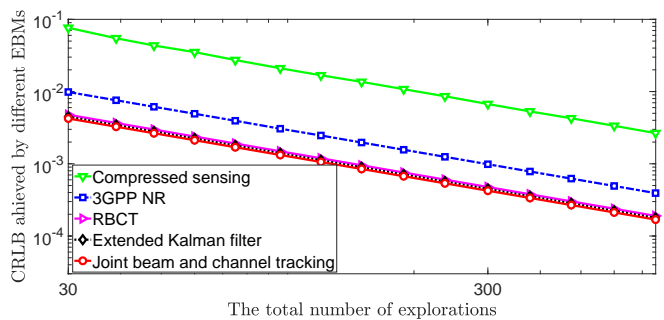}
	\vspace{-0mm}
	\caption{The CRLBs in Quasi-static Case.}
	\label{fig_bound_static_0dB}
\end{figure}
\begin{figure}[!t]
	\centering
	\vspace{-0mm}
	\includegraphics[width=8.6cm]{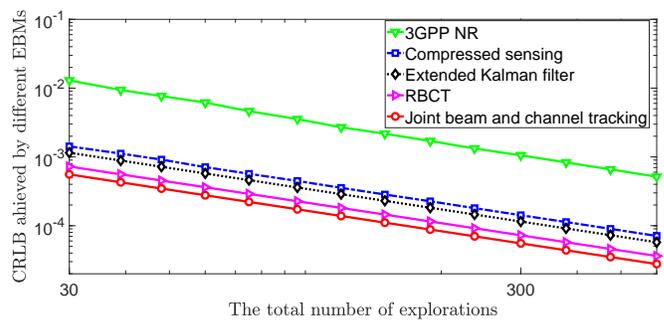}
	\vspace{-0mm}
	\caption{The CRLBs in Dynamic Case I.}
	\label{fig_bound_DI_0dB}
\end{figure}

\vspace{-0mm}
\subsection{Results of Tracking Accuracy}\label{subsec_simulation_accuracy}
\vspace{-0mm}
In this subsection, we will evaluate the tracking performance of our algorithms in the central direction region. The AoA ($\theta$,$\,\phi$) as defined in Section \ref{sec_model} is chosen evenly and randomly in $\theta \in \left[-\frac{\pi}{6},\frac{\pi}{6}\right],\phi \in \left[\frac{\pi}{3},\frac{2\pi}{3}\right]$. The corresponding antenna gain of each element varies from -5.2 dB to 0 dB via \eqref{eq_pattern}. 

1) \textbf{Quasi-static Case}

The channel gain $\beta^c$ is modeled as Rician fading with a K-factor $\kappa$=15 dB, according to the channel model in \cite{Samimi2016FadingModel}. The step-size is set as $b_{S,k} = \frac{1}{k}$.
 

As can be observed in Fig. \ref{fig_static_mse_0dB},
\begin{figure}[!t]
	\centering
	\vspace{-0mm}
	\includegraphics[width=8.5cm]{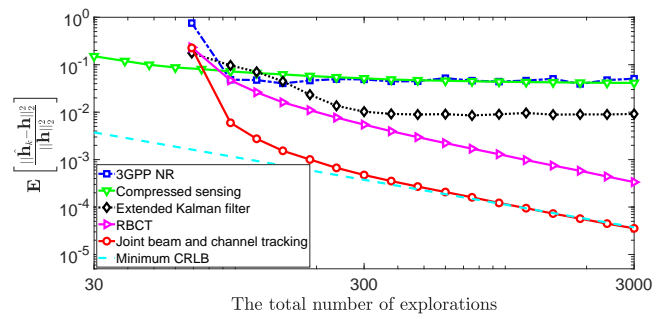}
	\vspace{-0mm}
	\caption{$\text{MSE}_\textbf{h}$ in Quasi-static Case when the AoA is in the central direction region.}
	\label{fig_static_mse_0dB}
\end{figure}the tracking accuracy of the compressed sensing algorithm and the 3GPP NR tracking algorithm gradually keeps steady with the increasing number of explorations, as both of the two algorithms are grids-of-beam based approaches and the performance is restricted by the codebook resolution. The EKF algorithm in Fig. \ref{fig_static_mse_0dB} also shows a similar feature, which results from the fact that the original EKF algorithm itself cannot efficiently track static parameters \cite{Dan2006Optimal}. Although the tracking error of the RBCT algorithm gradually reduces as the number of exploration increases, it cannot achieve the corresponding CRLB in Fig. \ref{fig_bound_static_0dB}. This is caused by tracking the horizontal and the vertical directions separately in the RBCT algorithm, causing loss as against joint tracking. Compared with the four reference algorithms, our proposed JBCT algorithm can approach the minimum CRLB quickly and achieve much lower tracking error.



2) \textbf{Dynamic Case I}

The channel gains between different ECCs are independent of each other and Rayleigh fading channels are adopted in each ECC with $\frac{\lvert \textbf{s} \rvert^2 \left(\sigma_{\beta}^c\right) ^2}{\sigma_z^2} = 0\,\text{dB}$. 
The step-size is set as $b_{DI,k} = \frac{1}{k}$. As the RBCT algorithm in \cite{JLiJoint2018ICCASP} does not support fast-fading channel tracking, it is excluded from the reference algorithms in Dynamic Case I.

Fig. \ref{fig_BeamTrackingDITVTRayleigh}
\begin{figure}[!t]
	\centering
	\vspace{-0mm}
	\includegraphics[width=8.5cm]{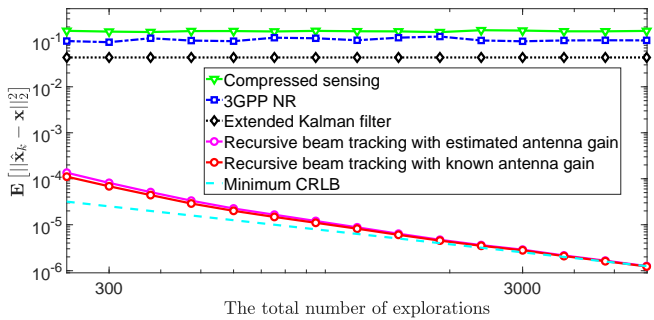}
	\vspace{-0mm}
	\caption{$\text{MSE}_{\textbf{x}}$ in Dynamic Case I for Rayleigh fading channel when the AoA is in the central direction region.}
	\label{fig_BeamTrackingDITVTRayleigh}
\end{figure}indicates that the DPV MSE of our proposed RBT algorithm can converge to the minimum CRLB if the antenna gain can be perfectly known. Even with the estimated antenna gain for tracking, our algorithm can also converge to the minimum CRLB and achieve much lower tracking error than other algorithms. Hence, the estimation error of the antenna gain has little influence on the tracking performance of our algorithm when the AoA is in the central direction region.

Further, we evaluate the robustness of the proposed algorithm for other types of time-varying channels. The channel gains between different ECCs are still independent of each other while Rician fading channels are adopted in each ECC with a K-factor $\kappa$=15 dB and $\frac{\lvert \textbf{s} \rvert^2  \left(\sigma_{\beta}^c\right) ^2}{\sigma_z^2} = 0\,\text{dB}$, where $\big(\sigma_{\beta}^c\big) ^2$ denotes the average energy gain of the Rician fading channel. The algorithm designed for the Rayleigh fading channel is adopted to track the Rician fading channel here. It can be observed in Fig. \ref{fig_BeamTrackingDITVTRician} that our algorithm can still converge and achieve much lower tracking error than existing algorithms. These results show that the proposed algorithm in Dynamic Case I is robust to different time-varying channels as long as the variance of the channel gain is known.
	
\begin{figure}[!t]
	\centering
	\vspace{-0mm}
	\includegraphics[width=8.5cm]{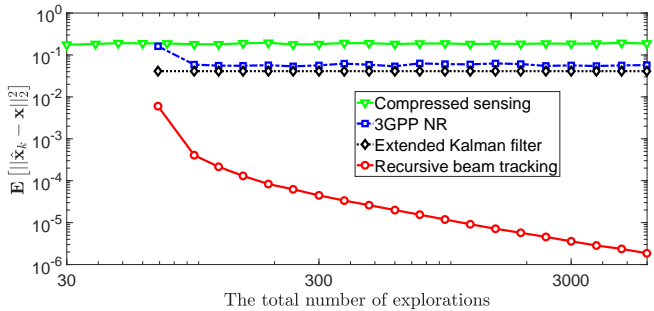}
	\vspace{-0mm}
	\caption{$ \text{MSE}_{\textbf{x}}$ in Dynamic Case I for Rician fading channel when the AoA is in the central direction region.}
	\label{fig_BeamTrackingDITVTRician}
\end{figure}

3) \textbf{Dynamic Case II}

In {Dynamic Case II}, the initial AoA ($\theta_0$,$\,\phi_0$) as defined in Section \ref{sec_model} is chosen evenly and randomly in $\theta_0 \in \left[-\frac{\pi}{6},\frac{\pi}{6}\right],\phi_0 \in \left[\frac{\pi}{3},\frac{2\pi}{3}\right]$. The AoA ($\theta_k$,$\,\phi_k$) is modeled as a random walk process with return, i.e., $\theta_{k+1} = \theta_ k+ \varpi_k^{\theta}\Delta \theta_k$, $\phi_{k+1} = \phi_k+ \varpi_k^{\phi}\Delta \phi_k$, where $\Delta \theta_k,\Delta \phi_k \sim \mathcal{N}(0,\delta_{{A}}^2)$, and $\varpi_k^{\theta},\varpi_k^{\phi} \in \left\{-1,1\right\}$ denote the rotation direction. The rotation direction $\varpi_k^{\theta},\varpi_k^{\phi}$ are chosen such that $\theta_k$ varies in $\left[-\frac{\pi}{6},\frac{\pi}{6}\right]$ and $\phi_k$ varies in $\left[\frac{\pi}{3},\frac{2\pi}{3}\right]$. The channel gain is modeled as a first-order Gaussian-Markov process, i.e., $\beta_{k+1}^c = \rho \beta_{k}^c+\gamma_{k}$, where $\gamma_{k} \sim \mathcal{CN}(0,1-\rho^2)$. We adopt $\rho = 0.995$ in simulation. As for the step-size, numerical results show that when $b_{DII,k} = 0.7$, the joint beam and channel tracking algorithm can track beams with higher velocity. Hence, the step-size is set as a constant $b_{DII,k} = 0.7$. 

Fig. \ref{fig_BeamTrackingDII}
\begin{figure}[!t]
	\centering
	\vspace{-0mm}
	\includegraphics[width=8.5cm]{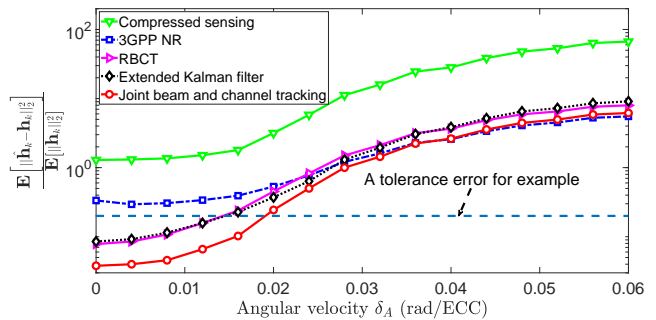}
	\vspace{-0mm}
	\caption{$\text{MSE}_{\textbf{h}_k}$ in Dynamic Case II when the AoA is in the central direction region.}
	\label{fig_BeamTrackingDII}
\end{figure}indicates the proposed JBCT algorithm in Dynamic Case II can achieve higher tracking accuracy than the other four algorithms. In addition, if we set a tolerance error in Fig. \ref{fig_BeamTrackingDII}, then our algorithm can support higher angular velocities.

\vspace{-0mm}
\subsection{The impact of the antenna pattern}\label{subsec_simulation_pattern}

In this subsection, we will evaluate the impact of the antenna pattern by setting the AoA in the edge direction region in Fig. \ref{fig_pattern3D}. Other parameters are the same as the setting in Section \ref{subsec_simulation_accuracy}.

1) \textbf{Quasi-static Case}

The AoA ($\theta$,\,$\phi$) as defined in Section \ref{sec_model} is chosen evenly and randomly in $\theta \in \left[\frac{\pi}{2}-\frac{\pi}{60},\frac{\pi}{2}\right],\phi \in \left[\pi-\frac{\pi}{60},\pi\right]$. The corresponding antenna gain of each element is -30 dB via \eqref{eq_pattern}.

\begin{figure}[!t]
	\centering
	\vspace{-0mm}
	\includegraphics[width=8.5cm]{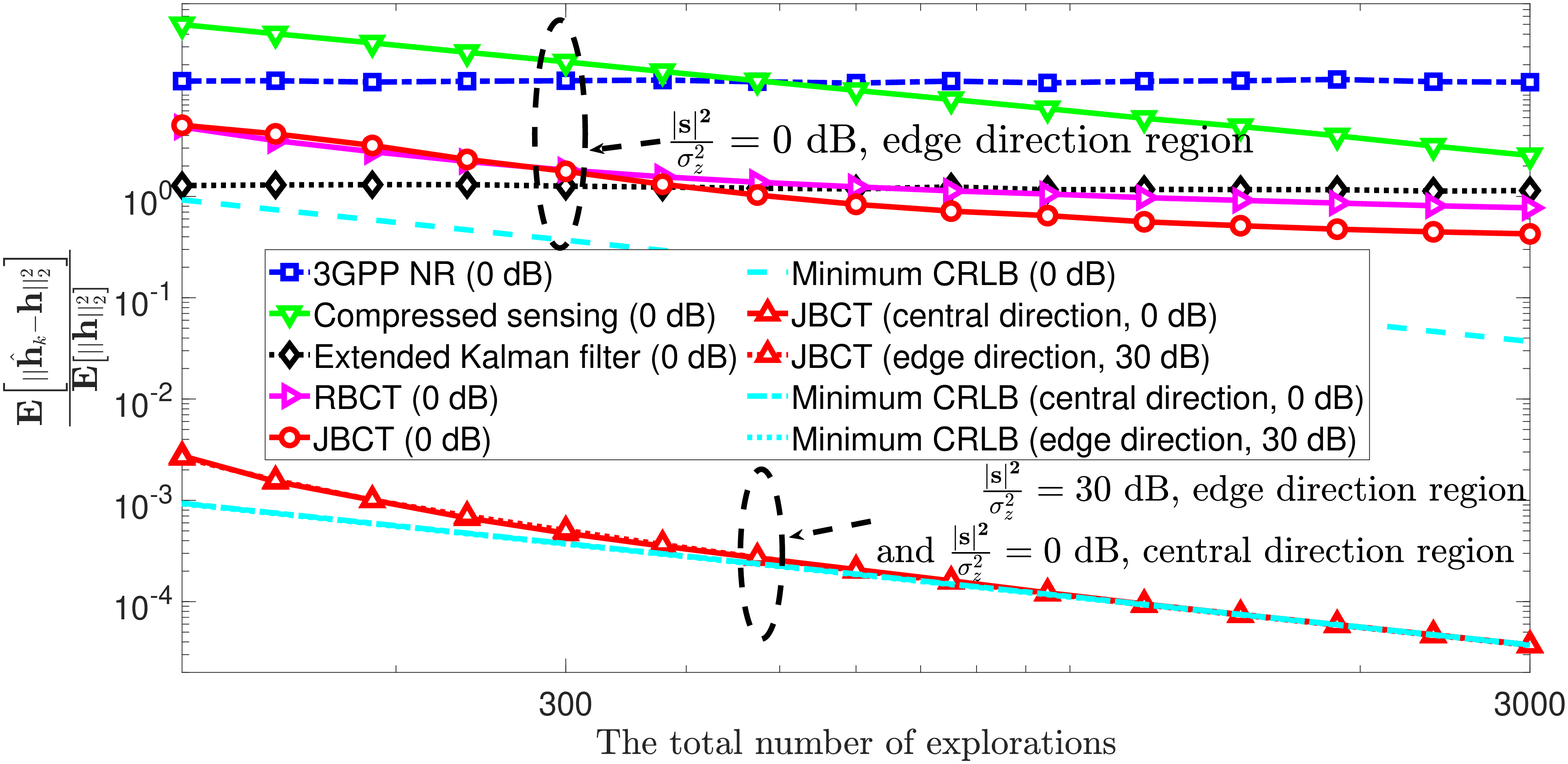}
	\vspace{-0mm}
	\caption{$\text{MSE}_\textbf{h}$ in Quasi-static Case when the AoA is in the edge direction region.}
	\label{fig_static_mse_edge}
\end{figure}

In Fig. \ref{fig_static_mse_edge}, it can be seen that our JBCT algorithm still outperforms existing algorithms when the AoA is in the edge direction region. Nevertheless, compared with Fig. \ref{fig_static_mse_0dB}, the minimum CRLB cannot be achieved any more and the performance of all algorithms deteriorate if the transmit power keeps unchanged, i.e., $\frac{ \lvert \textbf{s} \rvert^2}{\sigma_z^2}= 0\, \text{dB}$. This can be explained by the decrease of the equivalent SNR when the AoA is in the edge direction region. If we compensate the gain loss in the edge direction region by increasing the transmit power by 30 dB, i.e., $\frac{ \lvert \textbf{s} \rvert^2}{\sigma_z^2}= 30 \,\text{dB}$, then Fig. \ref{fig_static_mse_edge} demonstrates the performance can be greatly improved and the minimum CRLB can be achieved again. Furthermore, we compare the tracking performance in two cases:  1) $\frac{ \lvert \textbf{s} \rvert^2}{\sigma_z^2}= 30\, \text{dB}$ when $\theta \in \left[\frac{\pi}{2}-\frac{\pi}{60},\frac{\pi}{2}\right],\phi \in \left[\pi-\frac{\pi}{60},\pi\right]$, where the antenna gain of each element is -30 dB; 2) $\frac{ \lvert \textbf{s} \rvert^2}{\sigma_z^2}= 0\, \text{dB}$ when $\theta \in \left[-\frac{\pi}{120},\frac{\pi}{120}\right],\phi \in \left[\frac{\pi}{2}-\frac{\pi}{120},\frac{\pi}{2}+\frac{\pi}{120}\right]$, where the antenna gain of each element can be seen as 0 dB. It can be observed in Fig. \ref{fig_static_mse_edge} that the performance in these two cases is almost the same. This shows that the deterioration of the tracking performance in the edge direction region only results from the decrease of the equivalent SNR.

2) \textbf{Dynamic Case I}

The AoA ($\theta$,$\,\phi$) as defined in Section \ref{sec_model} is chosen evenly and randomly in $\theta \in \left[\frac{\pi}{3},\frac{\pi}{2}\right],\phi \in \left[\frac{5\pi}{6},\pi\right]$. The corresponding antenna gain of each element varies from $-30\,\text{dB}$ to -20.4 dB via \eqref{eq_pattern}. As can be observed in Fig. \ref{fig_BeamTrackingDIEdge}, our RBT algorithm still outperforms existing algorithms for Rayleigh fading channel when the AoA is in the edge direction region. Since the equivalent SNR decreases, the proposed algorithm cannot converge to the minimum CRLB as before. If we compensate the gain loss in the edge direction region by increasing the transmit power by 30 dB, i.e., $\frac{ \lvert \textbf{s} \rvert^2}{\sigma_z^2}= 30\,\text{dB}$, then our algorithm can still converge to the minimum CRLB and achieve the same performance as that in Fig. \ref{fig_BeamTrackingDITVTRayleigh} with the perfectly-known antenna gain.

With the estimated antenna gain, our RBT algorithm cannot converge to the minimum CRLB as before even when $\frac{ \lvert \textbf{s} \rvert^2}{\sigma_z^2}= \,30 \text{dB}$. This is caused by the larger slope in the edge direction region compared with the central edge direction region. Hence, a small estimation error of the AoA can result in a large deviation of the estimated equivalent channel gain parameter $\sigma_{\beta}^2$ in \eqref{eq_covariance_beta}, leading to the non-convergence when using estimated antenna gain in Fig. \ref{fig_BeamTrackingDIEdge}.

\begin{figure}[!t]
	\centering
	\vspace{-0mm}
	\includegraphics[width=8.5cm]{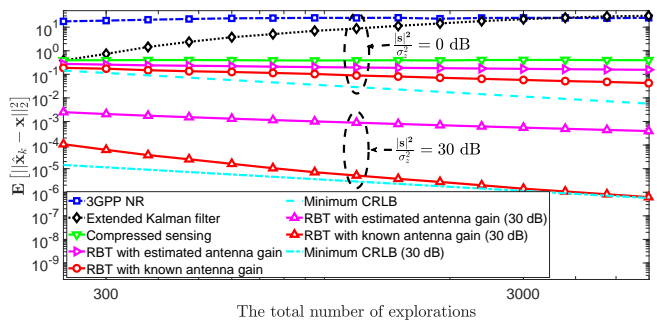}
	\vspace{-0mm}
	\caption{$\text{MSE}_{\textbf{x}}$ in Dynamic Case I for Rayleigh fading channel when the AoA is in the edge direction region.}
	\label{fig_BeamTrackingDIEdge}
\end{figure}

3) \textbf{Dynamic Case II}
\begin{figure}[!t]
	\centering
	\vspace{-0mm}
	\includegraphics[width=8.5cm]{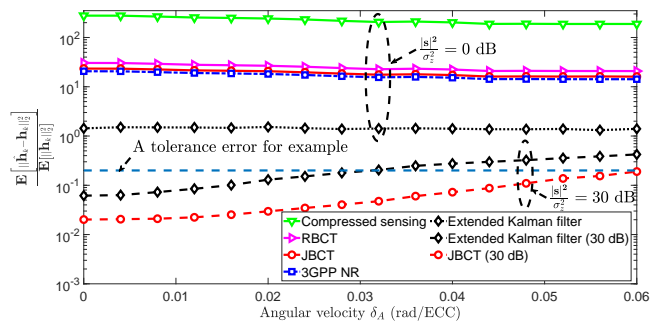}
	\vspace{-0mm}
	\caption{$\text{MSE}_{\textbf{h}_k}$ in Dynamic Case II when the AoA is in the edge direction region.}
	\label{fig_BeamTrackingDIIEdge}
\end{figure}

The initial AoA ($\theta_0$,$\,\phi_0$) as defined in Section \ref{sec_model} is chosen evenly and randomly in $\theta_0 \in \left[\frac{\pi}{3},\frac{\pi}{2}\right],\phi_0 \in \left[\frac{5\pi}{6},\pi\right]$. The rotation direction $\varpi_k^{\theta},\varpi_k^{\phi}$ are chosen such that $\theta_k$ varies in $\left[\frac{\pi}{3},\frac{\pi}{2}\right]$ and $\phi_k$ varies in $\left[\frac{5\pi}{6},\pi\right]$. The corresponding antenna gain of each element varies from $-30\,\text{dB}$ to -20.4 dB via \eqref{eq_pattern}}. As can be seen in Fig. \ref{fig_BeamTrackingDIIEdge}, all the algorithms cannot efficiently track the channels when the AoA varies in the edge direction region, since the equivalent SNR decreases sharply. If we compensate the gain loss in the edge direction region by increasing the transmit power by 30 dB, i.e., $\frac{ \lvert \textbf{s} \rvert^2}{\sigma_z^2}= 30 \,\text{dB}$, then our JBCT algorithm can still achieve lower tracking error and faster tracking speed.

\vspace{-1mm}
\subsection{Computational Complexity}\label{subsec_simulation_complexity}
\vspace{-0mm}
We then evaluate the computational complexity of our proposed algorithms. As can be seen in TABLE \ref{tab_complexity}, our algorithms require fewer complex operations than other algorithms except 3GPP NR. Compared with 3GPP NR, the proposed algorithms can achieve much more accurate tracking without greatly increasing the computational complexity.
\renewcommand\arraystretch{1.4}
\begin{table}[!t!]
	\vspace{-0mm}
	\centering
	\caption{\small {Number of required complex operations in each ECC.}}
	\vspace{2mm}
	\begin{tabular} {c|c|c|c|c}
		\hline
		\hline
		\tabincell{c}{3GPP \\NR} & Algorithm \ref{alg_DI} & \tabincell{c}{Algorithm \ref{alg_Static} and\\ Algorithm \ref{alg_DII}}&\tabincell{c}{Extended \\Kalman filter} & \tabincell{c}{Compressed \\sensing}\\
		\hline
		6 &
		28 & 45 & 1427 &129088\\ 
		\hline
		\hline
	\end{tabular}\label{tab_complexity}
\end{table}


\vspace{-0mm}
\section{Conclusion}\label{future work remarks}
\vspace{-0mm}
This paper focuses on fast accurate beam and channel tracking for 2D phased antenna arrays. We first give the minimum exploration overhead of joint 2D tracking in theory. Then three tracking algorithms are developed according to different practical time-varying channel models. 
	
In Quasi-static Case, the optimal exploration offsets are derived which are proved to a) be unrelated to the channel gain and the beam direction, b) be determined only by the array size, and c) approach constants as the array size goes to infinity. Also, a joint beam direction and channel gain tracking algorithm is proposed and the tracking error is proved to converge to the minimum CRLB. 

In Dynamic Case I, an algorithm for beam only tracking is proposed, and it is proved to converge and achieve the minimum CRLB on the beam direction. 

In Dynamic Case II, a joint tracking algorithm of beam direction and channel gain is proposed with faster and more accurate performance presented by simulation results.

This work is the first step to beam and channel tracking with 2D phased antenna arrays. In future work, we will further study the following problems: i) establishing the corresponding theorems in Dynamic Case II; ii) jointly tracking multipath channels; iii) tracking at both the transmitter and the receiver.
\vspace{-4mm}

%
%
%

\appendices
\section{Proof of Lemma~\ref{IndObservations}}\label{proof_IndObservations}
If the EBVs are of the steering vector forms, i.e., $\textbf{w}_{k,i} = \frac{1}{\sqrt{MN}}\textbf{a}\left(\boldsymbol{\omega}_{k,i}\right)$, where $\boldsymbol{\omega}_{k,i}=\left[\omega_{k,i1},\omega_{k,i2}\right]^\text{T}$ denotes the $i$-th exploring direction vector in $k$-th ECC, then the noiseless complex observation equation for the $i$-th observation is given in \eqref{eq_aObservationI},
     \begin{figure*}[!t]		
	\normalsize
	\vspace{-0mm}
	\begin{align}\label{eq_aObservationI}
{y_{k,i}} \!=\! \frac{\lvert\textbf{s}\rvert \beta\left(\textbf{x}_k\right)}{\sqrt{MN}}\textbf{a}\left(\boldsymbol{\omega}_{k,i}\right)^\text{H}\textbf{a}\left(\textbf{x}_k\right)\!&= \frac{{{\lvert\textbf{s}\rvert}\beta\left(\textbf{x}_k\right) }}{{\sqrt {MN} }}\sum\limits_{m = 1}^M {\sum\limits_{n = 1}^N {{e^{-j2\pi \left[\frac{{\left(m - 1\right)(\omega_{k,i1}-x_{k,1})}}{M} + \frac{{(n - 1)(\omega_{k,i2}-x_{k,2})}}{N}\right]}}} }\\
&= \frac{\lvert\textbf{s}\rvert\beta\left(\textbf{x}_k\right)}{{\sqrt {MN} }}\frac{{\sin \left[\pi (\omega_{k,i1}-x_{k,1})\right]}}{{\sin \left[ { \frac{\pi (\omega_{k,i1}-x_{k,1})}{M}} \right]}}\frac{{\sin \left[\pi (\omega_{k,i2}-x_{k,2})\right]}}{{\sin \left[ { \frac{\pi (\omega_{k,i2}-x_{k,2})}{N}} \right]}}{e^{-j\pi \left[ {\frac{{M - 1}}{M}(\omega_{k,i1}-x_{k,1}) + \frac{{N - 1}}{N}\left(\omega_{k,i2}-x_{k,2}\right)} \right]}}\nonumber\\
& \overset{(a)}{=}  \frac{\lvert\textbf{s}\rvert\beta\left(\textbf{x}_k\right)}{{\sqrt {MN} }} y_a\left(\boldsymbol{\omega}_{k,i}-\textbf{x}_k\right){e^{-j\pi \left[ {\frac{{M - 1}}{M}(\omega_{k,i1}-x_{k,1}) + \frac{{N - 1}}{N}\left(\omega_{k,i2}-x_{k,2}\right)} \right]}},\nonumber
\end{align}
	\hrulefill
	\vspace*{0pt}
\end{figure*}where Step (a) follows the definition of $y_a(\boldsymbol{\Delta})$:
\begin{equation}
y_a(\boldsymbol{\Delta}) \triangleq \frac{{\sin \left(\pi \delta_1\right)}}{{\sin \left( { \frac{\pi \delta_1}{M}} \right)}}\frac{{\sin \left(\pi \delta_2\right)}}{{\sin \left( { \frac{\pi \delta_2}{N}} \right)}}
\end{equation}with $\boldsymbol{\Delta} \triangleq \left[\delta_1,\delta_2\right]^\text{T}$. In our real tracking problem, the exploring direction vector $\boldsymbol{\omega}_{k,i}$ should be ensured within the main lobe of $\textbf{x}_k$ in \eqref{eq_MainLobe}, i.e., $\lvert{\omega}_{k,i1}-{x}_{k,1}\rvert < 1$ and $\lvert{\omega}_{k,i2}-{x}_{k,2}\rvert < 1$. Hence, we have that $y_a\left(\boldsymbol{\omega}_{k,i}-\textbf{x}_k\right)>0$.

The complex observation equation in \eqref{eq_aObservationI} contains two real equations, i.e., an amplitude equation and a phase angle equation. Therefore, $q$ amplitude equations and $q$ phase angle equations can be obtained after $q$ observations. If we set the first observation as a reference, then we can obtain $(q-1)$ relative amplitude equations and $(q-1)$ relative phase angle equations from the remaining $(q-1)$ observations. However, these phase angle equations are not independent, as to be explained below.

From \eqref{eq_aObservationI}, we can obtain the phase angle equation:
       \begin{small}
		\begin{align*}
		\angle({y_{k,i}})\!=\!\angle {\beta\left(\textbf{x}_k\right)}{\rm{ - }}\pi\! \left[\! {\frac{M \!-\! 1}{M}(\omega_{k,i1}\!-\!x_{k,1}) \!+\! \frac{N \!-\! 1}{N}{(\omega_{k,i2}\!-\!x_{k,2})}} \!\right]\!.\nonumber
		\end{align*}\end{small}Thus the relative phase angle equation of the $i$-th observation $y_{k,i}$ regarding the first observation $y_{k,1}\,(i \neq 1)$ can be obtained as below:
		\begin{align}\label{eq_aObservationIpr}
	&\angle({y_{k,i}})-\angle({y_{k,1}})\\
	=& \pi \left[ {\frac{M - 1}{M}(\omega_{k,11}-\omega_{k,i1}) + \frac{N - 1}{N}{(\omega_{k,12}-\omega_{k,i2})}} \right],\nonumber
	\end{align}where $\omega_{k,{11}} - {\omega_{k,i1}}$ and $\omega_{k,{12}} - {\omega_{k,i2}}$ are determined by the exploring direction vectors and unrelated to the channel parameter vector $\boldsymbol{\psi}_k$. From \eqref{eq_aObservationIpr}, we can know that once the exploring directions are determined, the relative phase angles are known constants unrelated to $\boldsymbol{\psi}_k$. In other words, the relative phase angle equations as revealed in \eqref{eq_aObservationIpr} cannot provide any information for estimating $\boldsymbol{\psi}_k$.

Following the conclusion above, we analyze the minimum exploration overhead in the following two cases:

1) If we want to obtain the unique solution of $\boldsymbol{\psi}_k$ within one ECC, at least 4 independent real equations with respect to $\boldsymbol{\psi}_k$ are needed since $\boldsymbol{\psi}_k$ contains four independent real variables (i.e., the real part $\text{Re}\left\{\beta\left(\textbf{x}_k\right)\right\}$, the imaginary part $\text{Im}\left\{\beta\left(\textbf{x}_k\right)\right\}$ of the equivalent channel gain $\beta_k$ and the two direction parameters ${x}_{k,1}, {x}_{k,2}$). After $q$ explorations in each ECC, we can obtain $q$ independent amplitude equations and only 1 independent phase angle equation, which is $q+1$ independent real equations with respect to $\boldsymbol{\psi}_k$ in total. Hence, at least 3 explorations are needed to obtain 4 independent real equations and estimate 4 independent real variables of $\boldsymbol{\psi}_k$.

2) If we only want to obtain the unique solution of $\textbf{x}_k$ within one ECC, at least 2 independent real equations with respect to $\textbf{x}_k$ are needed since $\textbf{x}_k$ contains two independent real variables (i.e., two direction parameters ${x}_{k,1}, {x}_{k,2}$). It seems that fewer explorations are sufficient. However, we cannot obtain any absolute amplitude and phase information with respect to $\textbf{x}_k$ from one observation in \eqref{eq_aObservationI} since $\beta\left(\textbf{x}_k\right)$ is unknown. In addition, the relative phase angles are constants unrelated to $\textbf{x}_k$. Thus, the phase angle equations are useless for estimating $\textbf{x}_k$. After $q$ explorations in each ECC, we can obtain $q-1$ independent relative amplitude equations with respect to $\textbf{x}_k$ in total. Hence, at least 3 explorations are needed to obtain 2 independent real equations and estimate 2 independent real variables of $\textbf{x}_k$.

Therefore, the proof is completed.

\section{Proof of Lemma~\ref{MSEOpt}}\label{proof_MSEOpt}
In problem \eqref{eq_problem}, the constraint \eqref{eq_constrant1} ensures that $\hat{\textbf{h}}_k$ is an unbiased estimate of ${\textbf{h}}$. Consider each element of the channel vector ${\textbf{h}}$, i.e., ${h_{mn}}({{\boldsymbol{\psi }}}) = \beta {e^{j2\pi \left( {\frac{{m - 1}}{M}{x_1} + \frac{{n - 1}}{N}{x_2}} \right)}}$. Immediately we have $\mathbb{E} \left[{h_{mn}}(\hat{{\boldsymbol{\psi }}}_k)\right]= {h_{mn}}({{\boldsymbol{\psi }}})$ since $\mathbb{E}\left[\hat{\textbf{h}}_k\right] = \textbf{h}$. According to Section 
3.8 of \cite{Sengijpta1993Fundamental}, if a function $f\left(\hat {\boldsymbol{\psi}}\right)$ is an unbiased estimate of $f\left(\boldsymbol{\psi}\right)$, i.e., $\mathbb{E} \left[f(\hat{{\boldsymbol{\psi}}})\right]= f({{\boldsymbol{\psi}}})$, then we can obtain that 
\vspace{-1mm}
\begin{equation}\label{eq_fLB}
\begin{aligned}
\operatorname{Var}[f(\hat {\boldsymbol{\psi}})] \ge \frac{{\partial f( {\boldsymbol{\psi}})}}{{\partial { {\boldsymbol{\psi}}^\text{T}}}}{\textbf{I}( {\boldsymbol{\psi}})^{ - 1}}{\left(\frac{{\partial f({ {\boldsymbol{\psi}}})}}{{\partial { {\boldsymbol{\psi}^\text{T}}}}}\right)^\text{H}},
\end{aligned}
\end{equation}where $\operatorname{Var}[f(\hat {\boldsymbol{\psi}})]$ denotes the variance of $f(\hat {\boldsymbol{\psi}})$ and $\textbf{I}(\boldsymbol{\psi})$ is the corresponding Fisher information matrix.

Combining \eqref{eq_problem} and \eqref{eq_fLB}, we have
\vspace{-1mm}
    \begin{small}
	\begin{align}\label{eq_CMMSEP}
	&\frac{1}{{MN}}\mathbb{E}\left[\left\|{\hat{\textbf{h}}_k} - {{ \textbf{h}}}\right\|_2^2\right]\nonumber\\
	=&\frac{1}{{MN}}\sum\limits_{m = 1}^M {\sum\limits_{n = 1}^N {\mathbb{E}\left[{\big| h_{mn}(\hat{\boldsymbol{\psi}})-h_{mn}({\boldsymbol{\psi}})\big|}^2  \right]} } \\
	\overset{(a)}{\ge} &\! \frac{1}{{MN}}\!\sum\limits_{m = 1}^M\! {\sum\limits_{n = 1}^N \!\left(\!{\frac{{\partial h_{mn}( {\boldsymbol{\psi}})}}{{\partial { {\boldsymbol{\psi}}^\text{T}}}}{{\!\left(\sum\limits_{l = 1}^k {{\textbf{I}_S}(\psi ,{{\bf{W}}_l})} \right)}^{ \!- 1}}\!{\left(\frac{{\partial h_{mn}({ {\boldsymbol{\psi}}})}}{{\partial { {\boldsymbol{\psi}^\text{T}}}}}\right)^\text{\!H}}}\!\right) } \nonumber\\
	= &\!\frac{1}{{MN}}\!\Tr\! \left\{\!\! {{{\left(\!\sum\limits_{l = 1}^k \!{{\textbf{I}_S}(\!\psi ,{{\bf{W}}_l}\!)} \!\!\right)}^{\!\!-\! 1}}\!\!\!\sum\limits_{m = 1}^M\! {\sum\limits_{n = 1}^N \!{\left(\!\left(\!\frac{{\partial h_{mn}({ {\boldsymbol{\psi}}})}}{{\partial { {\boldsymbol{\psi}^\text{T}}}}}\!\right)^\text{\!H}\!\frac{{\partial h_{mn}( {\boldsymbol{\psi}})}}{{\partial { {\boldsymbol{\psi}}^\text{T}}}}\!\!\right)}} } \!\!\right\}\nonumber\\
	= &\frac{1}{{MN}}\Tr \left\{ {{{\left(\sum\limits_{l = 1}^k {{\textbf{I}_S}(\psi ,{{\bf{W}}_l})} \right)}^{ - 1}}\left(\frac{{\partial \textbf{h}}}{{\partial { {\boldsymbol{\psi}^\text{T}}}}}\right)^\text{H}\frac{{\partial \textbf{h}}}{{\partial { {\boldsymbol{\psi}}^\text{T}}}}  } \right\},\nonumber\\
	\overset{(b)}{=}&\frac{1}{{MN}}\Tr \left\{ {{\left(\sum\limits_{l = 1}^k {{\textbf{I}_S}(\psi ,{{\bf{W}}_l})} \right)}^{ - 1}}\textbf{V}^\text{H}\textbf{V}   \right\},\nonumber
	\end{align}\end{small}where Step 
$(a)$ is obtained by substituting \eqref{eq_fLB} into \eqref{eq_CMMSEP} and Step $(b)$ is due to the definition of $\textbf{V}$ in \eqref{eq_Jacobian}.

As for the Fisher information matrix in \eqref{eq_fisher}, we can obtain $\frac {\partial \text{log} \, p_S \left(\textbf{y}_l |\boldsymbol{\psi},\textbf{W}_l \right)}{\partial {\beta^\text{re}}}$ as follows:
	\begin{align}
	\frac {\partial \text{log} \, p_S \left(\textbf{y}_l |\boldsymbol{\psi},\textbf{W}_l \right)}{\partial {\beta^\text{re}}} \!=\!& -\!\frac{1}{\sigma_z^2}\!\left(\textbf{y}_{l}\!-\!\lvert\textbf{s}\rvert\textbf{W}_{l}^\text{H}\textbf{h}\right)^{\!\text{H}}\!\left(\!\!-\lvert\textbf{s}\rvert\textbf{W}_{l}^\text{H} \frac{\partial \textbf{h}}{\partial \beta^\text{re}}\!\right)\nonumber\\
	&+\!\frac{1}{\sigma_z^2}\!\left(\!\lvert\textbf{s}\rvert\textbf{W}_{l}^\text{H} \frac{\partial \textbf{h}}{\partial \beta^\text{re}}\!\right)^{\!\text{H}}\left(\textbf{y}_{l}-\lvert\textbf{s}\rvert\textbf{W}_{l}^\text{H}\textbf{h}\right)\nonumber\\
	\!=\!&\frac{2\lvert\textbf{s}\rvert}{\sigma_z^2} \text{Re}\!\left\{\!\left(\textbf{y}_{l}\!-\!\lvert\textbf{s}\rvert\textbf{W}_{l}^\text{H}\textbf{h}\right)^{\!\text{H}}\left(\!\textbf{W}_{l}^\text{H} \frac{\partial \textbf{h}}{\partial \beta^\text{re}}\!\right)\!\right\}\nonumber\\
	=&\frac{2\lvert\textbf{s}\rvert}{\sigma_z^2} \text{Re}\left\{\textbf{z}_{l}^\text{H}\textbf{W}_{l}^\text{H} \frac{\partial \textbf{h}}{\partial \beta^\text{re}}\right\}.
	\end{align}Similarly, $\frac {\partial \text{log} \, p_S \left(\textbf{y}_l |\boldsymbol{\psi},\textbf{W}_l \right)}{\partial {\beta^\text{im}}}$, $\frac {\partial \text{log} \, p_S \left(\textbf{y}_l |\boldsymbol{\psi},\textbf{W}_l \right)}{\partial {x_1}}$, and $\frac {\partial \text{log} \, p_S \left(\textbf{y}_l |\boldsymbol{\psi},\textbf{W}_l \right)}{\partial {x_2}}$ are given as
\begin{equation}\label{eq_Bprop}
\begin{aligned}
\left\{\begin{array}{*{20}{l}}
\frac {\partial \text{log} \, p_S \left(\textbf{y}_l |\boldsymbol{\psi},\textbf{W}_l \right)}{\partial {\beta^\text{im}}} = \frac{2\lvert\textbf{s}\rvert}{\sigma_z^2} \text{Re}\left\{\textbf{z}_{l}^\text{H}\textbf{W}_{l}^\text{H} \frac{\partial \textbf{h}}{\partial \beta^\text{im}}\right\}\\
\frac {\partial \text{log} \, p_S \left(\textbf{y}_l |\boldsymbol{\psi},\textbf{W}_l \right)}{\partial {x_1}} = \frac{2\lvert\textbf{s}\rvert}{\sigma_z^2} \text{Re}\left\{\textbf{z}_{l}^\text{H}\textbf{W}_{l}^\text{H} \frac{\partial \textbf{h}}{\partial {x_1}}\right\}\\
\frac {\partial \text{log} \, p_S \left(\textbf{y}_l |\boldsymbol{\psi},\textbf{W}_l \right)}{\partial {x_2}} = \frac{2\lvert\textbf{s}\rvert}{\sigma_z^2} \text{Re}\left\{\textbf{z}_{l}^\text{H}\textbf{W}_{l}^\text{H} \frac{\partial \textbf{h}}{\partial {x_2}}\right\}
\end{array}\right..
\end{aligned}
\end{equation}Hence, the gradient of $\text{log} \, p_S \left(\textbf{y}_l |\boldsymbol{\psi},\textbf{W}_l \right)$ is obtained as follows:
	\begin{equation}\label{eq_gradient}
	\begin{aligned}
	\frac {\partial \text{log} \, p_S \left(\textbf{y}_l |\boldsymbol{\psi},\textbf{W}_l \right)}{\partial \boldsymbol{\psi}}
	=\frac{2\lvert\textbf{s}\rvert}{\sigma_z^2} \text{Re}\left\{\left[
	\begin{matrix}
	\textbf{z}_{l}^\text{H}\textbf{W}_{l}^\text{H} \frac{\partial \textbf{h}}{\partial \beta^\text{re}}\\
	\textbf{z}_{l}^\text{H}\textbf{W}_{l}^\text{H} \frac{\partial \textbf{h}}{\partial \beta^\text{im}}\\
	\textbf{z}_{l}^\text{H}\textbf{W}_{l}^\text{H} \frac{\partial \textbf{h}}{\partial x_1}\\
	\textbf{z}_{l}^\text{H}\textbf{W}_{l}^\text{H} \frac{\partial \textbf{h}}{\partial x_2}
	\end{matrix}\right]\right\}\\=\frac{2\lvert\textbf{s}\rvert}{\sigma_z^2} \text{Re}\left\{\left(
	\textbf{z}_{l}^\text{H}\textbf{W}_{l}^\text{H} \textbf{V}\right)^\text{T}\right\}.
	\end{aligned}
	\end{equation}With the help of \eqref{eq_gradient}, we can obtain that
\begin{equation}\label{eq_Tgradient}
\begin{aligned}
\frac {\partial \text{log} \, p_S \left(\textbf{y}_l |\boldsymbol{\psi},\textbf{W}_l \right)}{\partial \boldsymbol{\psi}^\text{T}} = \left(\frac {\partial \text{log} \, p_S \left(\textbf{y}_l |\boldsymbol{\psi},\textbf{W}_l \right)}{\partial \boldsymbol{\psi}}\right)^\text{T}\\=\frac{2\lvert\textbf{s}\rvert}{\sigma_z^2} \text{Re}\left\{
\textbf{z}_{l}^\text{H}\textbf{W}_{l}^\text{H} \textbf{V}\right\}.
\end{aligned}
\end{equation}

Substituting \eqref{eq_gradient} and \eqref{eq_Tgradient} into  \eqref{eq_fisher}, the Fisher information matrix is given as follows:
\begin{align}\label{eq_rrfisher}
\!\textbf{I}_S(\boldsymbol{\psi}, \textbf{W}_l)\triangleq \quad &  \! \mathbb{E}\!\left[\!\frac {\partial \text{log} \, p_S \left(\textbf{y}_l |\boldsymbol{\psi},\textbf{W}_l \right)}{\partial \boldsymbol{\psi}} \!\cdot\! \frac {\partial \text{log}\, p_S \left(\textbf{y}_l |\boldsymbol{\psi},\textbf{W}_l \right)}{\partial \boldsymbol{\psi}^\text{T}}\!\right]\nonumber\\
=\quad&\frac{4\lvert\textbf{s}\rvert^2}{\sigma_z^4}\mathbb{E}\left[\text{Re}\left\{\left(
\textbf{z}_{l}^\text{H}\textbf{W}_{l}^\text{H} \textbf{V}\right)^\text{T}\right\} \text{Re}\left\{
\textbf{z}_{l}^\text{H}\textbf{W}_{l}^\text{H} \textbf{V}\right\} \right]\nonumber\\
\overset{(c)}{=}\quad&\frac{2\lvert\textbf{s}\rvert^2}{\sigma_z^4}\mathbb{E}\left[\text{Re}\left\{\left(
\textbf{z}_{l}^\text{H}\textbf{W}_{l}^\text{H} \textbf{V}\right)^\text{T}
\textbf{z}_{l}^\text{H}\textbf{W}_{l}^\text{H} \textbf{V} \right\}\right]\nonumber\\
+&\frac{2\lvert\textbf{s}\rvert^2}{\sigma_z^4}\mathbb{E}\left[\text{Re}\left\{\left(
\textbf{z}_{l}^\text{H}\textbf{W}_{l}^\text{H} \textbf{V}\right)^\text{H}
\textbf{z}_{l}^\text{H}\textbf{W}_{l}^\text{H} \textbf{V} \right\}\right] \nonumber\\
\overset{(d)}{=}\quad &\frac{2\lvert\textbf{s}\rvert^2}{\sigma_z^4}\mathbb{E}\left[\text{Re}\left\{\left(
\textbf{z}_{l}^\text{H}\textbf{W}_{l}^\text{H} \textbf{V}\right)^\text{H}
\textbf{z}_{l}^\text{H}\textbf{W}_{l}^\text{H} \textbf{V} \right\}\right] \nonumber\\
\overset{(e)}{=}\quad&\frac {2 {\lvert\textbf{s}\rvert}^2} {\sigma_z^2}  \text{Re}\left\{\textbf{V}^\text{H} \textbf{W}_l \textbf{W}_l^\text{H} \textbf{V} \right\},
\end{align}where in Step $(c)$ we have used the following property of $\text{Re}\left\{\cdot\right\}$:
\begin{equation}\label{rp}
\text{Re}\left\{\textbf{u}\right\}\text{Re}\left\{\textbf{v}^\text{T}
\right\}=\frac{1}{2}\text{Re}\left\{\textbf{u}\textbf{v}^\text{T}\right\}+\frac{1}{2}\text{Re}\left\{\bar{\textbf{u}}\textbf{v}^\text{T}\right\}
\end{equation}with $\textbf{u},\,\textbf{v}$ denoting column vectors and $\bar{\textbf{u}}$ denoting the conjugate of $\textbf{u}$. 
Step $(d)$ is due to the exchangeability of $\mathbb{E}\left[\cdot\right]$ and $\text{Re}\left\{\cdot \right\}$:
\begin{equation}
\begin{aligned}\label{eq_stepb}
&\mathbb{E}\left[\text{Re}\left\{\left(
\textbf{z}_{l}^\text{H}\textbf{W}_{l}^\text{H} \textbf{V}\right)^\text{T}
\textbf{z}_{l}^\text{H}\textbf{W}_{l}^\text{H} \textbf{V} \right\}\right]\\
=& \text{Re}\left\{\mathbb{E}\left[\left(
\textbf{z}_{l}^\text{H}\textbf{W}_{l}^\text{H} \textbf{V}\right)^\text{T}
\textbf{z}_{l}^\text{H}\textbf{W}_{l}^\text{H} \textbf{V} \right]\right\}\\
=& \text{Re}\left\{\left(\textbf{W}_{l}^\text{H} \textbf{V}\right)^\text{T}\mathbb{E}\left[\left(
\textbf{z}_{l}^\text{H}\right)^\text{T}
\textbf{z}_{l}^\text{H}\right]\textbf{W}_{l}^\text{H} \textbf{V} \right\}\\
\overset{(f)}{=}&\textbf{0}.
\end{aligned}
\end{equation}
Step $(e)$ is due to the \emph{i.i.d.} circularly symmetric complex Gaussian property of each element of $\textbf{z}_l$, which means that $\mathbb{E}\left[\textbf{z}_l \textbf{z}_l^\text{H}\right]=\sigma_z^2\textbf{J}_3$, where $\textbf{J}_3$ is the 3-order identity matrix.
Step $(f)$ in \eqref{eq_stepb} results from the property of complex Gaussian noise:
\begin{equation}
\begin{aligned}\label{eq_GuassianProp}
\mathbb{E}\left[\left(
\textbf{z}_{l}^\text{H}\right)^\text{T}
\textbf{z}_{l}^\text{H}\right]=\textbf{0}.
\end{aligned}
\end{equation}

Therefore, the Fisher information matrix is derived in \eqref{eq_rrfisher} and Lemma \ref{MSEOpt} is proved in the end.

\section{Proof of Lemma~\ref{UnifiedOptShift}}\label{proof_UnifiedOptShift}
Lemma \ref{UnifiedOptShift} is proved in three steps:

\emph{\textbf{Step 1}: We prove that ${C}_S^{\min}(\boldsymbol{\psi})$ and $\left\{\boldsymbol{\Delta}_{S,1}^*,\boldsymbol{\Delta}_{S,2}^*,\boldsymbol{\Delta}_{S,3}^*\right\}$ are unrelated to the equivalent channel gain $\beta$.}

The basic method is block matrix inversion. We first rewrite the Jacobian matrix $\textbf{V}$ in \eqref{eq_Jacobian} as follows:
\vspace{-1mm}
\begin{align}\label{eq_rJacobian}
\textbf{V} = \left[\textbf{V}_1,\beta \textbf{V}_2\right],
\end{align}
\vspace{-1mm}where $\textbf{V}_1$ and $\textbf{V}_2$ are given by
\begin{equation}\label{eq_Vp}
\begin{aligned}
\left\{\begin{array}{*{20}{l}}
\textbf{V}_1 \triangleq \left[\textbf{a}\left(\textbf{x}\right),j\textbf{a}\left(\textbf{x}\right)\right]\\
\textbf{V}_2 \triangleq \left[ \frac {\partial \textbf{a}\left(\textbf{x}\right)}{\partial x_1},\frac {\partial \textbf{a}\left(\textbf{x}\right)}{\partial x_2}\right]
\end{array}\right..
\end{aligned}
\end{equation}It is clear that both $\textbf{V}_1$ and $\textbf{V}_2$ are unrelated to $\beta$. Besides, we can obtain the following properties of $\textbf{V}_1$:
\begin{equation}\label{eq_V1prop}
\begin{aligned}
\left\{\begin{array}{*{20}{l}}
\textbf{V}_1  \textbf{V}_1^\text{T}=\textbf{0}\\
\bar{\textbf{V}}_1 \textbf{V}_1^\text{H} = \textbf{0}
\end{array}\right.,
\end{aligned}
\end{equation}where $\bar{\textbf{V}}_1$ denotes the conjugate of ${\textbf{V}}_1$.

With the help of the Jacobian matrix $\textbf{V}$ in \eqref{eq_rJacobian}, the Fisher information matrix in \eqref{eq_fisher} can be divided into four $2 \times 2$ matrices as follows:
	\begin{align}\label{eq_FisherBlocks}
	&\textbf{I}_S(\boldsymbol{\psi}, \textbf{W})=\frac {2 {\lvert\textbf{s}\rvert}^2} {{\sigma}_z^2}  \text{Re}\left\{\textbf{V}^\text{H} \textbf{W} \textbf{W}^\text{H} \textbf{V} \right\}\nonumber\\
	&= \frac{2\lvert\textbf{s}\rvert^2}{{{\sigma_z ^2}}}\left[\!\!{\begin{array}{*{20}{c}}
		\text{Re}\left\{\textbf{V}_1^\text{H} \textbf{W} \textbf{W}^\text{H} \textbf{V}_1\right\} &\!\! \text{Re}\left\{\beta\textbf{V}_1^\text{H} \textbf{W} \textbf{W}^\text{H} \textbf{V}_2\right\}\\
		\text{Re}\left\{\bar{\beta}\textbf{V}_2^\text{H} \textbf{W} \textbf{W}^\text{H} \textbf{V}_1\right\} &\!\! \lvert\beta\rvert^2\text{Re}\left\{\textbf{V}_2^\text{H} \textbf{W} \textbf{W}^\text{H} \textbf{V}_2\right\}\\
		\end{array}} \!\!\!\right]\nonumber\\
	&= \frac{{2{{\lvert\textbf{s}\rvert}^2}}}{{{\sigma_z ^2}}}\left[\!{\begin{array}{*{20}{c}}
		{{\textbf{A}}}&\text{Re}\left\{\beta \textbf{B}\right\}\\
		\text{Re}\left\{\bar{\beta} \textbf{B}^\text{H}\right\}&\lvert \beta\rvert^2{{\textbf{D}}}
		\end{array}} \right],
	\end{align}where $\bar \beta$ denotes the conjugate of $\beta$ and $\textbf{A}$, $\textbf{B}$, $\textbf{D}$ are defined as:
\begin{align}\label{eq_Block}
\left\{\begin{array}{*{20}{l}}
{\textbf{A}} \triangleq \text{Re}\left\{\textbf{V}_1^\text{H} \textbf{W} \textbf{W}^\text{H} \textbf{V}_1\right\} = \left\|\textbf{W}^\text{H} \textbf{a}\left(\textbf{x}\right)\right\|_2^2 \textbf{J}_2\\
{\bf{B}} \triangleq \textbf{V}_1^\text{H} \textbf{W} \textbf{W}^\text{H} \textbf{V}_2. \\
{\textbf{D}} \triangleq \text{Re}\left\{\textbf{V}_2^\text{H} \textbf{W} \textbf{W}^\text{H} \textbf{V}_2\right\}
\end{array}\right.\!
\end{align}
\vspace{-0mm}with $\textbf{J}_2$ denoting the 2-order identity matrix. By combining \eqref{eq_V1prop} and \eqref{eq_Block}, we can obtain the properties of $\textbf{B}$:
\begin{equation}\label{eq_Bprop}
\begin{aligned}
\left\{\begin{array}{*{20}{l}}
\textbf{B}^\text{H}  \bar{\textbf{B}}=\textbf{0}\\
\textbf{B}^\text{T}  {\textbf{B}}=\textbf{0}\\
\textbf{B}^\text{H}  {\textbf{V}}_1^\text{T}=\textbf{V}_1 \bar {{\textbf{B}}}=\textbf{0}\\
\textbf{B}^\text{T} {\textbf{V}}_1^\text{H}=\bar{\textbf{V}}_1 {\textbf{B}}=\textbf{0}
\end{array}\right..
\end{aligned}
\end{equation}

By using the block matrix inversion method, the inverse of the Fisher information matrix in \eqref{eq_FisherBlocks} is given by
\vspace{-0mm}
	\begin{equation}\label{eq_InverseFisherBlocks}
	\begin{aligned}
	{\textbf{I}_S}\left(\boldsymbol{\psi} ,{{\textbf{W}}}\right)^{-1}= \frac{{{\sigma_z ^2}}}{{2{{\lvert\textbf{s}\rvert}^2}}}\left\{ {{{\bf{I}}_{i{p_1}}} + {{\bf{I}}_{i{p_2}}}\left(\beta \right)} \right\},
	\end{aligned}
	\end{equation}where ${{\textbf{I}}_{i{p_1}}}$ and ${{\textbf{I}}_{i{p_2}}}\left(\beta\right)$ are defined in \eqref{eq_Ip1} and \eqref{eq_Ip2}:
\begin{equation}\label{eq_Ip1}
\begin{aligned}
{{\textbf{{I}}}_{i{p_1}}} \triangleq \left[ \begin{matrix}
{{\bf{A}}^{-1}}&{\bf{0}}\\
{\bf{0}}&{\bf{0}}\end{matrix} \right],
\end{aligned}
\end{equation}
\begin{figure*}
	\normalsize
	\begin{equation}\label{eq_Ip2}
	{{\bf{I}}_{i{p_2}}}\left(\beta\right) \triangleq \left[ {\begin{matrix}
		{{\textbf{A}}{^{-1}}\text{Re}\left\{\beta \textbf{B}\right\}}\\
		{{\bf{ - }}{{\bf{J}}_2}}
		\end{matrix}} \right]{{\left( {\lvert \beta \rvert^2{\textbf{D}}-\text{Re}\left\{\bar{\beta} \textbf{B}^\text{H}\right\}{\bf{A}}{^{-1}}\text{Re}\left\{\beta \textbf{B}\right\}} \right)}^{-1}}\left[ {\begin{matrix}
		{\text{Re}\left\{\bar{\beta} \textbf{B}^\text{H}\right\}{\bf{A}}{^{-1}}}& {{\textbf{-}}{{\textbf{J}}_2}}
		\end{matrix}} \right].
	\end{equation}
	\end{figure*}The middle part of ${\bf{I}}_{i{p_2}}$, i.e., $( \lvert \beta \rvert^2{\textbf{D}}-\text{Re}\left\{\bar{\beta} \textbf{B}^\text{H}\right\}{\bf{A}}{^{-1}}$ $\text{Re}\left\{\beta \textbf{B}\right\} )$, can be rewritten as follows:
\begin{equation}\label{eq_midIip2}
\begin{aligned}
&{\lvert \beta \rvert^2{\textbf{D}}-\text{Re}\left\{\bar{\beta} \textbf{B}^\text{H}\right\}{\bf{A}}{^{-1}}\text{Re}\left\{\beta \textbf{B}\right\}}\\
 =&{\lvert \beta \rvert^2{\textbf{D}}-\frac{\bar{\beta} \textbf{B}^\text{H}+\beta \textbf{B}^\text{T}}{2}{\textbf{A}}{^{-1}}\frac{{\beta} \textbf{B}+\bar\beta \bar{\textbf{B}}}{2}}\\
\overset{(a)}{=}&{\lvert \beta \rvert^2{\textbf{D}}-\frac{\bar{\beta} \textbf{B}^\text{H}\textbf{A}^{-1}{\beta} \textbf{B}+\beta \textbf{B}^\text{T}\textbf{A}^{-1}\bar{\beta} \bar{\textbf{B}}}{4}}\\
\overset{(b)}{=}&{\lvert \beta \rvert^2{\textbf{D}}-\frac{\text{Re}\left\{\bar{\beta} \textbf{B}^\text{H}\textbf{A}^{-1}{\beta} \textbf{B}\right\}}{2}}\\
=&\lvert \beta \rvert^2 \left( {\textbf{D}}-\frac{ \text{Re}\left\{\textbf{B}^\text{H}\textbf{A}^{-1} \textbf{B}\right\}}{2}\right)\\
\overset{(c)}{=}&\lvert \beta \rvert^2 \textbf{I}_s,
\end{aligned}
\end{equation}where Step $(a)$ results from the properties of \textbf{B} in \eqref{eq_Bprop} and the definition of $\textbf{A}$ in \eqref{eq_Block}, Step $(b)$ is due to the fact that $\textbf{A}$ is a real matrix and  Step $(c)$ is due to the definition of $\textbf{I}_s$:
\begin{equation}
\textbf{I}_s\triangleq  {\textbf{D}}-\frac{ \text{Re}\left\{\textbf{B}^\text{H}\textbf{A}^{-1} \textbf{B}\right\}}{2}.
\end{equation}Therefore, we can rewrite $\textbf{I}_{ip_2}$ in \eqref{eq_Ip2} as follows:
    \begin{small}
	\begin{equation}\label{eq_rIp2}
	{{\bf{I}}_{i{p_2}}}\!\left(\beta\right) \!=\!\left[\! {\begin{matrix}
		{{\textbf{A}}{^{\!-\!1}}\text{Re}\left\{\beta \textbf{B}\right\}}\\
		{{\bf{ - }}{{\bf{J}}_2}}
		\end{matrix}} \!\right]{{\left(\! \lvert \beta \rvert^2 \textbf{I}_s\!\right)}^{\!-\!1}}\left[ {\begin{matrix}
		{\text{Re}\left\{\bar{\beta} \textbf{B}^\text{H}\right\}{\bf{A}}{^{\!\!-1}}}&\!\!\!\!\!-{{\textbf{J}}_2}
		\end{matrix}} \!\right]\!.\!
	\end{equation}\end{small}

By combining \eqref{eq_CMMSETemp} and \eqref{eq_InverseFisherBlocks}, we can obtain that
\vspace{-0mm}
	\begin{align}\label{eq_rCMMSETemp}
	&{C}_S(\boldsymbol{\psi},\textbf{W}) = \frac{1}{MN} \Tr\left\{ {{{\left( {{\bf{I}}_S(\psi ,{{\bf{W}}})} \right)}^{ - 1}}\textbf{V}^\text{H}\textbf{V}} \right\}\\
	&=\frac{1}{MN}\frac{{{\sigma_z ^2}}}{{2{{\lvert\textbf{s}\rvert}^2}}} \left(\Tr\left\{ {{{\bf{I}}_{i{p_1}}}}\textbf{V}^\text{H}\textbf{V}\right\}+\Tr\left\{ {{{\bf{I}}_{i{p_2}}}}\left(\beta\right)\textbf{V}^\text{H}\textbf{V}\right\}\right)\nonumber\\
	&\overset{(d)}{=}\frac{1}{MN}\frac{{{\sigma_z ^2}}}{{2{{\lvert\textbf{s}\rvert}^2}}} \left(\Tr\left\{ \textbf{A}^{-1}\textbf{V}_1^\text{H}\textbf{V}_1\right\}+\Tr\left\{ {{{\bf{I}}_{i{p_2}}}}\left(\beta\right)\textbf{V}^\text{H}\textbf{V}\right\}\right)\nonumber,
	\end{align}where Step $(d)$ is by substituting \eqref{eq_rJacobian} and \eqref{eq_Ip1} into \eqref{eq_rCMMSETemp}. Since both $\textbf{V}_1$ in \eqref{eq_Vp} and $\textbf{A}$ in \eqref{eq_Block} are unrelated to the equivalent channel gain $\beta$, the first part of \eqref{eq_rCMMSETemp}, i.e., $\Tr\left\{ \textbf{A}^{-1}\textbf{V}_1^\text{H}\textbf{V}_1\right\}$ are unrelated to $\beta$. By substituting \eqref{eq_rJacobian} and \eqref{eq_rIp2}, we can obtain the second part of \eqref{eq_rCMMSETemp}, i.e., $\Tr\left\{ {{{\bf{I}}_{i{p_2}}}}\left(\beta\right)\textbf{V}^\text{H}\textbf{V}\right\}$ in \eqref{eq_Imin2nd}, 
	\begin{figure*}
	\normalsize
	\begin{small}\begin{align}\label{eq_Imin2nd}
	\Tr\left\{ {{{\bf{I}}_{i{p_2}}}}\left(\beta\right)\textbf{V}^\text{H}\textbf{V}\right\}=&\quad \Tr\left\{\left[ {\begin{matrix}
		{{\textbf{A}}{^{-1}}\text{Re}\left\{\beta \textbf{B}\right\}}\\
		{{\bf{ - }}{{\bf{J}}_2}}
		\end{matrix}} \right]{\left( \lvert \beta \rvert^2 \textbf{I}_s\right)}^{-1}\left[ {\begin{matrix}
		{\text{Re}\left\{\bar{\beta} \textbf{B}^\text{H}\right\}{\bf{A}}{^{-1}}}& {{\textbf{-}}{{\textbf{J}}_2}}
		\end{matrix}}\right] \left[\begin{matrix}\textbf{V}_1^\text{H}\textbf{V}_1&\beta\textbf{V}_1^\text{H}\textbf{V}_2\\\bar{\beta}\textbf{V}_2^\text{H}\textbf{V}_1 & \lvert\beta\rvert^2 \textbf{V}_2^\text{H}\textbf{V}_2\end{matrix}\right]\right\}\nonumber\\
	=&\quad \Tr\left\{{\textbf{A}}{^{-1}}\text{Re}\left\{\beta \textbf{B}\right\}{\left( \lvert \beta \rvert^2 \textbf{I}_s\right)}^{-1}\left(\text{Re}\left\{\bar{\beta} \textbf{B}^\text{H}\right\}{\bf{A}}{^{-1}}\textbf{V}_1^\text{H}\textbf{V}_1-\bar{\beta}\textbf{V}_2^\text{H}\textbf{V}_1\right)\right\}\nonumber\\
	&+\Tr\left\{{\left( \lvert \beta \rvert^2 \textbf{I}_s\right)}^{-1}\left(\lvert\beta\rvert^2 \textbf{V}_2^\text{H}\textbf{V}_2-{\text{Re}\left\{\bar{\beta} \textbf{B}^\text{H}\right\}{\bf{A}}{^{-1}}}\beta\textbf{V}_1^\text{H}\textbf{V}_2\right)\right\}\nonumber\\
	=&\quad \Tr\left\{{\textbf{A}}{^{-1}}\frac{{\beta} \textbf{B}+\bar\beta \bar{\textbf{B}}}{2}{\left( \lvert \beta \rvert^2 \textbf{I}_s\right)}^{-1}\left(\frac{\bar{\beta} \textbf{B}^\text{H}+\beta \textbf{B}^\text{T}}{2}{\bf{A}}{^{-1}}\textbf{V}_1^\text{H}\textbf{V}_1-\bar{\beta}\textbf{V}_2^\text{H}\textbf{V}_1\right)\right\}\nonumber\\
	&+\Tr\left\{{\left( \lvert \beta \rvert^2 \textbf{I}_s\right)}^{-1}\left(\lvert\beta\rvert^2 \textbf{V}_2^\text{H}\textbf{V}_2-{\frac{\bar{\beta} \textbf{B}^\text{H}+\beta \textbf{B}^\text{T}}{2}{\bf{A}}{^{-1}}}\beta\textbf{V}_1^\text{H}\textbf{V}_2\right)\right\}\nonumber\\
	\overset{(e)}{=}&\quad \Tr\left\{{\textbf{A}}{^{-1}}\frac{{\beta} \textbf{B}+\bar\beta \bar{\textbf{B}}}{2}{\left( \lvert \beta \rvert^2 \textbf{I}_s\right)}^{-1}\left(\frac{\bar{\beta} \textbf{B}^\text{H}}{2}{\bf{A}}{^{-1}}\textbf{V}_1^\text{H}\textbf{V}_1-\bar{\beta}\textbf{V}_2^\text{H}\textbf{V}_1\right)\right\}\nonumber\\
	&+\Tr\left\{{\left( \lvert \beta \rvert^2 \textbf{I}_s\right)}^{-1}\left(\lvert\beta\rvert^2 \textbf{V}_2^\text{H}\textbf{V}_2-{\frac{\bar{\beta} \textbf{B}^\text{H}}{2}{\bf{A}}{^{-1}}}\beta\textbf{V}_1^\text{H}\textbf{V}_2\right)\right\}\nonumber\\
	=&\quad \Tr\left\{\left(\frac{\bar{\beta} \textbf{B}^\text{H}}{2}{\bf{A}}{^{-1}}\textbf{V}_1^\text{H}\textbf{V}_1-\bar{\beta}\textbf{V}_2^\text{H}\textbf{V}_1\right){\textbf{A}}{^{-1}}\frac{{\beta} \textbf{B}+\bar\beta \bar{\textbf{B}}}{2}{\left( \lvert \beta \rvert^2 \textbf{I}_s\right)}^{-1}\right\}\\
	&+\Tr\left\{\textbf{I}_s^{-1}\left( \textbf{V}_2^\text{H}\textbf{V}_2-\frac{ \textbf{B}^\text{H}{\bf{A}}{^{-1}}\textbf{V}_1^\text{H}\textbf{V}_2}{2}\right)\right\}\nonumber\\
	\overset{(f)}{=}&\quad \Tr\left\{\left(\frac{\bar{\beta} \textbf{B}^\text{H}}{2}{\bf{A}}{^{-1}}\textbf{V}_1^\text{H}\textbf{V}_1-\bar{\beta}\textbf{V}_2^\text{H}\textbf{V}_1\right){\textbf{A}}{^{-1}}\frac{{\beta} \textbf{B}}{2}{\left( \lvert \beta \rvert^2 \textbf{I}_s\right)}^{-1}\right\}\nonumber\\
	&+\Tr\left\{ \textbf{I}_s^{-1}\left( \textbf{V}_2^\text{H}\textbf{V}_2-\frac{ \textbf{B}^\text{H}{\bf{A}}{^{-1}}\textbf{V}_1^\text{H}\textbf{V}_2}{2}\right)\right\}\nonumber\\
	=&\quad \Tr\left\{\left(\frac{ \textbf{B}^\text{H}}{2}{\bf{A}}{^{-1}}\textbf{V}_1^\text{H}\textbf{V}_1-\textbf{V}_2^\text{H}\textbf{V}_1\right){\textbf{A}}{^{-1}}\frac{ \textbf{B}}{2}{\textbf{I}_s^{-1}}\right\}\nonumber\\
	&+\Tr\left\{{ \textbf{I}_s^{-1}}\left( \textbf{V}_2^\text{H}\textbf{V}_2-\frac{ \textbf{B}^\text{H}{\bf{A}}{^{-1}}\textbf{V}_1^\text{H}\textbf{V}_2}{2}\right)\right\}\nonumber\\
	=&\quad \Tr\left\{ \textbf{I}_s^{-1}\left(\frac{\textbf{B}^\text{H}{\bf{A}}{^{-1}}\textbf{V}_1^\text{H}\textbf{V}_1\textbf{A}^{-1}\textbf{B}}{4} +\textbf{V}_2^\text{H}\textbf{V}_2-\frac{ \textbf{B}^\text{H}{\bf{A}}{^{-1}}\textbf{V}_1^\text{H}\textbf{V}_2+\textbf{V}_2^\text{H}\textbf{V}_1 \textbf{A}^{-1}\textbf{B}}{2}\right)\right\}\nonumber,
	\end{align}\end{small}
\hrulefill
    \end{figure*}where Step $(e)$ and Step $(f)$ follow the properties of the $\textbf{B}$ in \eqref{eq_Bprop} and the definition of $\textbf{A}$ in \eqref{eq_Block}. It is clear that $\Tr\left\{ {{{\bf{I}}_{i{p_2}}}}\left(\beta\right)\textbf{V}^\text{H}\textbf{V}\right\}$ is also unrelated to $\beta$ because none of the matrix $\textbf{A},\textbf{B},\textbf{V}_1,\textbf{V}_2$ is related to $\beta$. Hence, ${C}_S(\boldsymbol{\psi},\textbf{W})$ is unrelated to $\beta$.

Since ${C}_S(\boldsymbol{\psi},\textbf{W})$ is unrelated to $\beta$, the minimum CRLB ${C}_S^{\min}(\boldsymbol{\psi})$ in \eqref{eq_CMMSE} and the optimal EBM $\textbf{W}_S^*$ are also unrelated to $\beta$. Hence, the optimal set of exploration offsets $\left\{\boldsymbol{\Delta}_{S,1}^*,\boldsymbol{\Delta}_{S,2}^*,\boldsymbol{\Delta}_{S,3}^*\right\}$ is unrelated to the equivalent channel gain $\beta$.

\emph{\textbf{Step 2}: We prove that ${C}_S^{\min}(\boldsymbol{\psi})$ and $\left\{\boldsymbol{\Delta}_{S,1}^*,\boldsymbol{\Delta}_{S,2}^*,\boldsymbol{\Delta}_{S,3}^*\right\}$ are unrelated to the DPV $\textbf{x}$.} 

Consider the CRLB in \eqref{eq_CMMSETemp}. we will first prove that the Fisher information matrix $\textbf{I}_S(\boldsymbol{\psi}, \textbf{W})$ is unrelated to the DPV $\textbf{x}$. Next, we will prove that $\textbf{V}^\text{H}\textbf{V}$ is also unrelated to $\textbf{x}$. Then it is clear that the minimum CRLB and the optimal set of exploration offsets $\left\{\boldsymbol{\Delta}_{S,1}^*,\boldsymbol{\Delta}_{S,2}^*,\boldsymbol{\Delta}_{S,3}^*\right\}$ are unrelated to $\textbf{x}$.

The Fisher information matrix in \eqref{eq_FisherBlocks} tells us that only $\textbf{W}^\text{H}\textbf{V}$ may be related to $\textbf{x}$, which is given by
    \begin{small}
	\begin{equation}\label{eq_WV}
	\textbf{W}^\text{H} \textbf{V} \!\!=\!\! \left[\!\textbf{W}^{
		\text{H}}\textbf{a}\!\left(\textbf{x}\right)\!,j\textbf{W}^\text{H}\textbf{a}\!\left(\textbf{x}\right)\!,\beta \textbf{W}^\text{H} \frac {\partial \textbf{a}\!\left(\textbf{x}\right)}{\partial x_1}\!,\beta \textbf{W}^\text{H} \frac {\partial \textbf{a}\!\left(\textbf{x}\right)}{\partial x_2}\!\right]\!
	\end{equation}\end{small}with $\textbf{W}^\text{H}\textbf{a}\left(\textbf{x}\right)$, $\textbf{W}^\text{H} \frac {\partial \textbf{a}\left(\textbf{x}\right)}{\partial x_1}$ and $\textbf{W}^\text{H} \frac {\partial \textbf{a}\left(\textbf{x}\right)}{\partial x_2}$ expanded as follows:
\begin{equation}\label{eq_Vpt}
\begin{aligned}
\left\{\begin{array}{*{20}{l}}
\textbf{W}^\text{H}\textbf{a}\left(\textbf{x}\right) = \left[\textbf{w}_{1}^\text{H}\textbf{a}\left(\textbf{x}\right),\textbf{w}_{2}^\text{H}\textbf{a}\left(\textbf{x}\right),\textbf{w}_{3}^\text{H}\textbf{a}\left(\textbf{x}\right)\right]^\text{T}\\\textbf{W}^\text{H} \frac {\partial \textbf{a}\left(\textbf{x}\right)}{\partial x_1}=\left[\textbf{w}_{1}^\text{H} \frac {\partial \textbf{a}\left(\textbf{x}\right)}{\partial x_1},\textbf{w}_{2}^\text{H} \frac {\partial \textbf{a}\left(\textbf{x}\right)}{\partial x_1},\textbf{w}_{3}^\text{H} \frac {\partial \textbf{a}\left(\textbf{x}\right)}{\partial x_1}\right]^\text{T}\\
\textbf{W}^\text{H} \frac {\partial \textbf{a}\left(\textbf{x}\right)}{\partial x_2}=\left[\textbf{w}_{1}^\text{H} \frac {\partial \textbf{a}\left(\textbf{x}\right)}{\partial x_2},\textbf{w}_{2}^\text{H} \frac {\partial \textbf{a}\left(\textbf{x}\right)}{\partial x_2},\textbf{w}_{3}^\text{H} \frac {\partial \textbf{a}\left(\textbf{x}\right)}{\partial x_2}\right]^\text{T}
\end{array}\right..
\end{aligned}
\end{equation}Since the EBVs are of the steering vector forms, i.e., $\textbf{w}_{i} = \frac{1}{\sqrt{MN}}\textbf{a}\left(\textbf{x}+\boldsymbol{\Delta}_{i}\right)$, where $\boldsymbol{\Delta}_{i}=\left[\delta_{i1},\delta_{i2}\right]^\text{T}$ denotes the $i$-th exploration offset, the  elements of $\textbf{W}^\text{H}\textbf{a}\left(\textbf{x}\right)$ and $\textbf{W}^\text{H} \frac {\partial \textbf{a}\left(\textbf{x}\right)}{\partial x_1}$ can be written in \eqref{eq_wa} and \eqref{eq_wpa}.
\vspace{-1mm}
\begin{align}\label{eq_wa}
\textbf{w}_{i}^\text{H}\textbf{a}\left(\textbf{x}\right) &=\frac{1}{{\sqrt {MN} }}\textbf{a}\left(\textbf{x}+\boldsymbol{\Delta}_{i}\right)^\text{H}\textbf{a}\left(\textbf{x}\right)\\
&=\frac{1}{{\sqrt {MN} }}\sum\limits_{m = 1}^M {\sum\limits_{n = 1}^N {{e^{-j2\pi \left[ {\frac{{(m - 1)\delta_{i1}}}{M} + \frac{{(n - 1)\delta_{i2}}}{N}} \right]}}} } \nonumber\\
&=\! \!\frac{1}{{\sqrt {M\!N} }}\frac{{\sin (\pi \delta_{i1})}}{{\sin \left( { \frac{\pi \delta_{i1}}{M}} \right)}}\frac{\sin (\pi \delta_{i2})}{{\sin \left( {\frac{\pi \delta_{i2}}{N}} \right)}}{e^{\!-\!j\pi \left( {\frac{{M \!-\! 1}}{M}\delta_{i1} \!+\! \frac{{N\!-\!1}}{N}\delta_{i2}} \right)}}\!, \nonumber
\end{align}
    \begin{figure*}
    \normalsize
	\begin{align}\label{eq_wpa}
	\textbf{w}_{i}^\text{H} \frac {\partial \textbf{a}\left(\textbf{x}\right)}{\partial x_1}&=  \frac{1}{{\sqrt {MN} }}\textbf{a}\left(\textbf{x}+\boldsymbol{\Delta}_{i}\right)^\text{H}\frac{{\partial {\bf{a}}({\bf{x}})}}{{\partial {x_1}}} = \frac{1}{{\sqrt {MN} }}\left( {\sum\limits_{m = 1}^M {\sum\limits_{n = 1}^N {j2\pi \frac{{m - 1}}{M}{e^{-j2\pi \left[ {\frac{{(m - 1)\delta_{i1}}}{M} + \frac{{(n - 1)\delta_{i2}}}{N}} \right]}}} } } \right)\\
	&= \frac{{j2\pi }}{{M\sqrt {MN} }}\left( {\frac{{\sin (\pi \delta_{i2})}}{{\sin \left( {\frac{\pi \delta_{i2} }{N}} \right)}}{e^{-j\pi \frac{{N - 1}}{N}\delta_{i2}}}\frac{{(M - 1){e^{-j2\pi \delta_{i1}}} - M{e^{-j2\pi \frac{{M - 1}}{M}\delta_{i1}}} + 1}}{{{{\left[ {1 - {e^{-j2\pi \frac{{\delta_{i1}}}{M}}}} \right]}^2}}}{e^{-j2\pi \frac{{\delta_{i1}}}{M}}}} \right).\nonumber
	\end{align}
	\end{figure*}As shown in \eqref{eq_wa} and \eqref{eq_wpa}, both $\textbf{w}_{i}^\text{H}\textbf{a}\left(\textbf{x}\right)$ and $\textbf{w}_{i}^\text{H} \frac {\partial \textbf{a}\left(\textbf{x}\right)}{\partial x_1}$ are unrelated to the DPV $\textbf{x}$. Similarly, $\textbf{w}_{i}^\text{H} \frac {\partial \textbf{a}\left(\textbf{x}\right)}{\partial x_2}$ is also unrelated to $\textbf{x}$. Therefore, $\textbf{W}^\text{H}\textbf{V}$ in \eqref{eq_WV} is unrelated to $\textbf{x}$. Hence, the whole Fisher information matrix in \eqref{eq_FisherBlocks} is invariant to $\textbf{x}$.

As for $\textbf{V}^\text{H}\textbf{V}$, we write it in \eqref{eq_VV},
       \begin{figure*}
       	\normalsize
		\begin{align}\label{eq_VV}
		\textbf{V}^\text{H}\textbf{V}
		&=\left[\begin{matrix}\textbf{a}\left(\textbf{x}\right)^\text{H}\\-j\textbf{a}\left(\textbf{x}\right)^\text{H}\\\bar{\beta}\frac {\partial \textbf{a}\left(\textbf{x}\right)^\text{H}}{\partial x_1}\\\bar{\beta}\frac {\partial \textbf{a}\left(\textbf{x}\right)^\text{H}}{\partial x_2}\end{matrix}\!\right] \left[\textbf{a}\left(\textbf{x}\right),j\textbf{a}\left(\textbf{x}\right),\beta \frac {\partial \textbf{a}\left(\textbf{x}\right)}{\partial x_1},\beta \frac {\partial \textbf{a}\left(\textbf{x}\right)}{\partial x_2}\right]\nonumber\\
		&=
		MN\left[ {\begin{matrix}
			{1}&{j}&{j\pi \beta \frac{M - 1}{M}}&{j\pi \beta \frac{N-1}{N}}\\
			{-j}&{1}&{\pi \beta \frac{M - 1}{M}}&{\pi \beta \frac{N-1}{N}}\\
			{ - j\pi \bar \beta \frac{M - 1}{M}}&{\pi \bar \beta \frac{M - 1}{M}}&{\frac{2}{3}{\pi ^2}{{\left| \beta  \right|}^2}\frac{{(M - 1)(2M - 1)}}{M^2}}&{{\pi ^2}{{\left| \beta  \right|}^2}\frac{(M - 1)(N - 1)}{MN}}\\
			{ - j\pi \bar \beta \frac{N-1}{N}}&{\pi \bar \beta \frac{N - 1}{N}}&{{\pi ^2}{{\left| \beta  \right|}^2}\frac{(M - 1)(N - 1)}{MN}}&{\frac{2}{3}{\pi ^2}{{\left| \beta  \right|}^2}M\frac{{(N - 1)(2N - 1)}}{N^2}}
			\end{matrix}} \right],
		\end{align}\hrulefill\end{figure*}which shows that $\textbf{V}^\text{H}\textbf{V}$ is unrelated to $\textbf{x}$.

Now it is clear that the CRLB in \eqref{eq_CMMSETemp}, i.e., ${C}_S(\boldsymbol{\psi},\textbf{W})$, is unrelated to $\textbf{x}$ because both the Fisher information matrix $\textbf{I}_S(\boldsymbol{\psi}, \textbf{W})$ and $\textbf{V}^\text{H}\textbf{V}$ are unrelated to $\textbf{x}$. Therefore, the minimum CRLB in \eqref{eq_CMMSE} and the optimal set of exploration offsets $\left\{\boldsymbol{\Delta}_{S,1}^*,\boldsymbol{\Delta}_{S,2}^*,\boldsymbol{\Delta}_{S,3}^*\right\}$ are invariant to the DPV $\textbf{x}$.

\emph{\textbf{Step 3}: We prove that} ${C}_S^{\min}(\boldsymbol{\psi})$ \emph{converges as} $\emph{M},\,\emph{N} \to +\infty$ \emph{and} 
	\vspace{-0mm}
	\begin{align}\label{eq_asymt} {\lim\limits_{M,N \to +\infty}}{C}_S(\boldsymbol{\psi},\widetilde{\textbf{W}}_S^*) = {\lim\limits_{M,N \to +\infty}}{C}_S^{\min}(\boldsymbol{\psi}),
	\nonumber
	\vspace{-0mm}
	\end{align}

Let us go into the asymptotic features of \eqref{eq_CMMSETemp}. According to \eqref{eq_wa} and \eqref{eq_wpa}, when the antenna number $\emph{M}$,\,$\emph{N} \to +\infty$, the limit of the $i$-th ($i=1,2,3$) element of $\textbf{W}^\text{H}\textbf{a}\left(\textbf{x}\right)$, $\textbf{W}^\text{H} \frac {\partial \textbf{a}\left(\textbf{x}\right)}{\partial x_1}$ and $\textbf{W}^\text{H} \frac {\partial \textbf{a}\left(\textbf{x}\right)}{\partial x_2}$ in \eqref{eq_Vpt} are given as follows:
\begin{small}
\begin{equation}\label{eq_walpl}
\begin{aligned}
\left\{\!\!\begin{array}{*{20}{l}}
	\mathop {\lim }\limits_{M.N \to  + \infty }\! \frac{\textbf{w}_{i}^\text{H}\textbf{a}\left(\textbf{x}\right)}{\sqrt {MN}}
&\!\!\!\!\!\!=\! \operatorname{Sa}\left[\pi \delta_{i1}\right]\operatorname{Sa}[\pi \delta_{i2}]{e^{{\rm{\! - }}j\pi \left(\delta_{i1}+\delta_{i2}\right)}}\\
\!\mathop {\lim }\limits_{M,N \to  + \infty }\! \frac{	\textbf{w}_{i}^\text{H} \frac {\partial \textbf{a}\left(\textbf{x}\right)}{\partial x_1}}{\sqrt {MN} }&\!\!\!\!\!\!=\! j2\pi  \operatorname{Sa}[\pi \delta_{i2}]{e^{{\rm{ - }}j\pi \delta_{i2}}}\frac{{{e^{{\rm{ \!- }}j2\pi \delta_{i1}}}\left(1{\rm{ + }}j2\pi \delta_{i1}\right) - 1}}{{{{\left(2\pi \delta_{i1}\right)}^2}}}\\
\!\mathop {\lim }\limits_{M,N \to  + \infty }\! \frac{	\textbf{w}_{i}^\text{H} \frac {\partial \textbf{a}\left(\textbf{x}\right)}{\partial x_2}}{\sqrt {MN} }&\!\!\!\!\!\!=\! j2\pi \operatorname{Sa}[\pi \delta_{i1}]{e^{{\rm{ - }}j\pi \delta_{i1}}}\frac{{{e^{{\rm{ \!- }}j2\pi \delta_{i2}}}\left(1{\rm{ + }}j2\pi \delta_{i2}\right) - 1}}{{{{\left(2\pi \delta_{i2}\right)}^2}}}
\end{array}\right.\!\!\!\!\!.
\end{aligned}
\end{equation}\end{small}where $\operatorname{Sa}\left[t\right] \triangleq \frac{\sin{t}}{t}$. Hence, each element of $\textbf{W}^\text{H}\textbf{V}/\sqrt{MN}$ in \eqref{eq_WV} converges when $M,\,N \to +\infty$, which results in that ${\textbf{I}}_S(\boldsymbol{\psi} ,\textbf{W})/{MN}$ in \eqref{eq_FisherBlocks} also converges. The limit is defined as follows:
\vspace{-0mm}
\begin{equation}\label{eq_Il}
\begin{aligned}
{\textbf{I}}_l(\boldsymbol{\psi} ,{{\bf{W}}}) \triangleq \lim\limits_{M,N \to  + \infty }  \frac{1}{MN}{\textbf{I}_S}(\boldsymbol{\psi} ,{{\bf{W}}}).
\end{aligned}
\end{equation}\vspace{-0mm}

The limit of $\textbf{V}^\text{H}\textbf{V}$ in \eqref{eq_VV} is given by
\vspace{-1mm}
\begin{align}\label{eq_VVl}
\mathop {\lim }\limits_{M,N \to  + \infty } \frac{1}{MN}\textbf{V}^\text{H}\textbf{V}&= \left[\! {\begin{matrix}
	1&j&{j\pi \beta }&{j\pi \beta }\\
	{ - j}&1&{\pi \beta }&{\pi \beta }\\
	{ - j\pi \bar \beta }&{\pi \bar \beta }&{\frac{4}{3}{\pi ^2}{{\left| \beta  \right|}^2}}&{{\pi ^2}{{\left| \beta  \right|}^2}}\\
	{ - j\pi \bar \beta }&{\pi \bar \beta }&{{\pi ^2}{{\left| \beta  \right|}^2}}&{\frac{4}{3}{\pi ^2}{{\left| \beta  \right|}^2}}
	\end{matrix}} \!\right]\nonumber\\
&\triangleq \textbf{H}_l.
\end{align}

By combining \eqref{eq_Il} and \eqref{eq_VVl}, we obtain the limit of $C_S(\boldsymbol{\psi},\textbf{W})$ in \eqref{eq_CMMSETemp} as $M,\,N \to + \infty$:
\vspace{-0mm}
\begin{equation}\label{eq_CMMSEL}
\begin{aligned}
&\mathop {\lim }\limits_{M,N \to  + \infty } \left( {MN \times C_S(\boldsymbol{\psi},\textbf{W})} \right)\\
=&\mathop {\lim }\limits_{M,N \to  + \infty }\Tr\left\{ {{{\left( {{\textbf{I}_S}(\psi ,{{\bf{W}}})} \right)}^{ - 1}}\textbf{V}^\text{H}\textbf{V}} \right\}\\
=&\mathop {\lim }\limits_{M,N \to  + \infty } \Tr\left\{ {{{\left( { M N{{\textbf{I}}_{l}(\boldsymbol{\psi} ,{{\bf{W}}})}} \right)}^{ - 1}}\textbf{V}^\text{H}\textbf{V}} \right\}\\
=&\mathop {\lim }\limits_{M,N \to  + \infty }\Tr\left\{ {{{\left( {{\textbf{I}}_{l}(\boldsymbol{\psi} ,{{\bf{W}}})} \right)}^{ - 1}}\frac{1}{MN}\textbf{V}^\text{H}\textbf{V}} \right\}\\
=&\Tr\left\{ {{{\left( {{\textbf{I}}_{l}(\boldsymbol{\psi} ,{{\bf{W}}})} \right)}^{ - 1}}}\textbf{H}_l \right\},
\end{aligned}
\end{equation}
\vspace{-1mm}which reveals that the CRLB in \eqref{eq_CMMSE}, i.e., $C_S(\boldsymbol{\psi},\textbf{W})$, converges. Hence, the minimum CRLB $C_S^{\min}(\boldsymbol{\psi})$ also converges. 

Let
\begin{align}\label{eq_Wtilde}
\widetilde{\textbf{W}}_S^* = \mathop{\arg\min}\limits_{\textbf{W}}\left(\mathop {\lim}\limits_{M,N \to  + \infty }C_S(\boldsymbol{\psi},\textbf{W})\right).
\end{align}Then we have 
\begin{align}
{\lim\limits_{M,N \to +\infty}}{C}_S^{\min}(\boldsymbol{\psi}) = \mathop {\lim}\limits_{M,N \to  + \infty }C_S(\boldsymbol{\psi},\textbf{W}^*)\\
\overset{(g)}{\ge} \mathop {\lim}\limits_{M,N \to  + \infty }C_S(\boldsymbol{\psi},\widetilde{\textbf{W}}_S^*),
\end{align}where Step (g) results from \eqref{eq_Wtilde}. On the other hand, we have
\begin{align}
{\lim\limits_{M,N \to +\infty}}{C}_S^{\min}(\boldsymbol{\psi}) 
\overset{(h)}{\le} \mathop {\lim}\limits_{M,N \to  + \infty }\left(C_S(\boldsymbol{\psi},\widetilde{\textbf{W}}_S^*)\right),
\end{align}where Step (h) results from \eqref{eq_CMMSE}. Hence, we can obtain that
\vspace{-0mm}
\begin{align}
{\lim\limits_{M,N \to +\infty}}{C}_S^{\min}(\boldsymbol{\psi}) = {\lim\limits_{M,N \to +\infty}}{C}_S(\boldsymbol{\psi},\widetilde{\textbf{W}}_S^*),
\nonumber
\vspace{-0mm}
\end{align}

Therefore, Lemma \ref{UnifiedOptShift} gets proved.

\section{Proof of Theorem~\ref{Converge to unique stable point}}\label{proof_Converge to unique stable point}
According to \eqref{eq_stoNew} and \eqref{eq_udp}, the tracking procedure can be rewritten as
	\vspace{-0mm}
	\begin{equation}\label{eq_rtracking}
	\hat {\boldsymbol{\psi}}_k = \hat {\boldsymbol{\psi}}_{k-1}+  b_{S,k} \left(\textbf{f}_{\boldsymbol{\psi}} \left(\hat {\boldsymbol{\psi}}_{k-1}\right) + \hat{\textbf{z}}_k \right).
	\vspace{-0mm}
	\end{equation}
And it can be derived in \eqref{eq_zini} that
\begin{align}\label{eq_z}
\hat{\textbf{z}}_k	= \frac{2\lvert \textbf{s} \rvert}{\sigma_z^2}\textbf{I}_S\left(\hat{\boldsymbol{\psi}}_{k-1}, \textbf{W}_k\right)^\text{-1}
\left[
\begin{matrix}
{\text{Re}\left\{ \textbf{e}_k^\text{H} \textbf{z}_k\right\}}\\
{\text{Im}\left\{ \textbf{e}_k^\text{H} \textbf{z}_k\right\}}\\
{\text{Re}\left\{ \tilde{\textbf{e}}_{k1}^\text{H} \textbf{z}_k\right\}}\\
{\text{Re}\left\{ \tilde{\textbf{e}}_{k2}^\text{H} \textbf{z}_k\right\}}
\end{matrix}
\right].
\end{align}

Since $\hat{\textbf{z}}_k \triangleq \left[\hat{z}_{k,1},\hat{z}_{k,2},\hat{z}_{k,3} \right]$ in \eqref{eq_z} is composed of three \emph{i.i.d.} circularly symmetric complex Gaussian random variables, the expectation of $\hat{\textbf{z}}_k$ is $\mathbb{E}\left[\hat{\textbf{z}}_k\right] = \textbf{0}$ and the covariance matrix is given in \eqref{eq_rzc},
\begin{figure*}
\normalsize
	\begin{equation}\label{eq_rzc}
	\begin{aligned}
	&~\mathbb{E}\left[ \left(\hat{\mathbf{z}}_k - \mathbb{E}\left[ \hat{\mathbf{z}}_k \right] \right) \left(\hat{\mathbf{z}}_k - \mathbb{E}\left[ \hat{\mathbf{z}}_k \right] \right)^\text{T}\right]\\ =&~\frac{4\lvert\textbf{s}\rvert^2}{\sigma_z^4}\textbf{I}_S\left(\hat{\boldsymbol{\psi}}_{k-1}, \textbf{W}_k\right)^\text{-1} \mathbb{E}\!\left\{\!\!\left[\begin{matrix} \operatorname{Re}\{{\mathbf{e}}_{k}^\text{H}\mathbf{z}_k\} \\
	\operatorname{Im}\{{\mathbf{e}}_{k}^\text{H}\mathbf{z}_k\} \\
	\operatorname{Re}\{\tilde{\mathbf{e}}_{k1}^\text{H}\mathbf{z}_k\}\\
	\operatorname{Re}\{\tilde{\mathbf{e}}_{k2}^\text{H}\mathbf{z}_k\}
	\end{matrix}\right]\!\!\cdot\!\!\left[\begin{matrix} \operatorname{Re}\{{\mathbf{e}}_{k}^\text{H}\mathbf{z}_k\} \\
	\operatorname{Im}\{{\mathbf{e}}_{k}^\text{H}\mathbf{z}_k\} \\
	\operatorname{Re}\{\tilde{\mathbf{e}}_{k1}^\text{H}\mathbf{z}_k\}\\
	\operatorname{Re}\{\tilde{\mathbf{e}}_{k2}^\text{H}\mathbf{z}_k\} \end{matrix}\right]^\text{\!\!T} \!\right\} \textbf{I}_S\left(\hat{\boldsymbol{\psi}}_{k-1}, \textbf{W}_k\right)^\text{-1}\\
	 = &~ \frac{4\lvert\textbf{s}\rvert^2}{\sigma_z^4}\textbf{I}_S\left(\hat{\boldsymbol{\psi}}_{k-1}, \textbf{W}_k\right)^\text{-1}\mathbb{E}\left\{\text{Re}\left\{\left[
	\begin{matrix}
	\textbf{z}_{k}^\text{H}\textbf{W}_{k}^\text{H} \frac{\partial \hat{\textbf{h}}_{k-1}}{\partial \hat{\beta}_{k-1}^\text{re}}\\
	\textbf{z}_{k}^\text{H}\textbf{W}_{k}^\text{H} \frac{\partial \hat{\textbf{h}}_{k-1}}{\partial \hat{\beta}_{k-1}^\text{im}}\\
	\textbf{z}_{k}^\text{H}\textbf{W}_{k}^\text{H} \frac{\partial \hat{\textbf{h}}_{k-1}}{\partial \hat{x}_{k-1,1}}\\
	\textbf{z}_{k}^\text{H}\textbf{W}_{k}^\text{H} \frac{\partial \hat{\textbf{h}}_{k-1}}{\partial \hat{x}_{k-1,2}}
	\end{matrix}\right]\right\}\text{Re}\left\{\left[
	\begin{matrix}
	\textbf{z}_{k}^\text{H}\textbf{W}_{k}^\text{H} \frac{\partial \hat{\textbf{h}}_{k-1}}{\partial \hat{\beta}_{k-1}^\text{re}}\\
	\textbf{z}_{k}^\text{H}\textbf{W}_{k}^\text{H} \frac{\partial \hat{\textbf{h}}_{k-1}}{\partial \hat{\beta}_{k-1}^\text{im}}\\
	\textbf{z}_{k}^\text{H}\textbf{W}_{k}^\text{H} \frac{\partial \hat{\textbf{h}}_{k-1}}{\partial \hat{x}_{k-1,1}}\\
	\textbf{z}_{k}^\text{H}\textbf{W}_{k}^\text{H} \frac{\partial \hat{\textbf{h}}_{k-1}}{\partial \hat{x}_{k-1,2}}
	\end{matrix}\right]^\text{T}\right\}\right\}\textbf{I}_S\left(\hat{\boldsymbol{\psi}}_{k-1}, \textbf{W}_k\right)^\text{-1}\\
     \buildrel  {(a)} \over =  &~ \frac{4\lvert\textbf{s}\rvert^2}{\sigma_z^4}\textbf{I}_S\left(\hat{\boldsymbol{\psi}}_{k-1}, \textbf{W}_k\right)^\text{-1}\mathbb{E}\left\{\frac{\sigma_z^2}{2\lvert\textbf{s}\rvert}\frac {\partial \text{log} \, p_S \left(\textbf{y}_k |\hat{\boldsymbol{\psi}}_{k-1},\textbf{W}_k \right)}{\partial \hat{\boldsymbol{\psi}}_{k-1}}\frac{\sigma_z^2}{2\lvert\textbf{s}\rvert}\frac {\partial \text{log} \, p_S \left(\textbf{y}_k |\hat{\boldsymbol{\psi}}_{k-1},\textbf{W}_k \right)}{\partial \hat{\boldsymbol{\psi}}_{k-1}^\text{T}}\right\}\textbf{I}_S\left(\hat{\boldsymbol{\psi}}_{k-1}, \textbf{W}_k\right)^\text{-1}\\	
     \buildrel  {(b)} \over =  &~ \textbf{I}_S\left(\hat{\boldsymbol{\psi}}_{k-1}, \textbf{W}_k\right)^\text{-1}\textbf{I}_S\left(\hat{\boldsymbol{\psi}}_{k-1}, \textbf{W}_k\right)\textbf{I}_S\left(\hat{\boldsymbol{\psi}}_{k-1}, \textbf{W}_k\right)^\text{-1}\\	
= 	&~ \textbf{I}_S\left(\hat{\boldsymbol{\psi}}_{k-1}, \textbf{W}_k\right)^\text{-1},
	\end{aligned}
	\end{equation}
    \hrulefill
    \end{figure*}where Step $(a)$ is the result of \eqref{eq_gradient} and Step $(b)$ follows the definition of the Fisher information matrix in \eqref{eq_fisher}.

Assume $\{ \mathcal{G}_k: k \ge 0 \}$ is an increasing sequence of $\sigma$-fields of $\{ \hat{\boldsymbol{\psi}}_{0}, \hat{\boldsymbol{\psi}}_{1}, \hat{\boldsymbol{\psi}}_{2}, \ldots \}$, i.e., $\mathcal{G}_{k-1}\!\subset\!\mathcal{G}_k$, where $\mathcal{G}_{0} \!\overset{\Delta}{=}\! \sigma(\hat{\boldsymbol{\psi}}_{0})$ and $\mathcal{G}_k \!\overset{\Delta}{=}\! \sigma(\hat{\boldsymbol{\psi}}_{0}, \hat{\mathbf{z}}_{1}, \ldots, \hat{\mathbf{z}}_{k}) $ for $k \ge 1$. Because the $\hat{\mathbf{z}}_k$'s are composed of \emph{i.i.d.} circularly symmetric complex Gaussian random variables with zero mean, $\hat{\mathbf{z}}_k$ is independent of $\mathcal{G}_{k-1}$, and $\hat{\boldsymbol{\psi}}_{k-1} \!\in\! \mathcal{G}_{k-1}$. Hence, we have
\begin{align}\label{eq_fil2}
&\mathbb{E} \left[ \left. \textbf{f}_{\boldsymbol{\psi}} \left(\hat {\boldsymbol{\psi}}_{k-1}\right) + \hat{\mathbf{z}}_k \right| \mathcal{G}_{k-1} \right]\nonumber\\
=& \mathbb{E} \left[ \left. \textbf{f}_{\boldsymbol{\psi}} \left(\hat {\boldsymbol{\psi}}_{k-1}\right) \right| \mathcal{G}_{k-1} \right] + \mathbb{E} \left[ \left. \hat{\mathbf{z}}_k \right| \mathcal{G}_{k-1} \right] \nonumber\\
=& \textbf{f}_{\boldsymbol{\psi}} \left(\hat {\boldsymbol{\psi}}_{k-1}\right),
\end{align}
for $k \ge 1$ and $\boldsymbol{\varsigma}_k = \textbf{f}_{\boldsymbol{\psi}} \left(\hat {\boldsymbol{\psi}}_{k-1}\right) + \hat{\textbf{z}}_k$ is also independent of $\mathcal{G}_{k-1}$.

Theorem 5.2.1 in \cite[Section 5.2.1]{kushner2003stochastic} gives the conditions that ensure $\hat{\boldsymbol{\psi}}_k$ converges to a unique point  with probability one when there are several stable points. Next, we will prove that if the step-size $b_{S,k}$ is given by \eqref{eq_stepsize} with any $\varepsilon_S > 0$ and $K_{S,0} \ge 0$, then the joint beam and channel tracking algorithm in \eqref{eq_Tracking} satisfies the corresponding conditions below:
\begin{itemize}
	\item[1)] Step-size requirements:
	\begin{equation}\left\{\!\begin{aligned}&b_{S,k} = \frac{\varepsilon_S}{k + K_{S,0}} \rightarrow 0, \\
	& \sum\limits_{k=1}^{+\infty} b_{S,k} = \sum\limits_{k=1}^{+\infty} \frac{\varepsilon_S}{k+K_{S,0}} = +\infty, \\
	& \sum\limits_{k=1}^{+\infty} b_{S,k}^2 \!=\! \sum\limits_{k=1}^{+\infty} \frac{\varepsilon_S^2}{(k\!+\!K_{S,0})^2} \!\le\! \sum\limits_{l=1}^{+\infty} \!\frac{\varepsilon_S^2}{l^2} \!<\! {+\infty}.\end{aligned}\right.\end{equation}
	
	\item[2)] It is necessary to prove that $\sup\nolimits_k \mathbb{E} \left[ \left\|\textbf{f}_{\boldsymbol{\psi}} \left(\hat {\boldsymbol{\psi}}_{k-1}\right) + \hat{\mathbf{z}}_k\right\|_2^2 \right] < +\infty$. \\
	From \eqref{eq_rtracking} and \eqref{eq_rzc}, we have
	\begin{align}\label{eq_expectation_yk}  &\mathbb{E} \left[ \left\|\textbf{f}_{\boldsymbol{\psi}} \left(\hat {\boldsymbol{\psi}}_{k-1}\right) + \hat{\mathbf{z}}_k\right\|_2^2 \right]\\
	= & \mathbb{E} \left[ \left\|\textbf{f}_{\boldsymbol{\psi}} \left(\hat {\boldsymbol{\psi}}_{k-1}\right)\right\|_2^2 + 2 \textbf{f}_{\boldsymbol{\psi}} \left(\hat {\boldsymbol{\psi}}_{k-1}\right)^\text{H} \hat{\mathbf{z}}_k + \left\|\hat{\mathbf{z}}_k\right\|_2^2 \right] \nonumber \\ \overset{(c)}{=} & \mathbb{E} \left[ \left\|\textbf{f}_{\boldsymbol{\psi}} \left(\hat {\boldsymbol{\psi}}_{k-1}\right)\right\|_2^2 \right] + \Tr\left\{\mathbf{I}_S(\hat{\boldsymbol{\psi}}_{k\!-\!1},\!\mathbf{W}_k)^{-1}\right\}, \nonumber\end{align}
	where Step $(c)$ is due to \eqref{eq_rzc} and that $\hat{\mathbf{z}}_k$ is independent of $\textbf{f}_{\boldsymbol{\psi}} \left(\hat {\boldsymbol{\psi}}_{k-1}\right)$. \\
	From \eqref{eq_fd}, we have
	\begin{small}\begin{align}\label{eq_ub_fx}
&\left\|\textbf{f}_{\boldsymbol{\psi}} \left(\hat {\boldsymbol{\psi}}_{k-1}\right)\right\|_2^2 \le \left\|\mathbf{I}_S(\hat{\boldsymbol{\psi}}_{k\!-\!1},\!\mathbf{W}_k)^{-1}\right\|_\text{F}^2 \\
&\cdot\left\|\frac{2{\lvert\textbf{s}\rvert}^2}{\sigma_z^2}\!\!\left[\begin{matrix} {\operatorname{Re}\left\{\textbf{e}_k^\text{H}\left(\beta \textbf{W}_k^\text{H} \textbf{a}\left(\textbf{x}\right) -\hat{\beta}_{k-1} \textbf{e}_k \right)\right\}}\\
	{\operatorname{Im}\left\{\textbf{e}_k^\text{H}\left(\beta \textbf{W}_k^\text{H} \textbf{a}\left(\textbf{x}\right)-\hat{\beta}_{k-1} \textbf{e}_k \right)\right\}}\\
	{\operatorname{Re}\left\{\tilde{\textbf{e}}_{k1}^\text{H}\left(\beta \textbf{W}_k^\text{H} \textbf{a}\left(\textbf{x}\right)-\hat{\beta}_{k-1} \textbf{e}_k \right)\right\}}\\
	{\operatorname{Re}\left\{\tilde{\textbf{e}}_{k2}^\text{H}\left(\beta \textbf{W}_k^\text{H} \textbf{a}\left(\textbf{x}\right)-\hat{\beta}_{k-1} \textbf{e}_k \right)\right\}} \end{matrix}\right]\right\|_2^2.\nonumber \end{align}\end{small}
	As the Fisher information matrix is invertible, we get
	\begin{align}\label{eq:proof1_1}\left\|\mathbf{I}_S(\hat{\boldsymbol{\psi}}_{k\!-\!1},\!\mathbf{W}_k)^{-1}\right\|_\text{F}^2 < +\infty.\end{align}
	Besides, $\mathbf{W}_{k}\!=\!\left[\mathbf{w}_{k,1},\mathbf{w}_{k,2}, \mathbf{w}_{k,3}\right]$, ${\mathbf{e}}_{k}\!=\!\mathbf{W}_k^\text{H}\mathbf{a}(\hat{\textbf{x}}_{k\!-\!1})$, $\tilde{\textbf{e}}_{k1} = \hat{\beta}_{k-1}\textbf{W}_k^\text{H} \frac{\partial \textbf{a}\left(\hat{\textbf{x}}_{k-1}\right)}{\partial x_1}$, $\tilde{\textbf{e}}_{k2} =\hat{\beta}_{k-1} \textbf{W}_k^\text{H} \frac{\partial \textbf{a}\left(\hat{\textbf{x}}_{k-1}\right)}{\partial x_2}$, hence we have
	\begin{align}
	&\left| {{\bf{w}}_{k,i}^{\rm{H}}{\bf{a}}({\bf{x}})} \right|\nonumber	\\&= \left| {\frac{1}{{\sqrt {MN} }}\sum\limits_{m = 1}^M {\sum\limits_{n = 1}^N {{e^{-j2\pi \left( {\frac{{(m - 1){\delta_{k,i1}}}}{M} + \frac{{(n - 1){\delta_{k,i2}}}}{N}} \right)}}} } } \right|\nonumber\\
	&\le \frac{1}{{\sqrt {MN} }}\sum\limits_{m = 1}^M {\sum\limits_{n = 1}^N {\left| {{e^{-j2\pi \left( {\frac{{(m - 1){\delta_{k,i1}}}}{M} + \frac{{(n - 1){\delta_{k,i2}}}}{N}} \right)}}} \right|} }\nonumber \\
	&= \sqrt {MN} < +\infty,
	\end{align}
	    \begin{small}
		\begin{align}
		&\left| {{\bf{w}}_{k,i}^{\rm{H}}\frac{{\partial {\bf{a}}({\bf{x}})}}{{\partial {x_1}}}} \right|\nonumber\\
		=\!& \left| {\frac{1}{{\sqrt {MN} }}\!\sum\limits_{m = 1}^M\! {\sum\limits_{n = 1}^N {j2\pi \frac{{m - 1}}{M}\!{{e^{-j2\pi \left(\! {\frac{{(m - 1){\delta_{k,i1}}}}{M} \!+\! \frac{{(n \!-\! 1){\delta_{k,i2}}}}{N}} \!\right)}}}} } } \!\right|\nonumber\\
		\le \!& \frac{{2\pi}}{{M\!\sqrt {MN} }}\!\sum\limits_{m = 1}^M\! {\sum\limits_{n = 1}^N {(m \!-\! 1)\left| {{{e^{-j2\pi \left(\! {\frac{{(m - 1){\delta_{k,i1}}}}{M} + \frac{{(n - 1){\delta_{k,i2}}}}{N}} \!\right)}}}} \!\right|} } \nonumber\\
		=& \sqrt {MN} \left( {M - 1} \right) < +\infty,
		\end{align}\end{small}and
		\begin{small}\begin{align}
		&\left| {{\bf{w}}_{k,i}^{\rm{H}}\frac{{\partial {\bf{a}}({\bf{x}})}}{{\partial {x_2}}}} \right|\nonumber\\
		=\!& \left|\! {\frac{1}{{\sqrt {MN} }}\!\sum\limits_{m = 1}^M\! {\sum\limits_{n = 1}^N {j2\pi \frac{{n \!-\! 1}}{N}{{e^{-j2\pi \left( {\frac{{(m - 1){\delta_{k,i1}}}}{M} + \frac{{(n - 1){\delta_{k,i2}}}}{N}} \right)}}}} } } \right|\nonumber\\
		\le& \frac{{2\pi }}{{N\!\sqrt {MN} }}\!\sum\limits_{m = 1}^M \!{\sum\limits_{n = 1}^N {(n - 1)\left| {{{e^{-j2\pi \left( {\frac{{(m - 1){\delta_{k,i1}}}}{M} + \frac{{(n \!-\! 1){\delta_{k,i2}}}}{N}} \right)}}}} \right|} } \nonumber\\
		=& \sqrt {MN} \left( {N - 1} \right) < +\infty,
		\end{align}\end{small}for $i = 1, 2,3$ and all possible $\textbf{w}_{k,i}$ and $\textbf{x}$, where $\left[\delta_{k,i1},\delta_{k,i2}\right]^\text{T} = \boldsymbol{\omega}_{k,i}-\textbf{x}$. Thus we can get
	\begin{align}\label{eq:proof1_2}
	\left\|\frac{2{\lvert\textbf{s}\rvert}^2}{\sigma_z^2}\!\!\left[\begin{matrix} {\operatorname{Re}\left\{\textbf{e}_k^\text{H}\left(\beta \textbf{W}_k^\text{H} \textbf{a}\left(\textbf{x}\right) -\hat{\beta}_{k-1} \textbf{e}_k \right)\right\}}\\
	{\operatorname{Im}\left\{\textbf{e}_k^\text{H}\left(\beta \textbf{W}_k^\text{H} \textbf{a}\left(\textbf{x}\right)-\hat{\beta}_{k-1} \textbf{e}_k \right)\right\}}\\
	{\operatorname{Re}\left\{\tilde{\textbf{e}}_{k1}^\text{H}\left(\beta \textbf{W}_k^\text{H} \textbf{a}\left(\textbf{x}\right)-\hat{\beta}_{k-1} \textbf{e}_k \right)\right\}}\\
	{\operatorname{Re}\left\{\tilde{\textbf{e}}_{k2}^\text{H}\left(\beta \textbf{W}_k^\text{H} \textbf{a}\left(\textbf{x}\right)-\hat{\beta}_{k-1} \textbf{e}_k \right)\right\}} \end{matrix}\right]\right\|_2^2 < +\infty.
	\end{align}
	Combining \eqref{eq:proof1_1} and \eqref{eq:proof1_2}, we have
	\begin{align}\label{eq:expectation_fx}\mathbb{E} \left[ \left\|\textbf{f}_{\boldsymbol{\psi}} \left(\hat {\boldsymbol{\psi}}_{k-1}\right)\right\|_2^2 \right] < +\infty. \end{align}
	According to  \eqref{eq:proof1_1}, it is clear that $ \Tr\left\{\mathbf{I}(\hat{\boldsymbol{\psi}}_{k-1},\mathbf{W}_k)^{-1}\right\}$ $  < +\infty$. Then, we can get that
	\begin{align}\sup\nolimits_k \mathbb{E} \left[ \left\|\textbf{f}_{\boldsymbol{\psi}} \left(\hat {\boldsymbol{\psi}}_{k-1}\right) + \hat{\mathbf{z}}_k\right\|_2^2 \right] < +\infty.\end{align}
	
	\item[3)] The function $\textbf{f}_{\boldsymbol{\psi}} \left(\hat {\boldsymbol{\psi}}_{k-1}\right)$ should be continuous with respect to $\hat{\boldsymbol{\psi}}_{k-1}$.\\
	By using \eqref{eq_fd}, we know that each element of $\textbf{f}_{\boldsymbol{\psi}} \left(\hat {\boldsymbol{\psi}}_{k-1}\right)$ is continuous with respect to $\hat{\boldsymbol{\psi}}_{k-1} = \left [\hat{\beta}_{k-1}^\text{re},\hat{\beta}_{k-1}^\text{im},\hat{x}_{k-1,1},\hat{x}_{k-1,2}\right]^\text{T}$. Therefore, $\textbf{f}_{\boldsymbol{\psi}} \left(\hat {\boldsymbol{\psi}}_{k-1}\right)$ is continuous with respect to $\hat{\boldsymbol{\psi}}_{k-1}$.
	
	\item[4)] Let $\boldsymbol\mu_k = \mathbb{E} \left[ \left.\textbf{f}_{\boldsymbol{\psi}} \left(\hat {\boldsymbol{\psi}}_{k-1}\right) + \hat{\mathbf{z}}_k\right| \mathcal{G}_{k-1} \right] - \mathbf{f}\left(\hat{\boldsymbol{\psi}}_{k-1}, \boldsymbol{\psi}\right)$. We need to prove that $\sum_{k=1}^{+\infty} \left\| b_{S,k} \boldsymbol\mu_k \right\|_2 < +\infty$ with probability one. \\
	From (\ref{eq_fil2}), we get $\boldsymbol\mu_k = \mathbf{0}$ for all $k \ge 1$. So we have $\sum_{k=1}^{+\infty} \left\| b_{S,k} \boldsymbol\mu_k \right\|_2 = 0 < +\infty$ with probability one.
	
\end{itemize}

By Theorem 5.2.1 in \cite{kushner2003stochastic}, $\hat{\boldsymbol{\psi}}_k$ converges to a unique stable point within the stable points set with probability one. 


\section{Proof of Theorem~\ref{Converge to real beam direction}}\label{proof_Converge to real beam direction}

Theorem \ref{Converge to real beam direction} is proven in three steps:

\emph{\textbf{Step 1:} Two continuous processes based on the discrete process $ \hat{\boldsymbol{\psi}}_k = [\hat{\beta}^\text{re}_{k}, \hat{\beta}^\text{im}_{k}, \hat{x}_{k,1},\hat{x}_{k,2}]^\text{\emph{T}}$ are established, i.e., $\bar{\boldsymbol{\psi}}(t) \!\overset{\Delta}{=}\! [\bar{\beta}^\text{re}(t), \bar{\beta}^\text{im}(t), \bar{x}_{1}(t),\bar{x}_{2}(t)]^\text{\emph{T}}$ and $\tilde{\boldsymbol{\psi}}^k(t) \!\overset{\Delta}{=}\! [\tilde{\beta}^{\text{re},k}(t), \tilde{\beta}^{\text{im},k}(t), \tilde{x}_{1}^k(t), \tilde{x}_{2}^k(t)]^\text{\emph{T}}$.}

The discrete time parameters are defined as: $t_{0} \overset{\Delta}{=} 0$, $t_k \overset{\Delta}{=} \sum_{l=1}^k b_{S,l}$, $k \ge 1$. The first continuous process $\bar{\boldsymbol{\psi}}(t), t \ge 0$ is constructed as the linear interpolation of the sequence $\hat{\boldsymbol{\psi}}_k, k\ge 0$, where $\bar{\boldsymbol{\psi}}(t_k) = \hat{\boldsymbol{\psi}}_k, k \ge 0$. Therefore, $\bar{\boldsymbol{\psi}}(t)$ is given by
\begin{small}
\begin{equation}\label{eq_continuous}
\begin{aligned}
\bar{\boldsymbol{\psi}}(t)\!=\!\bar{\boldsymbol{\psi}}(t_k)\!+\!\frac{(t\!-\!t_k)}{b_{S,k+1}}\left[\bar{\boldsymbol{\psi}}(t_{k+1})\!-\!\bar{\boldsymbol{\psi}}(t_k)\right], t\!\in\![t_k, t_{k+1}].
\end{aligned}
\end{equation}\end{small}

The second continuous process $\tilde{\boldsymbol{\psi}}^k(t)$ is the solution of the following ordinary differential equation (ODE):
\begin{align}\label{eq_ODE}
\frac{d \tilde{\boldsymbol{\psi}}^k(t)}{dt} = \mathbf{f}_{ \boldsymbol{\psi}}\left(\tilde{\boldsymbol{\psi}}^k(t)\right),
\end{align}
for $t \in [t_k, \infty)$, where $\tilde{\boldsymbol{\psi}}^k(t_k) = \bar{\boldsymbol{\psi}}(t_k) = \hat{\boldsymbol{\psi}}_k, k \ge 0$. Thus, $\tilde{\boldsymbol{\psi}}^k(t)$ can be given as
\begin{equation}\label{eq_ODE_new}
\begin{aligned}
\tilde{\boldsymbol{\psi}}^k(t) & = \bar{\boldsymbol{\psi}}(t_k) + \int_{t_k}^t \mathbf{f}_{ \boldsymbol{\psi}}\left(\tilde{\boldsymbol{\psi}}^k(v)\right) dv, t \ge t_k.
\end{aligned}
\end{equation}

\emph{\textbf{Step 2:} By using the two continuous processes $\bar{\boldsymbol{\psi}}(t)$ and $\tilde{\boldsymbol{\psi}}^k(t)$ constructed in Step 1, a sufficient condition for the convergence of the discrete process $\hat{\textbf{x}}_k$ is provided here.}

\begin{figure}[!t]
	\centering
	\includegraphics[width=6.5cm]{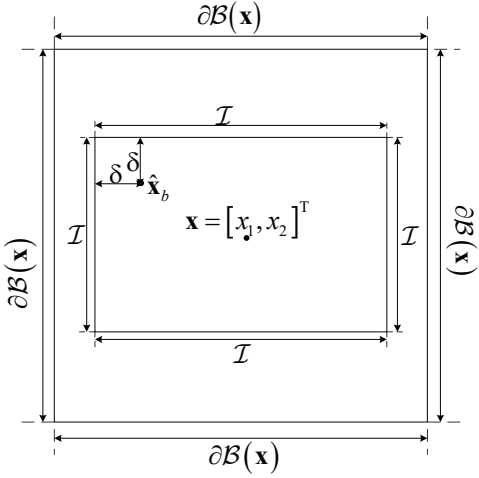}
	\caption{An illustration of the invariant set $\mathcal{I}$.}
	\label{fig_invariant_set}
\end{figure}

We first construct a time-invariant set $\mathcal{I}$ that includes the DPV $\textbf{x}$ within the main lobe, i.e., $\textbf{x} \in \mathcal{I} \subset \mathcal{B}(\textbf{x})$. Define $\tilde{\textbf{x}}^0(t) \triangleq \left[\tilde{x}_1^0(t),\tilde{x}_2^0(t)\right]^\text{T}$ and denote $\hat{\textbf{x}}_{\text{b}} = \tilde{\textbf{x}}^0(t_{\text{b}})$ as the beam direction of the process $\tilde{\boldsymbol{\psi}}^0(t)$ that is closest to the boundary of the main lobe, which is given by\footnote{\,The boundary of the set $\mathcal{B}(\textbf{x})$ is denoted by $\partial \mathcal{B}(\textbf{x})$.}
\begin{align}\label{eq:x_b}
\inf_{\textbf{v} \in \partial \mathcal{B}(x), t \ge 0} \!\left\| \textbf{v} \!-\! \tilde{\textbf{x}}^0(t) \right\|_2 \!=\! \inf_{\textbf{v} \in \partial \mathcal{B}(x)} \left\| \textbf{v} - \hat{\textbf{x}}_{\text{b}} \right\|_2  > 0.
\end{align}Then we pick $\delta$ such that
\begin{align}\label{eq:delta}
\min \!\left\{\!\inf_{\textbf{v} \in \partial \mathcal{B}(x)} \left\| \textbf{v} \!-\! \hat{\textbf{x}}_{\text{b}} \right\|_{-\infty}, \left\|\hat{\textbf{x}}_b - \textbf{x}\right\|_{-\infty} \!\right\} > \delta > 0,
\end{align}where $\left\|\textbf{u}\right\|_{-\infty} = \underset{l=1,2}{\min}{\left[\textbf{u}\right]_l}$ denotes the minimum element of $\textbf{u}$. Note that when $t \ge t_b$, the solution $\tilde{\boldsymbol{\psi}}^0(t)$ of the ODE (\ref{eq_ODE}) will approach the real equivalent channel gain $\beta$ and DPV $\textbf{x}$ monotonically as time $t$ increases.
Hence, we construct the invariant set $\mathcal{I}$ in \eqref{main_lobe_b}.
\begin{figure*}
\normalsize
\begin{small}
	\begin{equation}\label{main_lobe_b}
	\mathcal{I} = \Big( x_1 - |x_1 - \hat{x}_{1,\text{b}}| - \delta,~x_1 + |x_1 - \hat{x}_{1,\text{b}}| + \delta \Big)
	\times \Big( x_2 - |x_2 - \hat{x}_{2,\text{b}}| - \delta,~x_2 + |x_2 - \hat{x}_{2,\text{b}}| + \delta \Big) \subset \mathcal{B}(\textbf{x}).
	\end{equation}\end{small}
\hrulefill
\end{figure*}
An example of the invariant set $\mathcal{I}$ is shown in Fig. \ref{fig_invariant_set}.

Then, a sufficient condition will be established in Lemma \ref{le_sufficient} that ensures $\hat{\textbf{x}}_k\!\in\!\mathcal{I}~\text{for}~k\!\ge\!0$, and hence from Corollary 2.5 in \cite{borkar2008stochastic}, we can obtain that $\hat{\textbf{x}}_k$ converges to $\textbf{x}$.
Before giving Lemma \ref{le_sufficient}, let us provide some definitions first:
\begin{itemize}
	\item Pick $T > 0$ such that the solution $\tilde{\boldsymbol{\psi}}^0(t), t\ge 0$ of the ODE (\ref{eq_ODE}) with $\tilde{\boldsymbol{\psi}}^0(0)=\big[\hat{\beta}^\text{re}_{0}, \hat{\beta}^\text{im}_{0}, \hat{x}_{{0},1},$ $\hat{x}_{{0},2}\big]^\text{T}$ satisfies $\inf_{\textbf{v} \in \partial \mathcal{B}}\left| \textbf{v}\!-\!\tilde{\textbf{x}}^0(t) \right| \ge 2\delta$ for $t \ge T$. Since when $t \ge t_b$, $\tilde{\textbf{x}}^0(t)$ will approach the DPV $\textbf{x}$ monotonically as time $t$ increases, one possible $T$ is given by
	\begin{small}\begin{align}\label{eq_T}
	T= \arg\min\limits_{t\in[t_\text{b}, +\infty)}\left|~\!\!\left|\!\left[\int_{t_\text{b}}^t \mathbf{f}_{ \boldsymbol{\psi}}\left(\tilde{\boldsymbol{\psi}}^0(v)\right) dv\right]_3\right| - \delta\right|,
	\end{align}\end{small}where $[\cdot]_{i}$ denotes the $i$-th element of the vector.
	
	\item Let $T_0 \overset{\Delta}{=} 0$ and $T_{l+1} \overset{\Delta}{=} \min \left\{ t_i: t_i \ge T_l + T, i \ge 0 \right\}$ for $l \ge 0$. Then $T_{l+1} - T_l \in [T, T+b_{S,1}]$ and $T_l = t_{\tilde{k}(l)}$ for some $\tilde{k}(l) \uparrow +\infty$, where $\tilde{k}(0) = 0$. Let $\tilde{\boldsymbol{\psi}}^{\tilde{k}(l)}(t)$ denote the solution of ODE (\ref{eq_ODE}) for $t \in I_l \overset{\Delta}{=} \left[ T_l, T_{l+1} \right]$ with $\tilde{\boldsymbol{\psi}}^{\tilde{k}(l)}(T_l) = \bar{\boldsymbol{\psi}}(T_l)$, $l \ge 0$.
\end{itemize}
Hence, we can obtain the following lemma:
\begin{lemma}\label{le_sufficient}
	If $ \underset{t\in I_l}{\sup} \left\| \bar{\textbf{x}}(t) - \tilde{\textbf{x}}^{\tilde{k}(l)}(t)\right\|_2 \le \delta$ for all $l \ge 0$, then $\hat{\textbf{x}}_k \in \mathcal{I}~\text{for all}~k \ge 0$.
\end{lemma}
\begin{proof}
	If $\underset{t\in I_l}{\sup} \left\| \bar{\textbf{x}}(t) - \tilde{\textbf{x}}^{\tilde{k}(l)}(t)\right\|_2 \le \delta$ for all $l \ge 0$, then $\underset{t\in I_l}{\sup} \left| \bar{x}_1(t) - \tilde{{x}}_{1}^{\tilde{k}(l)}(t)\right|\le \delta$ and $\underset{t\in I_l}{\sup} \left| \bar{x}_2(t) - \tilde{{x}}_{2}^{\tilde{k}(l)}(t)\right| \le \delta$.
	
	According to Lemma 1 in \cite{JLiJoint2018}, $\hat{{x}}_{k,1} \in \mathcal{I}~\text{for all}~k \ge 0$ and $\hat{{x}}_{k,2} \in \mathcal{I}~\text{for all}~k \ge 0$. Hence, $\hat{\textbf{x}}_k \in \mathcal{I}~\text{for all}~k \ge 0$.
\end{proof}

\emph{\textbf{Step 3:} We will derive the probability lower bound for the condition in Lemma \ref{le_sufficient}, which is also a lower bound for $P\left( \left. \hat{\textbf{x}}_k\!\rightarrow\!\textbf{x} \right| \hat{\textbf{x}}_0\!\in\!\mathcal{B}\left(\textbf{x}\right) \right)$. }

We will derive the probability lower bound for the condition in Lemma \ref{le_sufficient}, which results in the following lemma:
\begin{lemma}\label{le_lower_bound}
	If (i) the initial point satisfies $\hat{\textbf{x}}_0 \in \mathcal{B}(\textbf{x})$, (ii) $b_{S,k}$ is given by (\ref{eq_stepsize}) with any $\epsilon_S > 0$,
	then there exist  $K_{S,0} \ge 0$ and $R>0$ such that
	\begin{equation}\label{eq_lock}
	\begin{aligned}
	P\left( \hat{\textbf{x}}_k \in \mathcal{I}, \forall k \ge 0 \right) \ge 1 - 8e^{-\frac{R\lvert\textbf{s}\rvert^2}{\epsilon_S^2\sigma_z^2}}.
	\end{aligned}
	\end{equation}
\end{lemma}
\begin{proof} See Appendix \ref{sec_proof_le_lower_bound}.
\end{proof}

Finally, by applying Lemma \ref{le_lower_bound} and Corollary 2.5 in \cite{borkar2008stochastic}, we can obtain
\begin{align}\label{eq_lock10}
P\left( \left. \hat{\textbf{x}}_k \rightarrow \textbf{x} \right| \hat{\textbf{x}}_0 \in \mathcal{B} \right) \ge &~ P\left( \hat{\textbf{x}}_k \in \mathcal{I}, \forall k \ge 0 \right) 
\ge  1 - 8e^{-\frac{R{\lvert\textbf{s}\rvert}^2}{\epsilon_S^2\sigma_z^2}},\nonumber
\end{align}
which completes the proof of Theorem \ref{Converge to real beam direction}.

\section{Proof of Theorem~ \ref{Converge to with minimum CRLB}}\label{proof_Converge to with minimum CRLB}

If the step-size $b_{S,k}$ is given by \eqref{eq_stepsize} with any $\varepsilon_S > 0$ and $K_{S,0} \ge 0$, the sufficient conditions are provided by Theorem 6.6.1 \cite[Section 6.6]{nevel1973stochastic} to prove the asymptotic normality of $\sqrt{k} \left( \hat{\textbf{x}}_k - \textbf{x} \right)$, i.e., $\sqrt{k} \left( \hat{\textbf{x}}_k - \textbf{x} \right) \overset{d}{\rightarrow}\mathcal{N}\left( 0, \Sigma_\textbf{x} \right)$. With the condition that $\hat{\boldsymbol{\psi}}_k \rightarrow \boldsymbol{\psi}$, we can prove that the beam and channel tracking algorithm satisfies the conditions above and obtain the variance $\boldsymbol{\Sigma}_{\textbf{x}}$ as follows:
\begin{itemize}
	
	\item[1)] Equation \eqref{eq_rtracking} is supposed to satisfy: (i) there exists an increasing sequence of $\sigma$-fields $\{\mathcal{F}_{k}: k \ge 0\}$ such that $\mathcal{F}_{l} \!\subset\!\mathcal{F}_{k}$ for $l\!<\!k$, and (ii) the random noise $\hat{\mathbf{z}}_k$ is $\mathcal{F}_{k}$-measurable and independent of $\mathcal{F}_{k-1}$.\\
	As is shown in Appendix \ref{proof_Converge to unique stable point}, there exists an increasing sequence of $\sigma$-fields $\{ \mathcal{G}_k : k \ge 0 \}$, where
	$\boldsymbol{\varsigma}_k$ is measurable with respect to $\mathcal{G}_{k-1}$ and independent of $\mathcal{G}_{k-1}$.
	
	\item[2)] $\hat{\textbf{x}}_k$ should converge to $\textbf{x}$ almost surely as $k \rightarrow +\infty$. \\
	We assume that $\hat{\boldsymbol{\psi}}_k \rightarrow \boldsymbol{\psi}$, hence $\hat{\textbf{x}}_k$ converges to $\textbf{x}$ almost surely when $k \rightarrow +\infty$.
	
	\item[3)] The stable condition:\\
	In \eqref{eq_fd}, we rewrite $\mathbf{f}_{ \boldsymbol{\psi}}\left(\hat{\boldsymbol{\psi}}_{k-1}\right)$ as follows:
	\begin{small}
	\begin{equation}
	\begin{aligned}
	\mathbf{f}_{ \boldsymbol{\psi}}\left(\!\hat{\boldsymbol{\psi}}_{k-1}\!\right)\!=\!\mathbf{D}_1 \left(\! \hat{\boldsymbol{\psi}}_{k-1} \!-\! \boldsymbol{\psi} \!\right) \!+\! \left[\!\begin{matrix} o(\| \hat{\boldsymbol{\psi}}_{k-1} - \boldsymbol{\psi} \|_2) \\ o(\| \hat{\boldsymbol{\psi}}_{k-1} - \boldsymbol{\psi} \|_2) \\ o(\| \hat{\boldsymbol{\psi}}_{k-1} - \boldsymbol{\psi} \|_2) \\ o(\| \hat{\boldsymbol{\psi}}_{k-1} - \boldsymbol{\psi} \|_2) \end{matrix}\!\right]\!,\!
	\end{aligned}
	\end{equation}\end{small}where $\mathbf{D}_1$ is given by
	\begin{equation}
	\begin{aligned}
	\!\!\!\mathbf{D}_1 \!\!=\!  \left.\frac{\partial \mathbf{f}_{ \boldsymbol{\psi}}\left(\hat{\boldsymbol{\psi}}_{k-1}\right)}{\partial \hat{\boldsymbol{\psi}}_{k-1}^\text{T}} \right|_{\hat{\boldsymbol{\psi}}_{k-1} = \boldsymbol{\psi}} \!\!\!=\! - \!\left[\!\begin{matrix} 1 & 0 & 0 & 0 \\ 0 & 1 & 0  & 0 \\ 0 & 0 & 1 & 0 \\ 0 &0 &0 &1 \end{matrix} \!\right]\!.\!
	\end{aligned}
	\end{equation}
	Then the stable condition is obtained that:
	\begin{align}
	\mathbf{E} = \mathbf{D}_1  \cdot \varepsilon_S + \frac{1}{2} =\left[\begin{smallmatrix}  \frac{1}{2}- \varepsilon_S  & 0 & 0 & 0 \\ 0 & \frac{1}{2}- \varepsilon_S& 0 & 0 \\ 0 & 0 & \frac{1}{2}- \varepsilon_S & 0\\ 0 & 0 & 0 &  \frac{1}{2}- \varepsilon_S& \end{smallmatrix} \right]  \prec 0,\nonumber
	\end{align}
	which leads to $\varepsilon_S > \frac{1}{2}$.
	
	\item[4)] The noise vector $\hat{\mathbf{z}}_k$ satisfies:\\
	\begin{equation}
	\mathbb{E}\left[\left\|\hat{\mathbf{z}}_k\right\|_2^2\right] = \operatorname{tr}\left\{\mathbf{I}_S(\hat{\boldsymbol{\psi}}_{k\!-\!1},\!\mathbf{W}_k)^{-1}\right\} < +\infty,
	\end{equation}
	and
	\begin{equation}
	\underset{v\rightarrow +\infty}{\lim}\ \ \underset{k\ge1}{\sup}\ \ \int\limits_{\left\| \hat{z}_k \right\|_2 > v} \left\| \hat{\mathbf{z}}_k \right\|_2^2 p(\hat{\mathbf{z}}_k) d\hat{\mathbf{z}}_k = 0.
	\end{equation}
	
\end{itemize}

Let
\begin{align}
\mathbf{F} =  \lim_{\begin{matrix} k \rightarrow +\infty \\ \hat{\boldsymbol{\psi}}_k \rightarrow \boldsymbol{\psi} \end{matrix}} \mathbb{E}\left[\hat{\mathbf{z}}_k \hat{\mathbf{z}}_k^\text{T}\right] 
&\overset{(a)}{=}\lim_{\begin{matrix} k \rightarrow +\infty \\ \hat{\boldsymbol{\psi}}_k \rightarrow \boldsymbol{\psi} \end{matrix}}\mathbf{I}_S(\hat{\boldsymbol{\psi}}_{k-1},\!\mathbf{W}_{k})^{-1} \\
&= \mathbf{I}_S(\boldsymbol{\psi}, \widetilde{\textbf{W}}_S^*)^{-1}, \nonumber
\end{align}
where Step $(a)$ is obtained from \eqref{eq_rzc}.

By Theorem 6.6.1 \cite[Section 6.6]{nevel1973stochastic}, we have
\begin{align*}\sqrt{k+K_{S,0}}\left( \hat{\boldsymbol{\psi}}_{k} - \boldsymbol{\psi} \right) \overset{d}{\rightarrow}\mathcal{N}\left( 0, \boldsymbol\Sigma_{\textbf{x}} \right), \end{align*}
where
\begin{equation}\label{eq_Sigma}\begin{aligned}
\boldsymbol\Sigma_{\textbf{x}} \!=\! \varepsilon_S^2  \cdot\! \!\int_0^\infty \!e^{\mathbf{E}v} \mathbf{F} e^{\mathbf{E}^\text{H}v} dv
\!=\! \frac{\varepsilon_S^2}{2\varepsilon_S \!-\!1}\mathbf{I}_S(\boldsymbol{\psi}, \widetilde{\textbf{W}}^*)^{\!-1}\!.\!
\end{aligned}\end{equation}
Due to that $\lim_{k\rightarrow +\infty}\sqrt{{(k+K_{S,0})}/{k}} = 1$, we have
\begin{align*}
\sqrt{k}\left( \hat{\boldsymbol{\psi}}_{k} \!-\! \boldsymbol{\psi} \right) \rightarrow \sqrt{k}\!\cdot\!\sqrt{\frac{k\!+\!K_{S,0}}{k}}\left( \hat{\boldsymbol{\psi}}_{k} \!-\! \boldsymbol{\psi} \right) \overset{d}{\rightarrow}\mathcal{N}\left( 0, \boldsymbol\Sigma_{\textbf{x}} \right),
\end{align*}
if $k\rightarrow +\infty$. Thus, we can get
\begin{align}
\sqrt{k}\left( \hat{\boldsymbol{\psi}}_{k} - \boldsymbol{\psi} \right) \overset{d}{\rightarrow}\mathcal{N}\left( 0, \boldsymbol\Sigma_{\textbf{x}} \right).
\end{align}
By adopting $\epsilon_S=1$ in \eqref{eq_Sigma}, we can obtain
\vspace{-1mm}
\begin{equation}\label{eq_asypsi}
\begin{aligned}
\sqrt{k}\left( \hat{\boldsymbol{\psi}}_{k} - \boldsymbol{\psi} \right) \overset{d}{\rightarrow}\mathcal{N}\left( 0, \mathbf{I}_S(\boldsymbol{\psi}, \widetilde{\textbf{W}}^*)^{-1} \right).
\end{aligned}
\end{equation}
\vspace{-0mm}Since $\hat{\boldsymbol{\psi}}_{k} \to \boldsymbol{\psi}$ as $k \to +\infty$, $\hat{\textbf{h}}_k - \textbf{h}$ is linear to $ \hat{\boldsymbol{\psi}}_{k} - \boldsymbol{\psi}$. Hence, $\hat{\textbf{h}}_k - \textbf{h}$ is also asymptotically Gaussian.

Combining \eqref{eq_fLB}, \eqref{eq_asypsi} and \eqref{eq_CMMSE}, we can conclude that
\begin{align}
\mathop {\lim }\limits_{k \to +\infty } \frac{k}{MN} \mathbb{E} \left[{\left\| \hat{\textbf{h}}_k - \textbf{h} \right\|}_2^2 \bigg| \hat{\boldsymbol{\psi}}_k \to \boldsymbol{\psi} \right] = {C}_S^{\min}(\boldsymbol{\psi}).
\end{align}
\section{Proof of Lemma~\ref{MSEOptDI}}\label{proof_MSEOptDI}
In problem \eqref{eq_problemDI}, the constraint \eqref{eq_constrant1DI} ensures that $\hat{\textbf{x}}_k$ is an unbiased estimate of ${\textbf{x}}$. According to Section 
3.7 of \cite{Sengijpta1993Fundamental}, if $\hat{\textbf{x}}$ is an unbiased estimate of $\textbf{x}$, then we can obtain that 
\vspace{-1mm}
\begin{equation}\label{eq_LBDI}
\begin{aligned}
\operatorname{Cov}\left(\hat{\textbf{x}}\right)-\textbf{I}^{-1}\left(\textbf{x}\right) \succeq \textbf{0},
\end{aligned}
\end{equation}where $\operatorname{Cov}\left(\hat{\textbf{x}}\right)$ denotes the covariance matrix of $\hat{\textbf{x}}$, $\textbf{I}\left(\textbf{x}\right)$ is the corresponding Fisher information matrix and $\textbf{A} \succeq \textbf{0}$ means that the matrix $\textbf{A}$ is nonnegative definite. From \eqref{eq_LBDI}, we can get that
\begin{equation}\label{eq_fLBDI}
\begin{aligned}
\operatorname{Cov}\left(\hat{\textbf{x}}_k\right)-\left(\sum\limits_{l=1}^k\textbf{I}_{DI}\left(\textbf{x},{{\bf{W}}_{l}}\right)\right) \succeq \textbf{0},
\end{aligned}
\end{equation}which implies that the diagonal elements of the matrix on the left side of '$\succeq$' are nonnegative because all matrices are $2 \times 2$ in \eqref{eq_fLBDI}. Therefore, we obtain that
\begin{equation}\label{eq_LBtmpDI}
\begin{aligned}
\Tr \left\{\operatorname{Cov}\left(\hat{\textbf{x}}_k\right)\right\}-\Tr \left\{\left(\sum\limits_{l=1}^k\textbf{I}\left(\textbf{x},{{\bf{W}}_{l}}\right)\right)\right\} \geq 0,
\end{aligned}
\end{equation}i.e.,
\begin{equation}\label{eq_MSELBTDI}
\begin{aligned}
\mathbb{E}\left[\left\| \hat{\textbf{x}}_k - \textbf{x} \right\|_2^2 \right]-\Tr \left\{\left(\sum\limits_{l=1}^k\textbf{I}_{DI}\left(\textbf{x},{{\bf{W}}_{l}}\right)\right)\right\} \geq 0,
\end{aligned}
\end{equation}which yields the result of \eqref{MSELBDI}.

Now we try to obtain the Fisher information matrix in \eqref{eq_fisherDI}. According to \eqref{eq_SigmaDI}, the determinant and the inverse of the covariance matrix can be written as follows:
\begin{equation}\label{eq_SigmaInverseDI}
\left\{\!
\begin{aligned}
&\!\lvert\boldsymbol{\Sigma}_{\textbf{y},k}\rvert \!=\! \sigma_z^4\left(\!\lvert\textbf{s}\rvert^2 \sigma_{\beta}^2\left(\textbf{W}_{k}^\text{H}\textbf{a}\left(\textbf{x}\right)\right)^\text{H}\textbf{W}_{k}^\text{H}\textbf{a}\left(\textbf{x}\right) \!+\!\sigma_z^2\!\right)\\
&\boldsymbol{\Sigma}_{\textbf{y},k}^{-1} \!=\!\frac{\textbf{J}_3}{\sigma_z^2}-\frac{\sigma_z^2 \lvert\textbf{s}\rvert^2 \sigma_{\beta}^2 \textbf{W}_{k}^\text{H}\textbf{a}\left(\textbf{x}\right)\left(\textbf{W}_{k}^\text{H}\textbf{a}\left(\textbf{x}\right)\right)^\text{H}}{\lvert\boldsymbol{\Sigma}_{\textbf{y},k}\rvert}
\end{aligned}\right.\!\!.\!\!
\end{equation}Based on the definition in \eqref{eq_partinfDI}, the determinant and the inverse of the covariance matrix in \eqref{eq_SigmaInverseDI} can be rewritten as
\begin{equation}\label{eq_SigmaInverserDI}
\left\{
\begin{aligned}
&\lvert\boldsymbol{\Sigma}_{\textbf{y},k}\rvert = \sigma_z^4 \left(\lvert\textbf{s}\rvert^2 \sigma_{\beta}^2 \textbf{g}_{k}^\text{H}\textbf{g}_{k}+\sigma_z^2\right)\\
&\boldsymbol{\Sigma}_{\textbf{y},k}^{-1} =\frac{\textbf{J}_3}{\sigma_z^2}-\frac{\sigma_z^2\lvert\textbf{s}\rvert^2 \sigma_{\beta}^2\textbf{g}_{k}\textbf{g}_{k}^\text{H}}{\lvert\boldsymbol{\Sigma}_{\textbf{y},k}\rvert}
\end{aligned}\right..
\end{equation}In addition, with the help of \eqref{eq_pdfDI}, we can obtain that
\begin{small}\begin{equation}\label{eq_pdfPDI}
\frac{\partial{log\,p_{DI}(\textbf{y}_k| \textbf{x}, \textbf{W}_k) }}{\partial{x_p}}\!=\!-\frac{1}{\lvert\boldsymbol{\Sigma}_{\textbf{y},k}\rvert}\frac{\partial\lvert\boldsymbol{\Sigma}_{\textbf{y},k}\rvert}{\partial{x}_p}\!-\!\textbf{y}_k^\text{H}\frac{\partial \boldsymbol{\Sigma}_{\textbf{y},k}^{-1}}{\partial{x}_p}\textbf{y}_k,
\end{equation}\end{small}where $\frac{\lvert\boldsymbol{\Sigma}_{\textbf{y},k}\rvert}{\partial{x}_p}$ and $\frac{\partial \boldsymbol{\Sigma}_{\textbf{y},k}^{-1}}{\partial{x}_p}$ are given by \eqref{eq_SigmaInverserPDI} according to \eqref{eq_SigmaInverserDI}:
\begin{equation}\label{eq_SigmaInverserPDI}
\left\{\!
\begin{aligned}
&\frac{\partial\lvert\boldsymbol{\Sigma}_{\textbf{y},k}\rvert}{\partial{x}_p} = \sigma_z^4 \lvert\textbf{s}\rvert^2 \sigma_{\beta}^2\frac{\partial \textbf{g}_{k}^\text{H}\textbf{g}_{k}}{\partial{x}_p}\\
&\frac{\partial \boldsymbol{\Sigma}_{\textbf{y},k}^{-1}}{\partial{x}_p} \!=\! -\sigma_z^2\lvert\textbf{s}\rvert^2 \sigma_{\beta}^2 \frac{\frac{\partial \textbf{g}_{k}\textbf{g}_{k}^\text{H}}{\partial{x}_p}\lvert\boldsymbol{\Sigma}_{\textbf{y},k}\rvert\!-\!\textbf{g}_{k}\textbf{g}_{k}^\text{H}\frac{\lvert\boldsymbol{\Sigma}_{\textbf{y},k}\rvert}{\partial{x}_p}}{\lvert\boldsymbol{\Sigma}_{\textbf{y},k}\rvert^2}
\end{aligned}\right..
\end{equation}

By combining \eqref{eq_SigmaDI}, \eqref{eq_pdfDI}, \eqref{eq_fisherDI} and \eqref{eq_pdfPDI}, we can obtain the $p$-th row, $j$-th column element of the Fisher information below:
\begin{small}
\begin{align}\label{eq_fisherDIt_ele}
&\left[\textbf{I}_{DI}\left(\textbf{x} ,{{\bf{W}}_{k}}\right)\right]_{p,j} \!=\! \mathbb{E}\!\left[\!\frac{\partial{log\,p_{DI}(\textbf{y}_k| \textbf{x},\! \textbf{W}_k) }}{\partial{x_p}}\frac{\partial{log\,p_{DI}(\textbf{y}_k| \textbf{x},\! \textbf{W}_k) }}{\partial{x_j}}\!\right]\nonumber\\
&=-\frac{1}{\lvert\boldsymbol{\Sigma}_{\textbf{y},k}\rvert}\frac{\partial \lvert\boldsymbol{\Sigma}_{\textbf{y},k}\rvert }{\partial x_p}\frac{\partial \lvert\boldsymbol{\Sigma}_{\textbf{y},k}\rvert }{\partial x_j}+2\lvert\textbf{s}\rvert^4 \sigma_{\beta}^4 \textbf{g}_{k}^\text{H}\frac{\partial \boldsymbol{\Sigma}_{\textbf{y},k}^{-1}}{\partial{x}_p}\textbf{g}_{k}\textbf{g}_{k}^\text{H}\frac{\partial\boldsymbol{\Sigma}_{\textbf{y},k}^{-1}}{\partial{x}_j}\textbf{g}_{k}\nonumber\\
&\quad+\sigma_z^2\lvert\textbf{s}\rvert^2 \sigma_{\beta}^2\textbf{g}_{k}^\text{H}\frac{\partial \boldsymbol{\Sigma}_{\textbf{y},k}^{-1}}{\partial{x}_p}\textbf{g}_{k}\Tr\left\{\frac{\partial \boldsymbol{\Sigma}_{\textbf{y},k}^{-1}}{\partial{x}_j}\right\}\nonumber\\
&\quad+\sigma_z^2\lvert\textbf{s}\rvert^2 \sigma_{\beta}^2\textbf{g}_{k}^\text{H}\frac{\partial \boldsymbol{\Sigma}_{\textbf{y},k}^{-1}}{\partial{x}_j}\textbf{g}_{k}\Tr\left\{\frac{\partial \boldsymbol{\Sigma}_{\textbf{y},k}^{-1}}{\partial{x}_p}\right\}\\
&\quad+\sigma_z^4 \Tr\left\{\frac{\partial \boldsymbol{\Sigma}_{\textbf{y},k}^{-1}}{\partial{x}_p}\right\}\Tr\left\{\frac{\partial \boldsymbol{\Sigma}_{\textbf{y},k}^{-1}}{\partial{x}_j}\right\}+\sigma_z^4 \Tr\left\{\frac{\partial \boldsymbol{\Sigma}_{\textbf{y},k}^{-1}}{\partial{x}_p}\frac{\partial \boldsymbol{\Sigma}_{\textbf{y},k}^{-1}}{\partial{x}_j}\right\}\nonumber\\
&\quad+\sigma_z^2 \lvert\textbf{s}\rvert^2 \sigma_{\beta}^2\textbf{g}_{k}^\text{H} \left(\frac{\partial \boldsymbol{\Sigma}_{\textbf{y},k}^{-1}}{\partial{x}_p}\frac{\partial \boldsymbol{\Sigma}_{\textbf{y},k}^{-1}}{\partial{x}_j}+\frac{\partial \boldsymbol{\Sigma}_{\textbf{y},k}^{-1}}{\partial{x}_j}\frac{\partial \boldsymbol{\Sigma}_{\textbf{y},k}^{-1}}{\partial{x}_p}\right) \textbf{g}_{k}\nonumber.
\end{align}\end{small}Then we substitute \eqref{eq_SigmaDI}, \eqref{eq_SigmaD}, \eqref{eq_partinfDI}, \eqref{eq_SigmaInverserDI}, \eqref{eq_SigmaInverserPDI} into \eqref{eq_fisherDIt_ele}, which yields the result of \eqref{eq_fisherDI_ele}.

Finally, the proof of Lemma~\ref{MSEOptDI} is completed.
\section{Proof of Lemma~\ref{UnifiedOptShiftDI}}\label{proof_UnifiedOptShiftDI}
The proof of property 1) in Lemma~\ref{UnifiedOptShiftDI} is similar to the proof of that in Lemma~\ref{UnifiedOptShift}. Hence, we focus on the proof of
property 2) and property 3) in Lemma~\ref{UnifiedOptShiftDI}.

Consider the $p$-th row, $j$-th column element of the Fisher information matrix $\textbf{I}_{DI}\left(\textbf{x} ,{{\bf{W}}_{k}}\right)$ in \eqref{eq_fisherDI_ele}. We can rewrite it in \eqref{eq_fisherDIr_ele},
\begin{figure*}
\normalsize
\begin{small}
	\begin{align}\label{eq_fisherDIr_ele}
	\left[\textbf{I}_{DI}\left(\textbf{x} ,{{\bf{W}}_{k}}\right)\right]_{p,j} &=\frac {\sigma_z^6\lvert\textbf{s}\rvert^6 \sigma_{\beta}^6} {{\lvert\boldsymbol{\Sigma}_{\textbf{y},k}\rvert}^2}\left\{
	-2 \lvert\textbf{g}_k\rvert^2 \tilde{{g}}_{k,p}\tilde{{g}}_{k,j}+\frac{\sigma_z^2}{\lvert \textbf{s}\rvert^2 \sigma_{\beta}^2}\Tr\left\{\textbf{G}_{k,p}\textbf{G}_{k,j}\right\}+ \textbf{g}_{k}^\text{H}\left(\textbf{G}_{k,p}\textbf{G}_{k,j}+\textbf{G}_{k,j}\textbf{G}_{k,p}\right)\textbf{g}_{k}
	\right\}\nonumber\\
	&\overset{(a)}{=}\frac {\sigma_z^6\lvert\textbf{s}\rvert^6 \sigma_{\beta}^6} {\sigma_z^8 \left(\lvert\textbf{s}\rvert^2 \sigma_{\beta}^2 \textbf{g}_{k}^\text{H}\textbf{g}_{k}+\sigma_z^2\right)^2}\left\{
	-2  \lvert\textbf{g}_k\rvert^2 \tilde{{g}}_{k,p}\tilde{{g}}_{k,j}+\frac{\sigma_z^2}{\lvert\textbf{s}\rvert^2 \sigma_{\beta}^2}\Tr\left\{\textbf{G}_{k,p}\textbf{G}_{k,j}\right\}+ \textbf{g}_{k}^\text{H}\left(\textbf{G}_{k,p}\textbf{G}_{k,j}+\textbf{G}_{k,j}\textbf{G}_{k,p}\right)\textbf{g}_{k}
	\right\}\\
	&=\frac {\lvert\textbf{s}\rvert^2 \sigma_{\beta}^2} {\sigma_z^2\left( \textbf{g}_{k}^\text{H}\textbf{g}_{k}+\frac{\sigma_z^2}{\lvert\textbf{s}\rvert^2 \sigma_{\beta}^2}\right)^2}\left\{
	-2  \lvert\textbf{g}_k\rvert^2 \tilde{{g}}_{k,p}\tilde{{g}}_{k,j}+\frac{\sigma_z^2}{\lvert\textbf{s}\rvert^2 \sigma_{\beta}^2}\Tr\left\{\textbf{G}_{k,p}\textbf{G}_{k,j}\right\}+ \textbf{g}_{k}^\text{H}\left(\textbf{G}_{k,p}\textbf{G}_{k,j}+\textbf{G}_{k,j}\textbf{G}_{k,p}\right)\textbf{g}_{k}
	\right\},\nonumber
	\end{align}\end{small}
     \hrulefill
     \end{figure*}where Step $(a)$ is obtained by substituting \eqref{eq_SigmaInverseDI} into \eqref{eq_fisherDIr_ele}. When $\frac{\lvert\textbf{s}\rvert^2  \sigma_{\beta} ^2}{\sigma_z^2} \to +\infty$, we can obtain the element of $\textbf{I}_{DI}\left(\textbf{x} ,{{\bf{W}}_{k}}\right)$ in \eqref{eq_fisherDIrl_ele},
 \begin{figure*}
 	\normalsize\begin{align}\label{eq_fisherDIrl_ele}
	\mathop {\lim }\limits_{\frac{\lvert\textbf{s}\rvert^2  \sigma_{\beta} ^2}{\sigma_z^2} \to  + \infty }
	\frac{\sigma_z^2}{\lvert\textbf{s}\rvert^2 \sigma_{\beta}^2}\left[\textbf{I}_{DI}\left(\textbf{x} ,{{\bf{W}}_{k}}\right)\right]_{p,j} =\frac {1} {\left( \textbf{g}_{k}^\text{H}\textbf{g}_{k}\right)^2}\left\{
	-2  \lvert\textbf{g}_k\rvert^2 \tilde{{g}}_{k,p}\tilde{{g}}_{k,j}+ \textbf{g}_{k}^\text{H}\left(\textbf{G}_{k,p}\textbf{G}_{k,j}+\textbf{G}_{k,j}\textbf{G}_{k,p}\right)\textbf{g}_{k}
	\right\},
	\end{align}\end{figure*}which reveals that $
\frac{\sigma_z^2}{\lvert\textbf{s}\rvert^2 \sigma_{\beta}^2}\textbf{I}_{DI}\left(\textbf{x} ,{{\bf{W}}_{k}}\right)$ converges as $\frac{\lvert\textbf{s}\rvert^2 \sigma_{\beta} ^2}{\sigma_z^2} \to +\infty$. Then the property 2) of Lemma \ref{UnifiedOptShiftDI} is proved.

Let us see the property 3) in Lemma \ref{UnifiedOptShiftDI}. Similar to Step 3 in Appendix \ref{proof_UnifiedOptShift}, we can obtain that ${C}_{DI}^{\min}(\boldsymbol{\psi})$ converge as $M,N \to +\infty$ and 
	\vspace{-0mm}
\begin{align}\label{eq_asymtDI} {\lim\limits_{M,N \to +\infty}}{C}_{DI}(\boldsymbol{\psi},\widetilde{\textbf{W}}_{DI}^*) = {\lim\limits_{M,N \to +\infty}}{C}_{DI}^{\min}(\boldsymbol{\psi}).
\vspace{-0mm}
\end{align}According to \eqref{eq_partinfDI} and \eqref{eq_walpl}, $\textbf{g}_k$ is $\Theta\left(\sqrt{MN}\right)$ while $\tilde{{g}}_{k,p}$ and $\textbf{G}_{k,p}$ are $\Theta\left(MN\right)$. Hence, $\sigma_z^2\Tr\left\{\textbf{G}_{k,p}\textbf{G}_{k,j}\right\}$ can be omitted since it is $\Theta\left({\left(MN\right)}^2\right)$ while other parts are $\Theta\left(\left(MN\right)^{\frac{5}{2}}\right)$. Then the $p$-th row, $j$-th column element of the Fisher information matrix in \eqref{eq_fisherDIr_ele} can be rewritten in \eqref{eq_fisherDIr2_ele},
\begin{figure*}
	\normalsize
	\begin{align}\label{eq_fisherDIr2_ele}
	\mathop {\lim }\limits_{M,N \to  + \infty }\frac{\left[\textbf{I}_{DI}\left(\textbf{x} ,{{\bf{W}}_{k}}\right)\right]_{p,j}}{\left(MN\right)^{5/2}} &=\frac {\sigma_z^6\lvert \textbf{s} \rvert^4 \sigma_{\beta}^4} {{\lvert\boldsymbol{\Sigma}_{\textbf{y},k}\rvert}^2}\left\{
	-2 \lvert \textbf{s} \rvert^2 \sigma_{\beta}^2 \frac{\lvert\textbf{g}_k\rvert^2 \tilde{{g}}_{k,p}\tilde{{g}}_{k,j}}{\left(MN\right)^{5/2}} + \lvert \textbf{s} \rvert^2 \sigma_{\beta}^2\frac{ \textbf{g}_{k}^\text{H}\left(\textbf{G}_{k,p}\textbf{G}_{k,j}+\textbf{G}_{k,j}\textbf{G}_{k,p}\right)\textbf{g}_{k}}{\left(MN\right)^{5/2}}
	\right\}\nonumber\\
	&=\frac {\sigma_z^6\lvert \textbf{s} \rvert^6 \sigma_{\beta}^6} {{\lvert\boldsymbol{\Sigma}_{\textbf{y},k}\rvert}^2}\left\{
	-2  \frac{\lvert\textbf{g}_k\rvert^2 \tilde{{g}}_{k,p}\tilde{{g}}_{k,j}}{\left(MN\right)^{5/2}} + \frac{ \textbf{g}_{k}^\text{H}\left(\textbf{G}_{k,p}\textbf{G}_{k,j}+\textbf{G}_{k,j}\textbf{G}_{k,p}\right)\textbf{g}_{k}}{\left(MN\right)^{5/2}}
	\right\},
	\end{align}\end{figure*}which reveals that $\left\{\widetilde{\boldsymbol{\Delta}}_{\begin{small}DI\end{small},1}^{*},\,\widetilde{\boldsymbol{\Delta}}_{\begin{small}DI\end{small},2}^{*},\widetilde{\boldsymbol{\Delta}}_{\begin{small}DI\end{small},3}^{*}\right\}$ is unrelated to $\frac{\lvert \textbf{s} \rvert^2  \sigma_{\beta} ^2}{\sigma_z^2}$.

Finally, the proof is completed.

\section{Proof of Lemma~\ref{lemma_complexity}}\label{proof_complexity}
We first analyze the computational complexity of Algorithm \ref{alg_Static}, which is composed of three steps:

\emph{\textbf{Step 1}: We evaluate the computational arithmetic operations of the Fisher information matrix inversion.}

The Fisher information matrix is obtained as follows:
\begin{equation}\label{eq_fisherCom}
\textbf{I}_S\left(\hat{\boldsymbol{\psi}}_{k-1}, \textbf{W}_k\right)=\frac {2 {\lvert \textbf{s} \rvert}^2} {{\sigma}_z^2}  \text{Re}\left\{\textbf{V}_k^\text{H} \textbf{W}_{k} \textbf{W}_{k}^\text{H} \textbf{V}_k \right\},
\end{equation}where $\textbf{V}_k$ is given by
\begin{align}\label{eq_Vk}
\textbf{V}_k&\!=\!\left[\!\textbf{a}\left(\hat{\textbf{x}}_{k-\!1}\right)\!,j\textbf{a}\left(\hat{\textbf{x}}_{k\!-\!1}\right)\!,\hat{\beta}_{k-1} \frac {\partial \textbf{a}\left(\hat{\textbf{x}}_{k-1}\right)}{\partial x_1}\!,\hat{\beta}_{k\!-\!1} \frac {\partial \textbf{a}\left(\hat{\textbf{x}}_{k\!-\!1}\right)}{\partial x_2}\!\right]\nonumber\\
&\overset{(a)}{=}\left[\textbf{V}_k^1,\hat{\beta}_{k-1}\textbf{V}_k^2\right]
\end{align}with Step (a) resulting from the definition of $\textbf{V}_k^1$ and $\textbf{V}_k^2$:
\begin{align}\label{eq_Vk12}
\textbf{V}_k^1 &\triangleq \left[\textbf{a}\left(\hat{\textbf{x}}_{k-1}\right),j\textbf{a}\left(\hat{\textbf{x}}_{k-1}\right)\right]\\
\textbf{V}_k^2 &\triangleq \left[\frac {\partial \textbf{a}\left(\hat{\textbf{x}}_{k-1}\right)}{\partial x_1}, \frac {\partial \textbf{a}\left(\hat{\textbf{x}}_{k-1}\right)}{\partial x_2}\right].
\end{align}

By combining 
\eqref{eq_BF}, \eqref{eq_wa} and \eqref{eq_wpa}, we can obtain that $\textbf{W}_k^\text{H}\textbf{V}_k^1$ and $\textbf{W}_k^\text{H}\textbf{V}_k^2$ are determined matrices that remain unchanged for different ECCs, given by
\begin{align}\label{eq_U}
\textbf{U}_1 &= \textbf{W}_k^\text{H}\textbf{V}_k^1\\
\textbf{U}_2 &= \textbf{W}_k^\text{H}\textbf{V}_k^2,
\end{align}where both $\textbf{U}_1$ and $\textbf{U}_2$ can be obtained by offline calculation. Hence, we can rewrite the Fisher information matrix in \eqref{eq_fisherCom} as:
\begin{small}\begin{align}\label{eq_FisherBlocksCom}
	\textbf{I}_S\left(\!\hat{\boldsymbol{\psi}}_{k\!-\!1},\!\textbf{W}_k\!\right)&\!=\! \frac{2\lvert\textbf{s}\rvert^2}{{{\sigma_z ^2}}}\!\left[\!\!\!{\begin{array}{*{20}{c}}
		\text{Re}\left\{\textbf{U}_1^\text{H}  \textbf{U}_1\right\} &\!\!\! \text{Re}\left\{\widetilde{\beta}_{k\!-\!1}\textbf{U}_1^\text{H}  \textbf{U}_2\right\}\\
		\text{Re}\left\{\bar{\beta}_{k\!-\!1}\textbf{U}_2^\text{H}\textbf{U}_1\right\} &\!\!\! \lvert\widetilde{\beta}_{k\!-\!1}\rvert^2\text{Re}\!\left\{\!\textbf{U}_2^\text{H} \textbf{U}_2\!\right\}\\
		\end{array}} \!\!\!\!\right]\nonumber\\
	&\!=\! \frac{{2{{\lvert\textbf{s}\rvert}^2}}}{{{\sigma_z ^2}}}\!\!\left[\!\!\!{\begin{array}{*{20}{c}}
		{{\widetilde{\textbf{A}}}}&\!\!\!\text{Re}\left\{\widetilde{\beta}_{k-1} \widetilde{\textbf{B}}\right\}\\
		\text{Re}\!\left\{\!\bar{\beta}_{k\!-\!1} \widetilde{\textbf{B}}^\text{H}\right\}&\!\!\!\lvert \widetilde{\beta}_{k\!-\!1}\rvert^2{\widetilde{\textbf{D}}}
		\end{array}} \!\!\!\right]\!,\!
	\end{align}\end{small}where $\bar \beta_{k-1}$ denotes the conjugate of $\hat{\beta}_{k-1}$ and $\widetilde{\textbf{A}}$, $\widetilde{\textbf{B}}$, $\widetilde{\textbf{D}}$ are defined as:
\begin{align}\label{eq_BlockCom}
\left\{\begin{array}{*{20}{l}}
{\widetilde{\textbf{A}}} \triangleq \text{Re}\left\{\textbf{U}_1^\text{H} \textbf{U}_1\right\} \\
{\widetilde{\bf{B}}} \triangleq \textbf{U}_1^\text{H} \textbf{U}_2. \\
\widetilde{\textbf{D}} \triangleq \text{Re}\left\{\textbf{U}_2^\text{H} \textbf{U}_2\right\}
\end{array}\right.\!
\end{align}Note that $\widetilde{\textbf{A}}$ is a diagonal matrix with the same diagonal elements and  the block matrices $\widetilde{\textbf{A}}$, $\widetilde{\textbf{B}}$, $\widetilde{\textbf{D}}$ can all be obtained by offline calculation.

Similar to the derivation in \eqref{eq_InverseFisherBlocks}, the inverse of the Fisher information matrix in \eqref{eq_FisherBlocksCom} can be calculated by using the block matrix inversion method, given by
\vspace{-0mm}
\begin{equation}\label{eq_InverseFisherBlocksCom}
\begin{aligned}
\textbf{I}_S\left(\hat{\boldsymbol{\psi}}_{k-1}, \textbf{W}_k\right)^{-1}= \frac{{{\sigma_z ^2}}}{{2{{\lvert\textbf{s}\rvert}^2}}}\left\{ {{{\widetilde{\bf{I}}}_{i{p_1}}} + {{\widetilde{\bf{I}}}_{i{p_2}}}\left(\hat{\beta}_{k-1} \right)} \right\},
\end{aligned}
\end{equation}where ${{\widetilde{\textbf{I}}}_{i{p_1}}}$ and ${{\widetilde{\textbf{I}}}_{i{p_2}}}\left(\hat{\beta}_{k-1}\right)$ are defined in \eqref{eq_Ip1Com} and \eqref{eq_Ip2Com}
\begin{equation}\label{eq_Ip1Com}
\begin{aligned}
{{\widetilde{\textbf{{I}}}}_{i{p_1}}} \triangleq \left[ \begin{matrix}
{{\widetilde{\bf{A}}}^{-1}}&{\bf{0}}\\
{\bf{0}}&{\bf{0}}\end{matrix} \right],
\end{aligned}
\end{equation}
\begin{figure*}
	\normalsize
	\begin{equation}\label{eq_Ip2Com}
	{{\widetilde{\bf{I}}}_{i{p_2}}}\left(\hat{\beta}_{k-1}\right) \triangleq\left[ {\begin{matrix}
		{{\widetilde{\textbf{A}}}{^{-1}}\text{Re}\left\{\hat{\beta}_{k-1} \widetilde{\textbf{B}}\right\}}\\
		{{\bf{ - }}{{\bf{J}}_2}}
		\end{matrix}} \right]{{\left( \lvert \hat{\beta}_{k-1} \rvert^2 \widetilde{\textbf{I}}_s\right)}^{-1}}\left[ {\begin{matrix}
		{\text{Re}\left\{\bar{{\beta}}_{k-1} \widetilde{\textbf{B}}^\text{H}\right\}{\widetilde{\bf{A}}}{^{-1}}}&-{{\textbf{J}}_2}
		\end{matrix}} \right]
	\end{equation}
\hrulefill\end{figure*}with $\widetilde{\textbf{I}}_s$ defined as follows:
\begin{equation}
\widetilde{\textbf{I}}_s\triangleq  {\widetilde{\textbf{D}}}-\frac{ \text{Re}\left\{\widetilde{\textbf{B}}^\text{H}\widetilde{\textbf{A}}^{-1} \widetilde{\textbf{B}}\right\}}{2}.
\end{equation}Since ${\widetilde{\textbf{A}}}{^{-1}}$ in \eqref{eq_Ip1Com} can be obtained by offline calculation, ${{\widetilde{\textbf{{I}}}}_{i{p_1}}}$ requires none online complex arithmetic operations.

As for ${{\widetilde{\textbf{{I}}}}_{i{p_2}}}$, we can further rewrite it as a block matrix:
\begin{equation}
\begin{aligned}
{{\widetilde{\bf{I}}}_{i{p_2}}}\left(\hat{\beta}_{k-1}\right) = \left[ \begin{matrix}
{{\widetilde{\bf{I}}}_{i{p_2}}}^{11}\left(\hat{\beta}_{k-1}\right)&{{\widetilde{\bf{I}}}_{i{p_2}}}^{12}\left(\hat{\beta}_{k-1}\right)\\
{{\widetilde{\bf{I}}}_{i{p_2}}}^{21}\left(\hat{\beta}_{k-1}\right)&{{\widetilde{\bf{I}}}_{i{p_2}}}^{22}\left(\hat{\beta}_{k-1}\right)\end{matrix} \right],
\end{aligned}
\end{equation}where the four block matrices are given by \eqref{eq_ip2BlockCom}.
\begin{figure*}
	\normalsize
	\begin{align}\label{eq_ip2BlockCom}
	\left\{\begin{array}{*{20}{l}}
	{{\widetilde{\bf{I}}}_{i{p_2}}}^{11}\left(\hat{\beta}_{k-1}\right) = {{\widetilde{\textbf{A}}}{^{-1}}\text{Re}\left\{\hat{\beta}_{k-1} \widetilde{\textbf{B}}\right\}}{{\left( \lvert \hat{\beta}_{k-1} \rvert^2 \widetilde{\textbf{I}}_s\right)}^{-1}}{\text{Re}\left\{\bar{{\beta}}_{k-1} \widetilde{\textbf{B}}^\text{H}\right\}{\widetilde{\bf{A}}}{^{-1}}} \\
	{{\widetilde{\bf{I}}}_{i{p_2}}}^{12}\left(\hat{\beta}_{k-1}\right) = -{\text{Re}\left\{\hat{\beta}_{k-1}\lvert \hat{\beta}_{k-1} \rvert^{-2} {\widetilde{\textbf{A}}}{^{-1}}\widetilde{\textbf{B}}\widetilde{\textbf{I}}_s^{-1}\right\}}\\
	{{\widetilde{\bf{I}}}_{i{p_2}}}^{21}\left(\hat{\beta}_{k-1}\right) = -{\text{Re}\left\{\bar{{\beta}}_{k-1}\lvert \hat{\beta}_{k-1} \rvert^{-2}\widetilde{\textbf{I}}_s^{-1} \widetilde{\textbf{B}}^\text{H}{\widetilde{\bf{A}}}{^{-1}}\right\}}=\left({{\widetilde{\bf{I}}}_{i{p_2}}}^{12}\left(\hat{\beta}_{k-1}\right)\right)^{\text{H}}\\
	{{\widetilde{\bf{I}}}_{i{p_2}}}^{22}\left(\hat{\beta}_{k-1}\right) = {\left( \lvert \hat{\beta}_{k-1} \rvert^{-2}\right)}\widetilde{\textbf{I}}_s^{-1}
	\end{array}\right..\!
	\end{align}\end{figure*}

Since $\widetilde{\textbf{A}}$, $\widetilde{\textbf{B}}$, $\widetilde{\textbf{D}},\widetilde{\textbf{I}}_s$ can all be obtained by offline calculation, ${{\widetilde{\bf{I}}}_{i{p_2}}}^{12}\left(\hat{\beta}_{k-1}\right)$ and ${{\widetilde{\bf{I}}}_{i{p_2}}}^{22}\left(\hat{\beta}_{k-1}\right)
$ only require 6 online complex arithmetic operations. In addition, ${{\widetilde{\bf{I}}}_{i{p_2}}}^{21}\left(\hat{\beta}_{k-1}\right)$ requires none online complex arithmetic operations as it can be obtained directly from ${{\widetilde{\bf{I}}}_{i{p_2}}}^{12}\left(\hat{\beta}_{k-1}\right)$. As for ${{\widetilde{\bf{I}}}_{i{p_2}}}^{11}\left(\hat{\beta}_{k-1}\right)$, we can convert it to \eqref{eq_Iip211r}
\begin{figure*}
	\normalsize
	\begin{equation}\label{eq_Iip211r}
	\begin{aligned}
	{{\widetilde{\bf{I}}}_{i{p_2}}}^{11}\left(\hat{\beta}_{k-1}\right) &\triangleq {{\widetilde{\textbf{A}}}{^{-1}}\text{Re}\left\{\hat{\beta}_{k-1} \widetilde{\textbf{B}}\right\}}{{\left( \lvert \hat{\beta}_{k-1} \rvert^2 \widetilde{\textbf{I}}_s\right)}^{-1}}{\text{Re}\left\{\bar{{\beta}}_{k-1} \widetilde{\textbf{B}}^\text{H}\right\}{\widetilde{\bf{A}}}{^{-1}}}\\
	&=\widetilde{\textbf{A}}^{-1}
	\left\{\frac{\hat{\beta}_{k-1} \widetilde{\textbf{B}}+\bar\beta_{k-1} \bar{\textbf{B}}}{2}{{\left( \lvert \hat{\beta}_{k-1} \rvert^2 \widetilde{\textbf{I}}_s\right)}^{-1}}\frac{\bar{\beta}_{k-1} \widetilde{\textbf{B}}^\text{H}+\hat{\beta}_{k-1} \widetilde{\textbf{B}}^\text{T}}{2}\right\}
	\widetilde{\bf{A}}^{-1}\\
	&=\widetilde{\textbf{A}}^{-1}
	\frac{\text{Re}\left\{\widetilde{\textbf{B}}\widetilde{\textbf{I}}_s^{-1}\widetilde{\textbf{B}}^\text{H}\right\}}{2}\widetilde{\bf{A}}^{-1}.
	\end{aligned}
	\end{equation}\hrulefill\end{figure*}. Finally, ${{\widetilde{\bf{I}}}_{i{p_2}}}^{11}\left(\hat{\beta}_{k-1}\right)$ requires none online complex arithmetic operations. Hence, the calculation of ${{\widetilde{\bf{I}}}_{i{p_2}}}\left(\hat{\beta}_{k-1}\right)$ in  \eqref{eq_ip2BlockCom} requires none online complex arithmetic operations for ${{\widetilde{\bf{I}}}_{i{p_2}}}^{11}\left(\hat{\beta}_{k-1}\right)$, 6 online complex arithmetic operations for ${{\widetilde{\bf{I}}}_{i{p_2}}}^{12}$, none online complex arithmetic operations for ${{\widetilde{\bf{I}}}_{i{p_2}}}^{21}\left(\hat{\beta}_{k-1}\right)$, and 5 online complex arithmetic operations for ${{\widetilde{\bf{I}}}_{i{p_2}}}^{11}\left(\hat{\beta}_{k-1}\right)$, which are 11 complex arithmetic operations in total.

In the end, the calculation of $\textbf{I}_S\left(\hat{\boldsymbol{\psi}}_{k-1}, \textbf{W}_k\right)^{-1}$ in  \eqref{eq_InverseFisherBlocksCom} requires 11 online complex arithmetic operations.

\emph{\textbf{Step 2}: We evaluate the computational arithmetic operations of $\frac {\partial \text{log} \, p_S \left(\textbf{y}_k |\boldsymbol{\psi},\textbf{W}_k \right)}{\partial \boldsymbol{\psi}}\bigg|_{\boldsymbol{\psi}=\hat{\boldsymbol{\psi}}_{k-1}}$}.

We write the $\frac {\partial \text{log} \, p_S \left(\textbf{y}_k |\boldsymbol{\psi},\textbf{W}_k \right)}{\partial \boldsymbol{\psi}}\bigg|_{\boldsymbol{\psi}=\hat{\boldsymbol{\psi}}_{k-1}}$ as follows:
\
\begin{equation}\label{eq_pCom}
\frac {\partial \text{log} \!\, p_S \left(\!\textbf{y}_k |\boldsymbol{\psi},\!\textbf{W}_k \!\right)}{\partial \boldsymbol{\psi}}\!\bigg|_{\boldsymbol{\psi}\!=\!\hat{\boldsymbol{\psi}}_{k-1}}\!=\!\left[\!
\begin{matrix}
{\text{Re}\left\{ \textbf{e}_k^\text{H}\left(\textbf{y}_k\!-\!\hat{\textbf{y}}_k \right)\right\}}\\
{\text{Im}\left\{ \textbf{e}_k^\text{H}\left(\textbf{y}_k\!-\!\hat{\textbf{y}}_k \right)\right\}}\\
{\text{Re}\left\{ \tilde{\textbf{e}}_{k1}^\text{H}\left(\textbf{y}_k\!-\!\hat{\textbf{y}}_k \right)\right\}}\\
{\text{Re}\left\{ \tilde{\textbf{e}}_{k2}^\text{H}\left(\textbf{y}_k\!-\!\hat{\textbf{y}}_k \right)\right\}}
\end{matrix}
\!\!\right]\!,\!
\end{equation}where ${\textbf{e}}_k = \textbf{W}_k^\text{H} \textbf{a}\left(\hat{\textbf{x}}_{k-1}\right)$, $\hat{\textbf{y}}_k = \lvert \textbf{s} \rvert \hat{\beta}_{k-1}\textbf{W}_k^\text{H} \textbf{a}\left(\hat{\textbf{x}}_{k-1}\right)$, $\tilde{\textbf{e}}_{k1} = \hat{\beta}_{k-1} \textbf{W}_k^\text{H} \frac{\partial \textbf{a}\left(\hat{\textbf{x}}_{k-1}\right)}{\partial x_1}$, $\tilde{\textbf{e}}_{k2} = \hat{\beta}_{k-1} \textbf{W}_k^\text{H} \frac{\partial \textbf{a}\left(\hat{\textbf{x}}_{k-1}\right)}{\partial x_2}$. Since $\textbf{W}_k^\text{H} \textbf{a}\left(\hat{\textbf{x}}_{k-1}\right),\,\textbf{W}_k^\text{H} \frac{\partial \textbf{a}\left(\hat{\textbf{x}}_{k-1}\right)}{\partial x_1},\,\textbf{W}_k^\text{H} \frac{\partial \textbf{a}\left(\hat{\textbf{x}}_{k-1}\right)}{\partial x_2}$ can all be obtained by offline calculation, $\hat{\textbf{y}}_k$ requires $3$ online complex arithmetic operations, ${\textbf{y}}_k-\hat{\textbf{y}}_k$ requires none complex arithmetic operations and $\tilde{\textbf{e}}_{k1},\,\tilde{\textbf{e}}_{k2}$ both require 3 complex arithmetic operations. Together with the inner-product calculation in \eqref{eq_pCom}, the final number of online complex arithmetic operations is $18$.

\emph{\textbf{Step 3}: We evaluate the total computational arithmetic operations.}

Considering the multiplication of $\textbf{I}_S\left(\hat{\boldsymbol{\psi}}_{k-1}, \textbf{W}_k\right)^{-1}$ and $\frac {\partial \text{log} \, p_S \left(\textbf{y}_k |\boldsymbol{\psi},\textbf{W}_k \right)}{\partial \boldsymbol{\psi}}\bigg|_{\boldsymbol{\psi}=\hat{\boldsymbol{\psi}}_{k-1}}$ (16 complex arithmetic operations), and the updating direction vector plus the previous estimate (none complex arithmetic operation), the final number of online complex arithmetic operations is 45 in each ECC.

By using a similar method,  the total number of complex computational arithmetic operations for Algorithm \ref{alg_DI} (Algorithm \ref{alg_DII}) is $28$ ($45$) in each ECC.

Therefore, Lemma~\ref{lemma_complexity} gets proved.

\section{Proof of Lemma \ref{le_lower_bound}}\label{sec_proof_le_lower_bound}
The following lemmas are introduced to prove Lemma \ref{le_lower_bound}.

\begin{lemma}[Lemma 3 \cite{JLiJoint2018}]\label{le_gronwall}
	Given $T$ by \eqref{eq_T} and
	\begin{align}\label{eq_nT}
	k_T \overset{\Delta}{=} \inf \left\{i \in \mathbb{Z}: t_{k+i} \ge t_k + T \right\}.
	\end{align}
	If there exists a constant $C>0$, which satisfies
	\begin{equation}\label{eq_gronwall1}
	\begin{aligned}
	&\left\| \bar{\boldsymbol{\psi}}(t_{k+l}) - \tilde{\boldsymbol{\psi}}^k(t_{k+l})\right\|_2 \\
	\le& L \sum_{i=1}^{l} b_{S,k+i} \left\| \bar{\boldsymbol{\psi}}(t_{k+i-1}) - \tilde{\boldsymbol{\psi}}^k(t_{k+i-1}) \right\|_2  + C,
	\end{aligned}
	\end{equation}
	for all $k \ge 0$ and $1 \le l \le k_T$, then
	\begin{small}
	\begin{equation}\label{eq_gronwall2}
	\begin{aligned}
	\underset{t\in\left[ t_k, t_{k+k_T} \right]}{\sup} \!\left\| \bar{\boldsymbol{\psi}}(t) \!-\! \tilde{\boldsymbol{\psi}}^k(t)\!\right\|_2 \!\le\!  \frac{C_{\mathbf{f}} b_{S,k+1}}{2} \!+\! C e^{L (T+b_{S,1})}\!,\!
	\end{aligned}
	\end{equation}\end{small}where $L$ and $C_{\mathbf{f}}$ are defined in \eqref{eq_Lip} and \eqref{eq_CT} separately.
\end{lemma}

\begin{lemma}[Lemma 4 \cite{JLiSuper2017}]\label{le_cheb}
	If $\{M_i: i = 1, 2, \ldots\}$ satisfies that: (i)  $M_i$ is Gaussian distributed with zero mean, and (ii) $M_i$ is a martingale in $i$, then
	\begin{equation}\label{eq_lock5}
	\begin{aligned}
	& P\left( \underset{0\le i \le k}{\sup}\left|M_i\right| > \eta \right) \le 2\exp\left\{-\frac{\eta^2}{2\operatorname{Var}\left[M_k\right]}\right\},
	\end{aligned}
	\end{equation}
	for any $\eta > 0$.
\end{lemma}

\begin{lemma}[Lemma 5 \cite{JLiSuper2017}]\label{le_sum}
	If given a constant $C > 0$, then
	\begin{equation}\label{eq_increasing}
	\begin{aligned}
	G(v) = \frac{1}{v}\exp\left[-\frac{C}{v}\right],
	\end{aligned}
	\end{equation}
	is increasing for all $0 < v < C$.
\end{lemma}
Let $\boldsymbol{\xi}_0 \overset{\Delta}{=} \mathbf{0}$ and $\boldsymbol{\xi}_k \overset{\Delta}{=} \sum_{l=1}^{k} b_{S,l} \mathbf{\hat{z}}_{l} $, $k \ge 1$, where $\mathbf{\hat{z}}_{l}$ is given in \eqref{eq_z}. With \eqref{eq_continuous} and \eqref{eq_ODE_new}, we have for $t_{k+l}, 1 \le l \le k_T$,
\begin{small}
\begin{align}\label{eq_seq_trace} \bar{\boldsymbol{\psi}}(t_{k+l}) \!=\! \bar{\boldsymbol{\psi}}(t_k)  \!+\!\! \sum_{i=1}^{l} \!b_{S,k+i} \mathbf{f}_{ \boldsymbol{\psi}}\left(\bar{\boldsymbol{\psi}}(t_{n+i\!-\!1})\right)\!+\! (\boldsymbol{\xi}_{k+l} \!-\! \boldsymbol{\xi}_{k})\!,\!
\end{align}\end{small}
and
\begin{align}\label{eq_ode_trace}
&\tilde{\boldsymbol{\psi}}^n(t_{k+l}) = \tilde{\boldsymbol{\psi}}^k(t_k) + \int_{t_k}^{t_{k+l}} \mathbf{f}_{ \boldsymbol{\psi}}\left(\tilde{\boldsymbol{\psi}}^k(v)\right) dv \nonumber\\
= &~\tilde{\boldsymbol{\psi}}^k(t_k) + \sum_{i=1}^{l} b_{S,k+i} \mathbf{f}_{ \boldsymbol{\psi}}\left(\tilde{\boldsymbol{\psi}}^k(t_{k+i-1})\right) \\
&+ \int_{t_k}^{t_{k+l}} \left[\mathbf{f}_{ \boldsymbol{\psi}}\left(\tilde{\boldsymbol{\psi}}^k(v)\right) - \mathbf{f}_{ \boldsymbol{\psi}}\left(\tilde{\boldsymbol{\psi}}^k(\underline{v})\right) \right]dv,  \nonumber
\end{align}
where $\underline{v} \overset{\Delta}{=} \max \left\{ t_k: t_k \le v, k \ge 0 \right\}$ for $v \ge 0$.

To bound $\int_{t_k}^{t_{k+l}} \left[\mathbf{f}_{ \boldsymbol{\psi}}\left(\tilde{\boldsymbol{\psi}}^k(v)\right) - \mathbf{f}_{ \boldsymbol{\psi}}\left(\tilde{\boldsymbol{\psi}}^k(\underline{v})\right) \right]dv$ on the RHS of (\ref{eq_ode_trace}), we obtain the Lipschitz constant of function $\mathbf{f}_{ \boldsymbol{\psi}}(\mathbf{v})$ considering the first variable $\mathbf{v}$, given by
\begin{equation}\label{eq_Lip}
L \overset{\Delta}{=} \underset{\mathbf{v}_1 \ne \mathbf{v}_2}{\sup} \frac{\left\| \mathbf{f}_{ \boldsymbol{\psi}}(\mathbf{v}_1) -  \mathbf{f}_{ \boldsymbol{\psi}}(\mathbf{v}_2) \right\|_2}{\left\| \mathbf{v}_1 - \mathbf{v}_2 \right\|_2}.
\end{equation}
Similar to \eqref{eq_ub_fx}, for any $t \ge t_k$, we can obtain that there exists a constant $0 < C_{\mathbf{f}} <+ \infty$ such that
\begin{equation}\label{eq_CT}
\begin{aligned}
\left\| \mathbf{f}_{ \boldsymbol{\psi}}\left(\tilde{\boldsymbol{\psi}}^k(t)\right) \right\|_2 \le C_{\mathbf{f}}.
\end{aligned}
\end{equation}
Hence, we have
\begin{align}\label{eq_int}
 &\left\| \int_{t_k}^{t_{k+m}} \left[\mathbf{f}_{ \boldsymbol{\psi}}\left(\tilde{\boldsymbol{\psi}}^k(v)\right) - \mathbf{f}_{ \boldsymbol{\psi}}\left(\tilde{\boldsymbol{\psi}}^k(\underline{v})\right) \right]dv \right\|_2 \nonumber\\\le & \int_{t_k}^{t_{k+l}} \left\| \mathbf{f}_{ \boldsymbol{\psi}}\left(\tilde{\boldsymbol{\psi}}^k(v)\right) - \mathbf{f}_{ \boldsymbol{\psi}}\left(\tilde{\boldsymbol{\psi}}^k(\underline{v})\right) \right\|_2 dv \nonumber\\
\overset{(a)}{\le} & \int_{t_k}^{t_{k+l}} L \left\| \tilde{\boldsymbol{\psi}}^k(v) - \tilde{\boldsymbol{\psi}}^k(\underline{v}) \right\|_2 dv \nonumber\\
\overset{(b)}{\le} & \int_{t_k}^{t_{k+l}} L \left\| \int_{\underline{v}}^{v} \mathbf{f}_{ \boldsymbol{\psi}}\left(\tilde{\boldsymbol{\psi}}^k(s)\right) ds \right\|_2 dv \nonumber\\
\le & \int_{t_k}^{t_{k+l}} \int_{\underline{v}}^{v} L \left\| \mathbf{f}_{ \boldsymbol{\psi}}\left(\tilde{\boldsymbol{\psi}}^k(s)\right) \right\|_2 ds dv \\
\overset{(c)}{\le} & \int_{t_k}^{t_{k+l}} \int_{\underline{v}}^{v} C_{\mathbf{f}} L ds dv =   \int_{t_k}^{t_{k+l}}C_{\mathbf{f}} L (v - \underline{v}) dv \nonumber\\
=  & \sum_{i=1}^{l} \int_{t_{k+i-1}}^{t_{k+i}}C_{\mathbf{f}} L (v - t_{k+i-1}) dv \nonumber\\
= & \sum_{i=1}^{l} \frac{C_{\mathbf{f}} L (t_{k+i} - t_{k+i-1})^2}{2}= \frac{C_{\mathbf{f}} L}{2} \sum_{i=1}^{l} b_{S,k+i}^2,\nonumber\end{align}
where Step $(a)$ is due to (\ref{eq_Lip}), Step $(b)$ is due to the definition in (\ref{eq_ODE_new}), and Step $(c)$ is due to (\ref{eq_CT}). Then, by subtracting $\tilde{\boldsymbol{\psi}}^k(t_{k+l})$ in \eqref{eq_ode_trace} from $\bar{\boldsymbol{\psi}}(t_{k+l})$  in \eqref{eq_seq_trace} and taking norms, the following inequality can be obtained from (\ref{eq_Lip}) and (\ref{eq_int}) for $k \ge 0, 1 \le l \le k_T$:
\begin{small}\begin{align}\label{eq_lock2}
 &\left\| \bar{\boldsymbol{\psi}}(t_{k+l}) - \tilde{\boldsymbol{\psi}}^k(t_{k+l})\right\|_2 \nonumber\\
\le & L \sum_{i=1}^{l} b_{S,k+i} \left\| \bar{\boldsymbol{\psi}}(t_{k+i-1}) - \tilde{\boldsymbol{\psi}}^k(t_{k+i-1}) \right\|_2 \nonumber\\
&+ \frac{C_{\mathbf{f}} L}{2} \sum_{i=1}^{l} b_{S,k+i}^2+ \left\|\boldsymbol{\xi}_{k+l} - \boldsymbol{\xi}_{k}\right\|_2 \\
\le & L \sum_{i=1}^{l} b_{S,k+i} \left\| \bar{\boldsymbol{\psi}}(t_{k+i-1}) - \tilde{\boldsymbol{\psi}}^k(t_{k+i-1}) \right\|_2 \nonumber \\
&+ \frac{C_{\mathbf{f}} L}{2} \sum_{i=1}^{k_T} b_{S,k+i}^2+ \underset{1 \le l\le k_T}{\sup}\left\|\boldsymbol{\xi}_{k+l} - \boldsymbol{\xi}_{k}\right\|_2.\nonumber
\end{align}\end{small}

Applying Lemma \ref{le_gronwall} to (\ref{eq_lock2}) and letting
\begin{align*}C = \frac{C_{\mathbf{f}} L}{2} \sum_{i=1}^{k_T} b_{S,k+i}^2+ \underset{1\le l\le k_T}{\sup}\left\|\boldsymbol{\xi}_{k+l} - \boldsymbol{\xi}_{k}\right\|_2,\end{align*}
yields
\begin{small}\begin{align}\label{eq_lock3}
&\underset{t\in\left[ t_k, t_{k+k_T} \right]}{\sup} \left\| \bar{\boldsymbol{\psi}}(t) - \tilde{\boldsymbol{\psi}}^k(t)\right\|_2\\
	 \le& C_e \!\left\{\! \frac{C_{\mathbf{f}} L}{2} \big[c(k) \!-\! c(k+k_T)\big] \right. \left. \!\!+\!\! \underset{1 \le l\le k_T}{\sup}\!\left\|\boldsymbol{\xi}_{k+l} \!-\! \boldsymbol{\xi}_{k}\right\|_2 \!\right\} \!\!+\! \frac{C_{\mathbf{f}} c_{k+1}}{2},\nonumber
\end{align}
\end{small}where $C_e \overset{\Delta}{=} e^{L (T+b_{S,1})}$, and $c(k) \overset{\Delta}{=} \sum_{i > k} b_{S,i}^2$.
Letting $k = \tilde{k}(l)$ in (\ref{eq_lock3}), we have $k + k_T = \tilde{k}(l+1)$ due to the definition of $T_{l+1} = t_{\tilde{k}(l+1)}$ in \emph{Step 2} of Appendix \ref{proof_Converge to real beam direction} and
\begin{small}\begin{align}\label{eq_lock3-2}
&\underset{t\in I_l}{\sup} \left\| \bar{\boldsymbol{\psi}}(t) - \tilde{\boldsymbol{\psi}}^{\tilde{k}(l)}(t)\right\|_2\\ \le &C_e \!\left\{\! \frac{C_{\mathbf{f}} L}{2}\! \big[ c(\tilde{k}(l)) \!-\! c(\tilde{k}(l+1)) \big] \!+\! \underset{\tilde{k}(l) \le p \le \tilde{k}(l+1)}{\sup}\!\left\|\boldsymbol{\xi}_{p} \!-\! \boldsymbol{\xi}_{\tilde{k}(l)}\!\right\|_2 \!\right\} \nonumber\\
+& \frac{C_{\mathbf{f}} b_{S,\tilde{k}(l)+1}}{2}.\nonumber
\end{align}\end{small}

Suppose that the step size $\{b_{S,k}: k > 0\}$ satisfies
\begin{equation}\label{eq_lock_constr}
C_e \frac{C_{\mathbf{f}} L}{2} \big[c(\tilde{k}(l)) - c(\tilde{k}(l+1))\big] +  \frac{C_{\mathbf{f}} b_{S,\tilde{k}(l)+1}}{2} < \frac{\delta}{2},
\end{equation}
for $l \ge 0$. Given $\underset{t\in I_l}{\sup} \left\| \bar{\textbf{x}}(t) - \tilde{\textbf{x}}^{\tilde{k}(l)}(t)\right\| \!>\! \delta$, we can obtain from \eqref{eq_lock3-2} and \eqref{eq_lock_constr} that
\begin{equation*}
\begin{aligned}
&~\underset{\tilde{k}(l)\le p \le \tilde{k}(l+1)}{\sup}\left\|\boldsymbol{\xi}_{p} - \boldsymbol{\xi}_{\tilde{k}(l)}\right\|_2 \\
\ge &~\frac{1}{C_e} \left( \underset{t\in I_l}{\sup} \left\| \bar{\boldsymbol{\psi}}(t) - \tilde{\boldsymbol{\psi}}^{\tilde{k}(l)}(t)\right\|_2 \right.\\
&\left.- \frac{C_{\mathbf{f}} L}{2} \big[ c(\tilde{k}(l)) \right. \left.  - c(\tilde{k}(l+1)) \big] - \frac{C_{\mathbf{f}} b_{S,\tilde{k}(l)+1}}{2}\right) \\
> &~\frac{1}{C_e}\left( \underset{t\in I_l}{\sup} \left| \bar{\textbf{x}}(t) - \tilde{\textbf{x}}^{\tilde{k}(l)}(t)\right| - \frac{\delta}{2} \right) \\
> &~\frac{\delta}{2C_e}.
\end{aligned}
\end{equation*}
Then, we get
\begin{equation}\label{eq_lock4}
\begin{aligned}
&~P\left( \left. \underset{t\in I_l}{\sup} \left\| \bar{\textbf{x}}(t) - \tilde{\textbf{x}}^{\tilde{k}(l)}(t)\right\|> \delta \right|\right.\\ 
&\left.\quad\quad\underset{t\in I_i}{\sup} \left\| \bar{\textbf{x}}(t) - \tilde{\textbf{x}}^{\tilde{k}(i)}(t)\right\| \le \delta, 0 \le i < l \right) \\
{\le} & P\left( \left. \underset{\tilde{k}(l)\le p \le \tilde{k}(l+1)}{\sup}\left\|\boldsymbol{\xi}_{p} - \boldsymbol{\xi}_{\tilde{k}(l)}\right\|_2 > \frac{\delta}{2C_e} \right|\right.\\
 &\left.\quad\quad\underset{t\in I_i}{\sup} \left\|
 \bar{\textbf{x}}(t) - \tilde{\textbf{x}}^{\tilde{k}(i)}(t)\right\| \le \delta, 0 \le i < l \right) \\
\overset{(d)}{=} &~P\left( \underset{\tilde{k}(l)\le p \le \tilde{k}(l+1)}{\sup}\left\|\boldsymbol{\xi}_{p} - \boldsymbol{\xi}_{\tilde{k}(l)}\right\|_2 > \frac{\delta}{2C_e} \right),
\end{aligned}
\end{equation}
where Step $(d)$ is due to the independence of noise, i.e., $ \boldsymbol{\xi}_{p}-\boldsymbol{\xi}_{\tilde{k}(l)} , \tilde{k}(l) \le p \le \tilde{k}(l+1)$ are independent of $\hat{\textbf{x}}_k, 0 \le k \le \tilde{k}(l)$.

%

The lower bound of the probability that the sequence $\{\hat{\textbf{x}}_k: k \ge 0\}$ remains in the invariant set $\mathcal{I}$ is given by
\begin{align}\label{eq_p_invariant}
 &P\left( \hat{\textbf{x}}_k \in \mathcal{I}, \forall k \ge 0 \right) \nonumber\\
 \overset{(e)}{\ge} & P\left( \underset{t\in I_l}{\sup} \left\| \bar{\textbf{x}}(t) - \tilde{\textbf{x}}^{\tilde{k}(l)}(t)\right\| \le \delta, \forall l \ge 0 \right)  \nonumber\\
\overset{(f)}{\ge} & 1 - \sum_{l\ge 0} P\left(  \underset{t\in I_l}{\sup} \left\| \bar{\textbf{x}}(t) - \tilde{\textbf{x}}^{\tilde{k}(l)}(t)\right\| > \delta \bigg|\right.\\
&\left.\quad\underset{t\in I_i}{\sup} \left\| \bar{\textbf{x}}(t) - \tilde{\textbf{x}}^{\tilde{k}(i)}(t)\right\| \le \delta, 0 \le i < l \right)  \nonumber\\
\overset{(g)}{\ge} & 1 - \sum_{l\ge 0} P\Bigg( \underset{\tilde{k}(l)\le p \le \tilde{k}(l+1)}{\sup}\left\|\boldsymbol{\xi}_{p} - \boldsymbol{\xi}_{\tilde{k}(l)}\right\|_2 > \frac{\delta}{2C_e} \Bigg),\nonumber
\end{align}
where Step $(e)$ is due to Lemma \ref{le_sufficient}, Step $(f)$ is due to Lemma 4.2 in \cite{borkar2008stochastic}, and Step $(g)$ is due to \eqref{eq_lock4}. Let $\left\|\cdot\right\|_{\infty}$ denote the max-norm, i.e., $\left\|\mathbf{u}\right\|_{\infty} = \max_{l} |[\mathbf{u}]_l|$. Note that for $\mathbf{u} \in \mathbb{R}^{D}$, $\left\|\mathbf{u}\right\|_{2} \le \sqrt{D} \left\|\mathbf{u}\right\|_{\infty}$. Hence we have
\begin{align}\label{eq_lock6-0}
&~ P\left( \underset{\tilde{k}(l)\le p \le \tilde{k}(l+1)}{\sup}\left\|\boldsymbol{\xi}_{p} - \boldsymbol{\xi}_{\tilde{k}(l)}\right\|_2 > \frac{\delta}{2C_e} \right) \nonumber\\
\le &~ P\left( \underset{\tilde{k}(l)\le p \le \tilde{k}(l+1)}{\sup} \left\|\boldsymbol{\xi}_{p} - \boldsymbol{\xi}_{\tilde{k}(l)}\right\|_{\infty} > \frac{\delta}{4C_e} \right)  \\
= &~ P\left( \underset{\tilde{k}(l)\le p \le \tilde{k}(l+1)}{\sup} \max_{1 \le j \le 4} \left|\big[\boldsymbol{\xi}_{p}\big]_j - \big[\boldsymbol{\xi}_{\tilde{k}(l)}\big]_j \right| > \frac{\delta}{4C_e} \right) \nonumber\\
= &~ P\left( \max_{1 \le j \le 4} \underset{\tilde{k}(l)\le p \le \tilde{k}(l+1)}{\sup} \left|\big[\boldsymbol{\xi}_{p}\big]_j - \big[\boldsymbol{\xi}_{\tilde{k}(l)}\big]_j \right| > \frac{\delta}{4C_e} \right) \nonumber\\
\le &~ \sum_{j = 1}^4 P\left( \underset{\tilde{k}(l)\le p \le \tilde{k}(l+1)}{\sup} \left|\big[\boldsymbol{\xi}_{p}\big]_j - \big[\boldsymbol{\xi}_{\tilde{k}(l)}\big]_j \right| > \frac{\delta}{4C_e} \right). \nonumber
\end{align}

With the increasing $\sigma$-fields $\{\!\mathcal{G}_k\!:\!k\!\ge\!0\!\}$ defined in Appendix \ref{proof_Converge to unique stable point}, we have for $k \ge 0$,
\begin{itemize}
	\item[1)] $\boldsymbol{\xi}_k \!=\! \sum_{l=1}^{k} b_{S,l} \hat{\mathbf{z}}_{l} \sim \mathcal{N}(0, \sum_{l=1}^k b_{S,k}^2 \mathbf{I}_S(\hat{\boldsymbol{\psi}}_{l\!-\!1},\!\mathbf{W}_l)^{-1})$,
	
	\item[2)] $\boldsymbol{\xi}_k$ is $\mathcal{G}_k$-measurable, i.e., $\mathbb{E} \left[ \left. \boldsymbol{\xi}_k \right| \mathcal{G}_k \right] = \boldsymbol{\xi}_k$,
	
	\item[3)] $\mathbb{E} \left[ \left\| \boldsymbol{\xi}_k \right\|^2_2 \right] = \sum_{l=1}^k b_{S,k}^2 \operatorname{tr}\left\{\mathbf{I}_S(\hat{\boldsymbol{\psi}}_{l\!-\!1},\!\mathbf{W}_l)^{-1}\right\}< +\infty$,
	
	\item[4)] $\mathbb{E} \left[ \left. \boldsymbol{\xi}_k \right| \mathcal{G}_l \right] = \boldsymbol{\xi}_l$ for all $0 \le l < k$.
\end{itemize}
Therefore, $\left[\boldsymbol{\xi}_k\right]_j, j = 1,2,3,4$ is a Gaussian martingale with respect to $\mathcal{G}_k$, and satisfies
\begin{align}\label{eq_lock6-1}
\operatorname{Var}\left[\big[\boldsymbol{\xi}_{k+l}\big]_j - \big[\boldsymbol{\xi}_{k}\big]_j\right] = &~\sum_{i = k+1}^{k+l} b_{S,i}^2 \left[\mathbf{I}_S(\hat{\boldsymbol{\psi}}_{i\!-\!1},\!\mathbf{W}_i)^{-1}\right]_{j,j} \nonumber\\
\le &~\sum_{i = k+1}^{k+l} b_{S,i}^2 \frac{C_{\mathbf{I}}\sigma_z^2}{\lvert\textbf{s}\rvert^2} \\
= &~\frac{C_{\mathbf{I}}\sigma_z^2}{\lvert\textbf{s}\rvert^2} \big[c(k) - c(k+l)\big],\nonumber
\end{align}
where $C_{\mathbf{I}} \!\overset{\Delta}{=}\! \max_{s} \max_{i \ge 1} \frac{\lvert\textbf{s}\rvert^2}{\sigma_z^2}\big[\mathbf{I}(\hat{\boldsymbol{\psi}}_{i\!-\!1},\!\mathbf{W}_i)^{-1}\big]_{j,j}$. Let $\eta \!=\! \frac{\delta}{4C_e}$, $M_i \!=\! \big[\boldsymbol{\xi}_{\tilde{k}(l)+i}\big]_j - \big[\boldsymbol{\xi}_{\tilde{k}(l)}\big]_j, j\!=\! 1, 2, 3 ,4$ and $p = {\tilde{k}(l+1) - \tilde{k}(l)}$ in Lemma \ref{le_cheb}, then from \eqref{eq_lock6-0} and \eqref{eq_lock6-1}, we can obtain
\begin{align}\label{eq_lock6}
&~ P\left( \underset{\tilde{k}(l)\le p \le \tilde{k}(l+1)}{\sup} \left|\big[\boldsymbol{\xi}_{p}\big]_j - \big[\boldsymbol{\xi}_{\tilde{k}(l)}\big]_j \right| > \frac{\delta}{4C_e} \right) \nonumber \\
\le & ~2\exp\left\{-\frac{\delta^2}{32C_e^2\operatorname{Var}\left[\big[\boldsymbol{\xi}_{\tilde{k}(l)+i}\big]_j - \big[\boldsymbol{\xi}_{\tilde{k}(l)}\big]_j\right]}\right\} \\
\le & ~2\exp\left\{-\frac{\delta^2{\lvert\textbf{s}\rvert}^2}{32C_{\mathbf{I}}C_e^2\big[c(\tilde{k}(l)) - c(\tilde{k}(l+1))\big]\sigma_z^2}\right\}.\nonumber
\end{align}
Combining \eqref{eq_p_invariant}, \eqref{eq_lock6-0} and \eqref{eq_lock6}, we have
\begin{align}\label{eq_lock7}
&P\left( \hat{\textbf{x}}_k \in \mathcal{I}, \forall k \ge 0 \right)\\
\ge& 1 -  8\sum_{l \ge 0} \exp\left\{-\frac{\delta^2\lvert\textbf{s}\rvert^2}{32C_{\mathbf{I}}C_e^2\big[c(\tilde{k}(l)) - c(\tilde{k}(l+1))\big]\sigma_z^2}\right\}. \nonumber
\end{align}

To use Lemma \ref{le_sum}, we assume that the step-size $b_{S,k}$ satisfies
\begin{equation}\label{eq_lock_constr2}
c(0) = \sum_{i > 0} b_{S,i}^2 \le \frac{\delta^2\lvert\textbf{s}\rvert^2}{32C_{\mathbf{I}}C_e^2\sigma_z^2}.
\end{equation}
Then, from Lemma \ref{le_sum}, we can obtain
\begin{equation*}
\begin{aligned}
\frac{\exp\left\{\!-\frac{\delta^2\lvert\textbf{s}\rvert^2}{32C_{\mathbf{I}}C_e^2\big[c(\tilde{k}(l)) \!-\! c(\tilde{k}(l+1))\big]\sigma_z^2}\!\right\}}{c(\tilde{k}(l)) - c(\tilde{k}(l+1))} 
\!\le\! \frac{\exp\left\{\!-\frac{\delta^2\lvert\textbf{s}\rvert^2}{32C_{\mathbf{I}}C_e^2 c(0)\sigma_z^2}\!\right\} }{c(0)}
\end{aligned}
\end{equation*}
for $c(\tilde{k}(l)) - c(\tilde{k}(l+1)) < c(\tilde{k}(l)) \le c(0)$. Hence, we have
\begin{align}\label{eq_lock7-2}
&\sum_{l\ge 0} \exp\left\{-\frac{\delta^2\lvert\textbf{s}\rvert^2}{32C_{\mathbf{I}}C_e^2\big[c(\tilde{k}(l)) - c(\tilde{k}(l+1))\big]\sigma_z^2}\right\} \\
\le &\sum_{l \ge 0} \left[ c(\tilde{k}(l)) - c(\tilde{k}(l+1))\right] \cdot \frac{\exp\left\{-\frac{\delta^2\lvert\textbf{s}\rvert^2}{32C_{\mathbf{I}}C_e^2 c(0)\sigma_z^2}\right\}}{c(0)} \nonumber \\
= &c(0)\! \cdot\! \frac{\exp\left\{\!-\frac{\delta^2\lvert\textbf{s}\rvert^2}{32C_{\mathbf{I}}C_e^2c(0)\sigma_z^2}\!\right\} }{c(0)} \!=\! \exp\left\{-\frac{\delta^2\lvert\textbf{s}\rvert^2}{32C_{\mathbf{I}}C_e^2c(0)\sigma_z^2}\right\}.\nonumber
\end{align}
As $C_e = e^{L (T+b_{S,1})}$, $c(0) = \sum_{i > 0} b_{S,i}^2$, and $b_{S,k}, T, L$ are given by \eqref{eq_stepsize}, \eqref{eq_T}, \eqref{eq_Lip} respectively, we can obtain
\begin{align}\label{eq_exponential}
\frac{\delta^2\lvert\textbf{s}\rvert^2}{32C_{\mathbf{I}}C_e^2c(0)\sigma_z^2} &=  \frac{\delta^2\lvert\textbf{s}\rvert^2}{32C_{\mathbf{I}} e^{2L (T+\frac{\epsilon_S}{K_{S,0}+1})} \sigma_z^2\sum\limits_{i \ge 1} \frac{\epsilon_S^2}{(i+K_{S,0}|)^2}}\nonumber\\
&=\frac{\delta^2}{\sum\limits_{i \ge 1} \frac{32C_{\mathbf{I}} e^{2L (T+\frac{\epsilon_S}{K_{S,0}+1})}}{(i+K_{S,0})^2}} \!\cdot\!\frac{\lvert\textbf{s}\rvert^2}{\epsilon_S^2 \sigma_z^2}.
\end{align}
In \eqref{eq_exponential}, $0 < \delta < \inf_{\textbf{v} \in \partial \mathcal{B}} \left\| \textbf{v} - \hat{\textbf{x}}_\text{b} \right\|$, (\ref{eq_lock_constr}) and (\ref{eq_lock_constr2}) should be satisfied, where a sufficiently large $K_{S,0} \ge 0$ can make both (\ref{eq_lock_constr}) and (\ref{eq_lock_constr2}) true.

To ensure that $\hat{\textbf{x}}_0 + b_{S,1}\left[\mathbf{f}_{ \boldsymbol{\psi}}\left(\hat{\boldsymbol{\psi}}_0\right)\right]_{3,4}$ does not exceed the main lobe $\mathcal{B}(\textbf{x})$, i.e., the first step-size $b_{S,1}$ satisfies
\begin{align*}&\left|\hat{x}_{0,1} + b_{S,1}\left[\mathbf{f}_{ \boldsymbol{\psi}}\left(\hat{\boldsymbol{\psi}}_0\right)\right]_3 - x_1\right| < 1\\
&\left|\hat{x}_{0,2} + b_{S,1}\left[\mathbf{f}_{ \boldsymbol{\psi}}\left(\hat{\boldsymbol{\psi}}_0\right)\right]_4 - x_2\right| < 1,
\end{align*}
we can obtain the maximum $\epsilon_S$ as follows
\begin{small}
	\begin{align}
	\epsilon_{S,\max} &= {\min}\frac{(K_{S,0}+1)}{\left|\left[\!\mathbf{f}_{ \boldsymbol{\psi}}\left(\hat{\boldsymbol{\psi}}_0\right)\right]_3\right|}\left\{1 - \lvert x_1-\hat{x}_{0,1} \rvert, 1 - \lvert x_2-\hat{x}_{0,2}\rvert \right\}\nonumber\\
	&\leq \frac{(K_{S,0}+1)}{\left|\!\left[\!\mathbf{f}_{ \boldsymbol{\psi}}\left(\hat{\boldsymbol{\psi}}_0\right)\!\right]_3\!\right|}
	 \triangleq \epsilon_b.
	\end{align}
\end{small}
Hence, from \eqref{eq_exponential}, we have
\begin{small}
\begin{align}\label{eq_exponential-2}
\frac{\delta^2\lvert\textbf{s}\rvert^2}{32C_{\mathbf{I}}C_e^2c(0)\sigma_z^2} \!\cdot\! \frac{\epsilon_S^2\sigma_z^2}{\lvert\textbf{s}\rvert^2} \! \ge\! \frac{\delta^2}{\sum\limits_{i \ge 1}\! \frac{32C_{\mathbf{I}} e^{2L (T+\frac{\epsilon_{b}}{K_{S,0}+1})}}{(i+K_{S,0})^2}} \overset{\Delta}{=} R.
\end{align}\end{small}

Combining \eqref{eq_lock7}, \eqref{eq_lock7-2} and \eqref{eq_exponential-2}, yields
\begin{equation*}
\begin{aligned}
P\left( \hat{\textbf{x}}_k \in \mathcal{I}, \forall k \ge 0 \right) \ge 1 - 8e^{-\frac{R\lvert\textbf{s}\rvert^2}{\epsilon_S^2\sigma_z^2}},
\end{aligned}
\end{equation*}
which completes the proof.

\bibliographystyle{IEEEtran}
\bibliography{IEEEabrv,reference}
\IEEEpeerreviewmaketitle

\end{document}